\pdfoutput=1

\documentclass[11pt]{article}
\usepackage{jheppub}

\usepackage{amsmath}
\usepackage{amssymb}
\usepackage{graphicx,color}

\renewcommand{\boldmath}{}
\usepackage{ifpdf}

\newcommand{\bZ}{\mathbb{Z}}
\newcommand{\bC}{\mathbb{C}}

\newcommand{\bR}{\mathbb{R}}
\newcommand{\bP}{\mathbb{P}}
\newcommand{\cA}{\mathcal{A}}
\newcommand{\cB}{\mathcal{B}}
\newcommand{\cC}{\mathcal{C}}
\newcommand{\cE}{\mathcal{E}}
\newcommand{\cF}{\mathcal{F}}

\newcommand{\cN}{\mathcal{N}}
\newcommand{\cO}{\mathcal{O}}

\newcommand{\ov}{\overline}

\newcommand{\dsz}[2]{\left\langle#1,#2\right\rangle}

\newcommand{\CP}{\ensuremath{\mathop{\null {\mathbb{P}}}}\nolimits}
\newcommand{\Kcone}{\ensuremath{\mathcal{K}}}
\newcommand{\dP}{\ensuremath{{dP_0}}}
\newcommand{\MP}{\ensuremath{{M_{dP_0}}}}
\newcommand{\Y}{\ensuremath{{Y^{(3,0)}}}}
\newcommand{\M}{\ensuremath{{M_{Y^{(3,0)}}}}}
\newcommand{\dPt}{\ensuremath{{\left(dP_0\right)^3}}}
\newcommand{\MPt}{\ensuremath{{M_{\left(dP_0\right)^3}}}}

\newcommand{\Osheaf}{\ensuremath{\mathcal{O}}}
\DeclareMathOperator{\Cl}{Cl}
\DeclareMathOperator{\conv}{conv}
\DeclareMathOperator{\Span}{span}
\DeclareMathOperator{\Vol}{Vol}

\DeclareMathOperator{\Hom}{Hom}
\DeclareMathOperator{\ch}{ch}
\DeclareMathOperator{\Td}{Td}
\DeclareMathOperator{\Ext}{Ext}

\def\Label#1{\label{#1}%
  \smash{\hbox to0pt{\raise1ex\hbox{\tiny[#1]}\hss}}}
\def\noLabels{\let\Label=\label}
\def\nobbibitem{\let\bbibitem=\bibitem}
 \def\noBibitem{\let\Bibitem=\bibitem}

\title{Global embeddings for branes at toric singularities}

\author[a]{Vijay Balasubramanian,}
\author[b]{Per Berglund,}
\author[c]{Volker Braun,}
\author[d]{and I\~naki Garc\'ia-Etxebarria}
\affiliation[a]{Department of Physics and Astronomy, University of Pennsylvania,\\
  Philadelphia, PA 19104-6396, USA}
\affiliation[b]{Department of Physics, University of New Hampshire,\\
  Durham, NH 03824, USA}
 \affiliation[c]{Dublin Institute for Advanced Studies, Dublin,
   Ireland}
\affiliation[d]{Theory Group, Physics Department, CERN,\\CH-1211,
  Geneva 23, Switzerland}
\emailAdd{vijay@physics.upenn.edu}
\emailAdd{per.berglund@unh.edu}
\emailAdd{vbraun@dias.ie}
\emailAdd{inaki@mail.cern.ch}

\abstract{We describe how local toric singularities, including the
  Toric Lego construction, can be embedded in compact Calabi-Yau
  manifolds. We study in detail the addition of D-branes, including
  non-compact flavor branes as typically used in semi-realistic model
  building. The global geometry provides constraints on allowable
  local models. As an illustration of our discussion we focus on $D3$
  and $D7$-branes on (the partially resolved) \dPt\ singularity, its
  embedding in a specific Calabi-Yau manifold as a hypersurface in a
  toric variety, the related type IIB orientifold compactification, as
  well as the corresponding F-theory uplift. Our techniques generalize
  naturally to complete intersections, and to a large class of
  F-theory backgrounds with singularities.}

\begin{document}
\noLabels 
\nobbibitem 
\noBibitem 

\makeatletter
\let\old@fpheader\@fpheader
\renewcommand{\@fpheader}{\old@fpheader\hfill CERN-PH-TH/2012-019}
\makeatother

\maketitle

\newpage

\section{Introduction}

A convenient bottom-up approach to model building in string theory is
to consider D-branes placed at local geometric singularities in a
compactification manifold
\cite{Douglas:1996sw,Klebanov:1998hh,Morrison:1998cs,Aldazabal:2000sa,Berenstein:2001nk,Verlinde:2005jr}.
The local properties of the singularities, together with the D-branes,
determine the matter and gauge content of the theory. In a previous
paper \cite{Balasubramanian:2009tv} (see also
\cite{GarciaEtxebarria:2006aq,GarciaEtxebarria:2006rw}) a construction
called Toric Lego was described, in which local toric singularities
supporting different desirable sectors of a low-energy field theory
(e.g. a standard model sector, a supersymmetry breaking sector, and a
dark matter sector) can be glued together geometrically to construct a
larger parent singularity, resulting in a consistent, anomaly free,
field theory encompassing the different sectors.  The sizes of
collapsing cycles connecting the different daughter singularities fix
the masses of messengers communicating between the different
low-energy field theory sectors.  This sort of modular model building
is especially convenient because it makes it possible to separately
engineer field theories realizing specific useful properties, which
can then be glued together to form a complete model.

Any local construction of this kind faces a basic question --- can the
desired local singularity be embedded consistently in a globally
well-defined compact Calabi-Yau manifold? In particular, global
tadpole cancellation conditions in general require the
introduction of orientifold planes with negative D-brane charges, 
which may be incompatible with the local structure and
embedding we have chosen, as we will see in some later examples.

In this paper we make progress towards an algorithm for systematically
producing tadpole-free global embeddings of toric singularities in
type IIB compactifications, e.g. those in the Toric Lego models
\cite{Balasubramanian:2009tv}.  The natural place to look for such
global realizations is in terms of Calabi-Yau manifolds given as
hypersurfaces in toric varieties. This task is made easier thanks to
Kreuzer and Skarke's classification of toric varieties in terms of
reflexive polyhedra in four dimensions~\cite{Kreuzer:2000xy}.  The
class of global models that we study can all be described as follows:
consider a Calabi-Yau manifold, $M$, described in terms of a
hypersurface constraint in a four dimensional toric variety
$\cA_\nabla$ obtained from the four dimensional polytope $\nabla$. The
global embedding of the local toric singularity is then obtained by
determining whether one of the three dimensional cones obtained in a
given (fine) triangulation of $\nabla$ is the cone over the two
dimensional toric diagram of the local singularity.

Failure to find a global realization through our procedure does not
imply that the local singularity can have no globally well-defined
embedding --- it simply means that there is no embedding within the
class of models considered.  For example, there may be an embedding
into a Calabi-Yau constructed as a complete intersection within a
higher dimensional toric variety. This class of embeddings can also be
studied using a natural extension of our methods -- see the discussion
in section~\ref{sec:conclusions-CI}.

As mentioned above, the introduction of D-branes in the global
realization of the toric singularities means that we naturally have to
include orientifold planes as well. Since we are mainly interested in
gauge theories with $U(N)$ factors only, we focus on $\bZ_2$
permutation involutions in which pairs of branes at singularities are
exchanged. These type IIB compactifications are then uplifted to
(singular) Calabi-Yau fourfolds in F-theory. One can in fact avoid the
step of constructing a IIB orientifold background, and directly
construct an F-theory compactification. We comment further on this
possibility in section~\ref{sec:conclusions-F}.

\medskip

Naturally, there are other important constraints that a realistic
global model must satisfy beyond tadpole cancellation, such as moduli
stabilization at the desired values for the local model, and realistic
supersymmetry breaking at a local minimum of the potential. In this
paper we content ourselves with providing a way to find large classes
of embeddings which satisfy $D7$ tadpole cancellation, with the
expectation that in the class of global models that we find, some
models also have working moduli stabilization and supersymmetry
breaking.

\subsubsection*{Relation to previous work}
There are various approaches to the problem of bottom-up model
building in the literature. In the following we quickly review the
main similarities and differences of the existing constructions and our
approach.

A recent class of F-theory models
\cite{Donagi:2008ca,Beasley:2008dc,Hayashi:2008ba,Beasley:2008kw,Donagi:2008kj}
embeds a (typically $SU(5)$) GUT brane wrapping a small but not
collapsed cycle into F-theory. The gauge dynamics comes from the
$SU(5)$ stack, and the breaking into the standard model comes from
fluxes living on the brane. In the quiver models we study, on the
other hand, the branes wrap zero size cycles, and the breaking into
factors happens due to $\alpha'$ corrections, which modify the
stability conditions of the branes from those at large volume. Thus, an
essential distinction is that the models in
\cite{Donagi:2008ca,Beasley:2008dc,Hayashi:2008ba,Beasley:2008kw,Donagi:2008kj}
live in a regime where $\alpha'$ corrections can be ignored, while
these $\alpha'$ corrections are essential for us. A related important
difference with the models that we analyze in this paper is that
generally one does not want the orientifold to intersect the quiver
locus, while this is unavoidable (and desirable) in realistic F-theory
GUT models. Similar comments apply to large volume GUTs in IIB, as in
\cite{Blumenhagen:2008zz}.

Previous works \cite{Diaconescu:2005pc,Buican:2006sn} deal with the
IIB/F-theory embedding of $dP_k$ singularities with $k>3$, which are
non-toric. In addition to the toric vs. non-toric distinction, a more
important difference with these studies is whether one looks for the
singular locus in K\"ahler or complex structure moduli space. We
consider singularities in the K\"ahler moduli space of the ambient
space, which are then inherited by the Calabi-Yau hypersurface. It is
certainly possible, and necessary in the case of non-toric
singularities, to obtain the singularity from a degeneration of the
Calabi-Yau hypersurface, appearing at particular loci in the complex
structure moduli space of the Calabi-Yau. One advantage of our choice
is that we gain a way of formulating the search in purely
combinatorial terms. Thus, the analysis can be performed in a
computer, and allows us to find a plethora of possible embeddings.

Another difference with previous studies is that we discuss in detail
flavor D7 branes. This introduces a number of complications, which we
 analyze in section~\ref{sec:branes}, but it is necessary if one
wants to embed many of the semi-realistic models in the literature
(see \cite{Dolan:2011qu} for the state of the art).

Branes at toroidal orbifolds (see \cite{Blumenhagen:2005mu} for a
review with further references), while superficially very different
from the models that we consider in this paper, can in fact often be
incorporated into our framework, as we discuss further in
section~\ref{sec:conclusions-CI}. Along the same lines, the topic of
landscape scans for realistic physics has been explored in detail in
the context of intersecting branes on toroidal orbifolds
\cite{Kumar:2005hf,Blumenhagen:2004xx,Gmeiner:2005vz,Douglas:2006xy,Gmeiner:2007zz,Gmeiner:2008xq}. These
papers focus on the open string sector, while we focus on the closed
string sector. Doing a combined search would definitely be desirable,
and ties together nicely with the question of general F-theory
embeddings discussed in section~\ref{sec:conclusions-F}.

\subsubsection*{Embedding procedure and layout of this paper}
The embedding prescription we propose proceeds in four steps:
\begin{enumerate}
\item[\bf 1.] One first finds an embedding of the singularity into a
  compact Calabi-Yau (or F-theory base, in which case one can skip step
  {\bf 3} below). Rather than searching for the singular point in the
  Calabi-Yau itself, we search for a curve of singularities in an
  ambient toric space. By dimensional counting the Calabi-Yau
  hypersurface  intersects the curve of singularities at a
  point. We describe how to do this in some particular examples in
  section~\ref{sec:globalembed}.

  In addition to just looking for particular models, we can use our
  method to get an idea of how generic it is to find semi-realistic
  singular loci in the landscape of Calabi-Yau spaces. We do such a
  scan in section~\ref{sec:stats}, focusing on the list of models
  produced by Kreuzer and Skarke \cite{Kreuzer:2000xy}.

\item[\bf 2.] Once we have the geometry, we introduce branes. The
  discussion of the local model at the quiver locus is most easily
  given in terms of quiver representations, while global
  considerations such as tadpole cancellation are most easily studied
  in terms of the algebraic geometry of sheaves. We thus need a
  dictionary between both languages. We review the known results for
  the gauge nodes in section~\ref{sec:branes}, and extend them to also
  cover flavor branes.

\item[\bf 3.] Tadpole cancellation forces us to introduce
  orientifolds. In section~\ref{sec:orientifold} we review how
  orientifolds can be introduced into our class of constructions, and
  how to construct the resulting quotient of the Calabi-Yau.

\item[\bf 4.]  The quotient constructed in the previous step can be
  used as a basis for a F-theory compactification. One advantage of
  lifting to F-theory is that this  automatically takes care of the
  7-brane tadpoles. Formulating the discussion in this way also allows
  for easy generalization to compactifications in which  the
  quiver sector is only locally weakly coupled. An important constraint is
  that the discriminant reproduces the local flavor structure close to
  the quiver theory. We illustrate how to satisfy this constraint in a
  particular example in section~\ref{sec:hybrid}.
\end{enumerate}

We summarize our results in section~\ref{sec:conclusions}, together
with some technically simple but physically interesting extensions of
our work.

\section{Global embedding of toric singularities}
\Label{sec:globalembed}

We  now describe how a given local toric singularity is
embedded in a compact Calabi-Yau manifold.  There are
constraints on which local geometries can be realized in a global
model in this way. After a brief general discussion in
section~\ref{sec:globalembed-general}, we  illustrate the method
with a few  examples. The first and simplest local geometry
is $\bC^3/\bZ_3$, and is studied in section~\ref{sec:MP}. In
section~\ref{sec:hyperconifold} we consider the embedding of the
hyperconifold\footnote{The name hyperconifold was first introduced in \cite{Davies:2009ub,Davies:2011is}.} $Y^{3,0}$, which can also be seen as two $\bC^3/\bZ_3$
singularities joined by a collapsed $\bP^1$. After these warm-up examples, in
section~\ref{sec:M30}, we analyze the main example in this paper, denoted
 \dPt. In the spirit of the Toric Lego construction
\cite{Balasubramanian:2009tv}, this geometry is obtained by joining
three copies of $\bC^3/\bZ_3$.

We assume some rudimentary knowledge of the basics of toric geometry,
good references are \cite{fulton,Hori:2003ic,Bouchard:2007ik} and the
excellent recent book \cite{CLS}. The discussion in
sections~\ref{sec:MP} to \ref{sec:M30} is technical. We encourage
those readers who are not interested in the details to
proceed directly to the discussion in section~\ref{sec:branes}, after
 a quick overview of our method in
section~\ref{sec:globalembed-general}.

\subsection{Generalities}
\Label{sec:globalembed-general}

Toric Calabi-Yau varieties are necessarily non-compact~\cite{fulton}.
Unfortunately, this makes a direct toric description of the compact
Calabi-Yau embedding impossible. Nevertheless, Calabi-Yau spaces
\emph{can} be obtained as hypersurfaces (or, more generally, complete
intersections) in ambient toric varieties (which are compact, and thus
not Calabi-Yau) associated to reflexive\footnote{In technical terms,
  we consider crepant partial resolutions of Fano toric
  varieties. This condition ensures that a generic Calabi-Yau
  threefold hypersurface is a smooth manifold.} polytopes. A famous
example is the quintic Calabi-Yau threefold as a degree 5 hypersurface
in the toric space $\bP^4$.

In this paper we focus on Calabi-Yau spaces which are constructed in
the above way. In order to make the discussion concrete, we restrict
ourselves to the list of Calabi-Yau hypersurfaces constructed by
Kreuzer and Skarke \cite{Kreuzer:2000xy}, although the discussion can
be generalized with little effort to the case of complete
intersections in toric spaces, see section~\ref{sec:conclusions-CI}.

\medskip

In order to introduce singularities, we have two possibilities:
\begin{itemize}
\item Specialize the embedding equation to be singular. For example,
  take the quintic at the conifold point. The singular loci can, in
  principle, be read off from the Picard-Fuchs equations. But actually
  doing so is computationally difficult unless the Hodge number $h^{21}$ is very
  small. Moreover, identifying the position and type of singularities
  can be a difficult computational problem.
\item Leave the embedding equation generic but make the ambient toric
  variety singular. The types of singularities are then determined
  by the singularities in the ambient space, which can be
  read off by the comparatively simple combinatorics of the associated toric fan.
\end{itemize}
Since it is computationally much easier, we  follow the second
approach in this paper and construct curves of singularities in the
ambient toric variety that then intersect the Calabi-Yau hypersurface
in points. Our general search strategy follows from the following
simple observation: consider a reflexive polytope $\nabla$ describing
the ambient toric space $\cA_\nabla$. Let us denote by $\cF_2$ the
toric diagram describing our local Calabi-Yau singularity
$X$.\footnote{We remind the reader that every toric, non-compact,
  Calabi-Yau threefold can be completely specified by a
  two-dimensional convex diagram, called the \emph{toric diagram}. We
  will see explicit examples below.} Now assume that $\nabla$ has
$\cF_2$ as one of its two-dimensional faces, i.e., the toric fan for
$X$ is included as a 3d cone in $\nabla$. This implies that there is a
patch of $\cA_\nabla$ that looks like $X\times\bC^*$. Now, if we
consider a Calabi-Yau hypersurface, due to simple dimension counting
we expect the hypersurface to intersect the curve of singularities at
a copy of $X$. Once we find  a candidate ambient space
$\cA_\nabla$ we can compute the topology of the divisors in $\cF_2$,
and make sure that they agree with what one expects from the local
structure of $X$.  Since we have a toric description of the whole
setup, this can be done straightforwardly and we  show this
explicitly in some of the examples below.  The search strategy is then
rather obvious, and easily implemented on a computer: we  go
through the 473,800,776~\cite{Kreuzer:2000xy} reflexive polytopes in
4d, and compare each of the 2d faces of each polytope with the
toric diagram of the singularity that we wish to embed.

A similar argument would apply if we search for two-dimensional faces
\emph{containing} the toric diagram for the singularity, in such a way
that the local geometry can be partially resolved to the geometry of
interest. The $(dP_0)^n$ examples with $n>1$ that we study below
provide for an illustration of this idea: upon partial resolution they
give rise to local $\bC^3/\bZ_3$ singularities. Although we will not
focus much on this possibility in this paper, it is a consistent
search strategy, and leads to interesting models.

As a technical remark, let us emphasize that since we are interested
in singular spaces the cones in our fan are not necessarily
simplicial. It is nevertheless technically easier to work with spaces
with at most orbifold singularities. Thus, in our examples we 
perform a partial resolution, and then make sure that the blown-up
cycles can be contracted to zero size by moving in K\"ahler moduli
space without making the volume of the whole Calabi-Yau space vanish.

Once we have the local geometry embedded in a compact Calabi-Yau
manifold we can turn to the open string sector and the inclusion of
D-branes. Before we do so in sections 3 and onwards, however, we
illustrate the embedding procedure in some simple examples.  These
examples are obtained by performing the scan over reflexive polytopes
mentioned above. In order to make the discussion as clear as possible
we have hand-picked some particularly simple reflexive polytopes
having the singularities that we want to analyze. We leave the
exhaustive scan over 4d reflexive polytopes to \autoref{sec:stats}.

\subsection{$\dP\to\MP$}
\Label{sec:MP}

We  start by considering the embedding of a particularly simple
example of a local geometry, namely $\bC^3/\bZ_3$ with the $\bZ_3$
orbifold action on the $\bC^3$ coordinates $(x,y,z)$ given by
\begin{align}
  (x,y,z) \to (\omega x, \omega y, \omega z)
  ,\qquad
  \omega=e^{2\pi i/3}.
\end{align}
The $\bC^3/\bZ_3$ singularity can be described as the singular limit
of the local $\bP^2$ Calabi-Yau geometry, that is, the total space of
the $\cO_{\bP^2}(-3)$ bundle. $\bP^2$ is a del Pezzo surface, known in
this context as $dP_0$, hence our notation.\footnote{Note that $dP_0$
  is also referred to as a del Pezzo surface of degree $9$. We 
  follow the physics tradition and denote by $dP_k$ a del Pezzo
  surface of degree $d=9-k$.} When the $dP_0$ surface is contracted to
zero size we obtain the $\bC^3/\bZ_3$ orbifold. We  embed this
local singularity into an elliptic fibration over $\bP^2$, which we
call $M_{dP_0}$.
\begin{figure}
  \centering
  \includegraphics[width=0.3\textwidth]{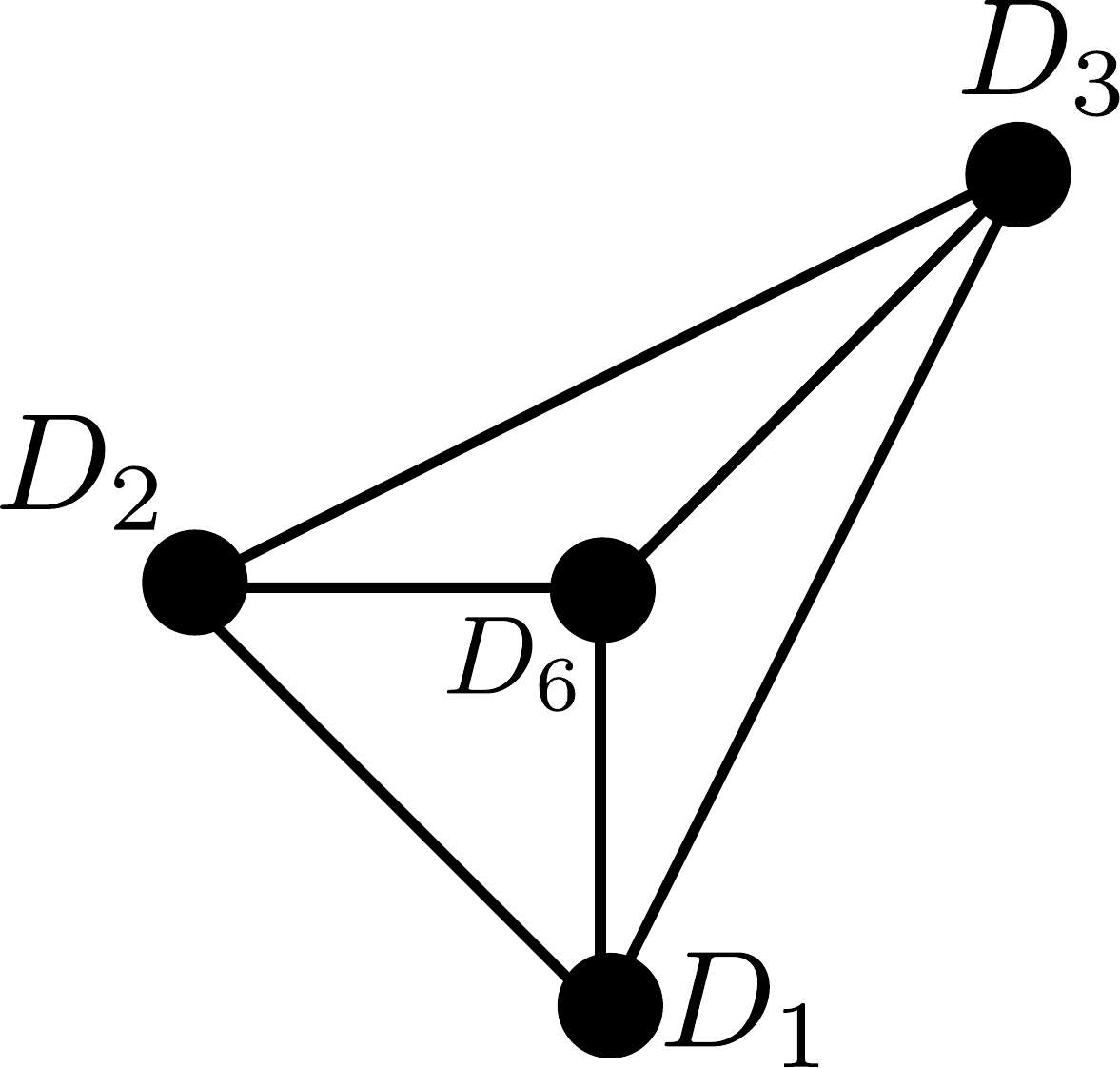}
  \caption{The polytope for $dP_0$, or equivalently, the toric diagram
    for the (resolved) $\bC^3/\bZ_3$. The particular assignment of
    labels to the vertices follows from the ones in the global
    embedding.}
  \label{fig:dP0-toric-diagram}
\end{figure}

The three-dimensional toric variety $\bC^3/\bZ_3$ has a very simple
description in terms of a toric diagram, see
\autoref{fig:dP0-toric-diagram}. By lifting the toric diagram into its
corresponding 3d cone by adding an extra coordinate, we have that the
coordinates of the vertices in a particular coordinate system are
given by $v_1=(0,-1,1)$, $v_2=(-1,0,1)$ and $v_3=(1,1,1)$. The
interior point of the diagram is also important, and we  denote it
by $v_6=(0,0,1)$. We can associate a (complex) coordinate $x_i$ to
each $v_i$, where $x_i=0$ gives the divisor $D_i$.  Hence, to each
vertex in the toric diagram we associate a 4-cycle in the non-compact
Calabi-Yau manifold, and the 2-cycles are given by the lines
connecting the vertices. In the singular $\bC^3/\bZ_3$ case we remove
the $v_6$ ray, and the singular point in $\bC^3/\bZ_3$ is represented
by the interior face of the polytope. These coordinates of the
vertices $v_i$ satisfy the linear relation
$v_1+v_2+v_3-3v_6=0$. Turning the coefficients of the relation into
charges under a $\bC^*$, we thus obtain a gauged linear $\Sigma$ model
given by the gauge symmetry
\begin{equation}
  (x_1,x_2,x_3,x_6) \to (\lambda x_1, \lambda x_2, \lambda x_3,
  \lambda^{-3} x_6)
  .
\end{equation}

We embed this local geometry in the Calabi-Yau hypersurface $M_{dP_0}$
in the resolved weighted projective space $\bP^4_{(1,1,1,6,9)}$.  This
Calabi-Yau threefold is also known as the elliptic fibration over
$\bP^2$ with Hodge numbers $(h^{11},
h^{21})=(2,272)$~\cite{Hosono:1993qy,Candelas:1994hw,Morrison:1996pp,Braun:2011ux}. To
understand how the local geometry of the resolved $\bC^3/\bZ_3$
singularity is embedded in $M_{dP_0}$, we describe the latter as a
hypersurface in a four dimensional toric variety $\cA_{dP_0}$.  The
ambient space $\cA_{dP_0}$ has an associated reflexive polytope with
coordinates
\begin{equation}
  \begin{array}{rrrrrr}
      x_1 & x_2 & x_3 & x_4 & x_5 & x_6\\
      \hline
      0 & -1 & \phantom{+}1 & 0 & 0 & \phantom{+}0 \\
      -1 & 0 & 1 & 0 & 0 & 0 \\
      2 & 2 & 2 & -1 & 0 & 2 \\
      3 & 3 & 3 & 0 & -1 & 3
    \end{array}
    \label{eq:dP0-polytope}
\end{equation}
Notice how the polytope has a face in the $(\bullet,\bullet,2,3)$
plane given by the diagram in \autoref{fig:dP0-toric-diagram}, which
we recognize as the toric diagram of $dP_0$. This is thus a candidate
embedding of our desired local singularity, as we  verify in more
detail momentarily.

In order to establish that the above Calabi-Yau hypersurface is indeed
the resolution of a compact Calabi-Yau variety with a $\bC^3/\bZ_3$
singularity, a few more details have to be checked. Since the ambient
space has to account for the local singularity of interest, there
exist in general many Calabi-Yau phases realizing the local geometry,
depending on how the singularity is resolved~\cite{Witten:1993yc,
  Aspinwall:1993yb}. Given a particular resolution of the ambient
singularity, corresponding to a fine triangulation of the associated
polytope, we compute the Mori cone, that is, the cone in
$H_2(\cA_\dP,\bZ)=H_2(M_\dP,\bZ)=\bZ^2$ spanned by holomorphic
curves. For $\cA_\dP$ there is a unique fine
triangulation,\footnote{In general the global geometry have multiple
  fine triangulations, each of which gives rise to a different
  Calabi-Yau manifold, once the hypersurface condition has been
  imposed. As we will see in sections~\ref{sec:hyperconifold} and
  \ref{sec:M30}, it is not always the case that all Calabi-Yau phases
  preserve the local structure of the partially resolved singularity
  one would like to embed.} resulting in the Mori cone given in
\autoref{table:P^2-fibration}.
\begin{table}[t]
  \centering
  \begin{tabular}{c|cccccc}
   $\ell\cdot D$& $x_1$ & $x_2$ & $x_3$ & $x_4$ & $x_5$ & $x_6$\\
    \hline
    $\ell_1$ & 1 & 1 & 1 & 0 & 0 & -3\\
    $\ell_2$ & 0 & 0 & 0 & 2 & 3 & 1
  \end{tabular}
  \caption{Mori cone for the ambient space in which $M_{dP_0}$, an
    elliptic fibration over $\bP^2$, is defined as a hypersurface. 
    Note that the Mori cones for the ambient toric variety, $\cA_\dP$,  and 
    the Calabi-Yau manifold, $M_{dP_0}$, are identical 
    in this case, unlike the  examples in sections \ref{sec:hyperconifold} and \ref{sec:M30}.}
  \label{table:P^2-fibration}
\end{table}
The fan of the ambient toric variety for this unique triangulation is
spanned by the $9$ generating cones
\begin{equation}
  \label{eq:dP0fan}
  \begin{split}
    \mathcal{F}(\cA_\dP)
    =
    \smash{\Big<}
    &
\langle x_1 x_2 x_4 x_5\rangle, 
\langle x_1 x_2 x_4 x_6\rangle, 
\langle x_1 x_2 x_5 x_6\rangle, 
\langle x_1 x_3 x_4 x_5\rangle, 
\langle x_1 x_3 x_4 x_6\rangle,
\\ &
\langle x_1 x_3 x_5 x_6\rangle, 
\langle x_2 x_3 x_4 x_5\rangle, 
\langle x_2 x_3 x_4 x_6\rangle, 
\langle x_2 x_3 x_5 x_6\rangle
    \smash{\Big>}.
  \end{split}
\end{equation}
The Stanley-Reisner ideal for this triangulation is
\begin{align}
  \label{eq:elliptic-P^2-SRI}
  SR(\cA_\dP) = \langle x_1x_2x_3, x_4x_5x_6 \rangle
  .
\end{align}
The K\"ahler cone of $\MP$, which is the dual of the Mori cone for
$\MP$, agrees with the K\"ahler cone of the ambient space $\cA_\dP$
\begin{equation}
  \begin{split}
    \Kcone(\cA_\dP) =
    \Kcone(\MP) 
    =&\; \Big\{ D \in \Cl(\MP) ~\Big|~ D\cdot \ell_i >0 ~,
    i=1,2 \Big\}
    \\
    =&\; \Span\Big\{ D_1 ,~D_6+3D_1 \Big\}\, .
  \end{split}
\end{equation}
As we will see in sections~\ref{sec:hyperconifold} and \ref{sec:M30}
it is in general not the case that the K\"ahler cones (and hence the
Mori cones) agree between the ambient space and the Calabi-Yau
hypersurface~\cite{Berglund:1995gd}, but for this simple example they do.

We can use this information to write down the hypersurface equation of
the Calabi-Yau manifold.  Since the entries for each of the Mori
generators also gives the scaling behavior of the $x_i$ homogeneous
coordinate under the corresponding $\bC^*$ in
\autoref{table:P^2-fibration}, we find that
\begin{equation}
  x_5^2=x_4^3+x_4 x_6^4 f_{12}(x_1,x_2,x_3)+x_6^6g_{18}(x_1,x_2,x_3)
  \label{eq:weierstrass}
\end{equation}
This is indeed the Weierstrass form of an elliptic fibration over
$\bP^2$. Note that the generic elliptic fiber of the ambient space
$\cA_\dP$ is the weighted projective plane $\bP^2_{(2,3,1)}$ with
coordinates $(x_4,x_5,x_6)$.  It is a general fact for such toric
fibrations that the Calabi-Yau hypersurface is automatically in
Weierstrass form, which is often used for computational
simplicity. This particular Calabi-Yau hypersurface has also been used
for Large Volume Scenario models in type IIB flux
compactifications~\cite{Balasubramanian:2005zx}.\footnote{For earlier
  studies of this Calabi-Yau manifold, see
  \cite{Hosono:1993qy,Candelas:1994hw,Morrison:1996pp,Braun:2011ux}.}

We now parametrize the K\"ahler form $\omega(t)$ by $t_1$ and $t_2\geq 0$,
that is,
\begin{equation}
  \begin{split}
    \omega =&\;
    t_1 D_1 +
    t_2 (3D_1 + D_6)    .
  \end{split}
\end{equation}
In these coordinates, the volume of $M_\dP$ is
\begin{equation}
  \smash{\int_\MP} \omega^3 = 3t_1^2t_2 + 9t_1t_2^2 + 9t_2^3\, .
\end{equation}
The volumes $\Vol(D_i) = \tfrac{1}{3} \partial_{t_i} \Vol(\MP)$ of the
dual $4$-cycles are
\begin{equation}
  \begin{split}
    \tau_1 &= 2t_1t_2 + 3t_2^2
    \\
    \tau_2 &= t_1^2 + 6t_1t_2 + 9 t_2^2\, .
     \end{split}
\end{equation}
In order to see the local geometry, we would like to identify the
contracting $dP_0$. Clearly, this is the divisor $D_6 \subset M_\dP$
associated to the additional ray required to refine the face fan (the
most coarse triangulation) to the smooth resolution. That this is
indeed a $dP_0$ surface is illustrated by the Chern numbers
\begin{equation}
  \int_{D_6} c_1^2(TD_6) = 9
  , \qquad 
  \int_{D_6} c_2(TD_6) = 3
  .
\end{equation}
Its volume is given by
\begin{align}
  \int_{D_6} \omega^2 = t_1^2\, ,
\end{align}
so we can indeed contract $D_6$ to zero volume by sending $t_1\to 0$
while the Calabi-Yau volume stays finite as long as we keep $t_2$
finite.

\subsection{The hyperconifold \boldmath $Y^{3,0}\to M_{Y^{3,0}}$}
\Label{sec:hyperconifold}

The previous example serves as a good introduction to our methods, but
it is too simple --- it is a well-known fact that one can contract a
del Pezzo surface in a Calabi-Yau manifold to zero size. In this section we
present an example with a non-del Pezzo singularity. This geometry
serves as a simple illustration of the techniques in a slightly more
involved case than the one studied in the previous section. However,
this example is not particularly useful as a perturbative IIB
orientifold background since it does not have a $\bZ_2$ permutation involution
acting on the ambient polytope. Hence, we will not try to put an open
string sector on it in coming sections.

\begin{figure}
  \centering
  \includegraphics[width=0.3\textwidth]{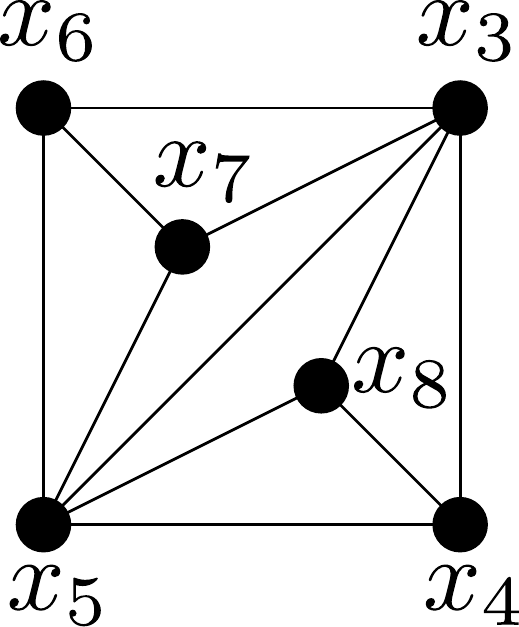}
  \caption{$Y^{3,0}$ resolved into two $\dP$'s
    separated by a $\bP^1$. We have skewed the perspective slightly to
    ease visualization. The actual position of the points 
    in the two-dimensional plane can be
    obtained by forgetting the last two coordinates in the polytope in
    eq.~\protect\eqref{eq:Y30-polytope}.
  }
  \label{fig:Y30}
\end{figure}

Let us consider a local model with two sectors, each of which is
locally $\bC^3/\bZ_3$. This local geometry, described by the toric
diagram in \autoref{fig:Y30}, is the $\bZ_3$ orbifold of the conifold,
also known as the $Y^{3,0}$ singularity.
A possible global embedding of the local geometry is given by the
reflexive polytope with vertices
\begin{equation}
  \begin{array}{rrrrrrrrr}
      x_1 & x_2 & x_3 & x_4 & x_5 & x_6 & x_7 & x_8 & x_9\\
      \hline
      0 & 0 & \phantom{+}1 & \phantom{+}0 & -1 & \phantom{+}0 &
      \phantom{+}0 & \phantom{+}0 & \phantom{+}0 \\
      0 & 0 & 1 & -1 & 0 & 2 & 1 & 0 & 1 \\
      -1 & 0 & 2 & 2 & 2 & 2 & 2 & 2 & 1 \\
      0 & -1 & 3 & 3 & 3 & 3 & 3 & 3 & 1
    \end{array}
  \label{eq:Y30-polytope}
\end{equation}
Notice that we have to include an extra point ($x_9$), along the edge
given by $x_2$ and $x_6$, in addition to the two points that were
added to the polytope that was used to describe $M_{dP_0}$.  The
resulting Calabi-Yau manifold, $M_{Y^{3,0}}$, has Hodge numbers
$(h_{11},h_{21})=(5,215)$.  The number of K\"ahler deformations can be
accounted for the in following way: the local geometry of $Y^{3,0}$
has three divisors corresponding to i) resolving $Y^{3,0}\to
dP_0\times dP_0$ and ii) resolving each individual $dP_0$. In the
global embedding, the elliptic fiber (which may or may not be realized
in a given phase) contributes one additional K\"ahler
modulus. Finally, the above mentioned divisor, $D_9$, associated to
the point $x_9$, gives the last modulus in the Calabi-Yau manifold.

We find seven different fine star triangulations of this polytope, two
of which give rise to an elliptic fibration. Four of the remaining
five triangulations reproduce the triangulation of the local geometry
given in \autoref{fig:Y30}.  We  focus on one of the latter four
triangulations. In this case, unlike the aforementioned elliptic
fibration $M_\dP\to \bP^2$, the actual Mori cone of the Calabi-Yau
hypersurface $M_\Y$ is strictly smaller than the Mori cone of the
ambient space $\cA_\Y$, because some of the curves in $\cA_\Y$ do not
lie in $\M$.

Let us show this in detail. The chosen triangulation gives rise to a
fan spanned by the following $19$ cones
\begin{equation}
  \label{eq:Y30fan}
  \begin{split}
    \mathcal{F}(\cA_\Y)
    =
    \smash{\Big<}
    &
    \langle x_1 x_2 x_3 x_4 \rangle,
    \langle x_1 x_2 x_3 x_9 \rangle,
    \langle x_1 x_2 x_4 x_5 \rangle,
    \langle x_1 x_2 x_5 x_9 \rangle,
    \langle x_1 x_3 x_4 x_8 \rangle,
    \\ &
    \langle x_1 x_3 x_5 x_7 \rangle,
    \langle x_1 x_3 x_5 x_8 \rangle,
    \langle x_1 x_3 x_6 x_7 \rangle,
    \langle x_1 x_3 x_6 x_9 \rangle,
    \langle x_1 x_4 x_5 x_8 \rangle,
    \\ &
    \langle x_1 x_5 x_6 x_7 \rangle,
    \langle x_1 x_5 x_6 x_9 \rangle,
    \langle x_2 x_3 x_4 x_8 \rangle,
    \langle x_2 x_3 x_5 x_8 \rangle,
    \langle x_2 x_3 x_5 x_9 \rangle,
    \\ &
    \langle x_2 x_4 x_5 x_8 \rangle,
    \langle x_3 x_5 x_7 x_9 \rangle,
    \langle x_3 x_6 x_7 x_9 \rangle,
    \langle x_5 x_6 x_7 x_9 \rangle
    \smash{\Big>}
  \end{split}
\end{equation}
and its Stanley-Reisner ideal is
\begin{multline}
  SR(\cA_\Y) = 
  \big<
  x_{2} x_{6}, x_{2} x_{7}, x_{4} x_{6}, x_{4} x_{7}, x_{4} x_{9},
  x_{8} x_{9}, x_{6} x_{8}, x_{7} x_{8}, x_{1} x_{2} x_{8}, 
  \\
  x_{3} x_{4} x_{5}, x_{1} x_{7} x_{9}, x_{3} x_{5} x_{6}, 
  x_{1} x_{2} x_{3} x_{5}, x_{1} x_{3} x_{5} x_{9}
  \big>
\end{multline}
reproducing the local geometry given in \autoref{fig:Y30}; for
example, $x_7=x_8=0$ has no solution. In other words, the associated
variety $\hat D_7 \cap \hat D_8 =\emptyset$ \footnote{We write $\hat
  D_i$ for the divisor $\{x_i=0\}\subset \cA_\Y$ and $D_i$ for the
  corresponding divisor $\{x_i=0\}\subset \M$.}  is empty instead of
the expected codimension two. The Mori cone for $\cA_\Y$, given the
above triangulation \ref{eq:Y30fan}, is then given by
\begin{equation}
  \renewcommand{\arraystretch}{1.3}
  \begin{array}{c|rrrrrrrrrr}
    \hat\ell \cdot \hat D 
    & x_1 & x_2 & x_3 & x_4 & x_5 & x_6 & x_7 & x_8 & x_9 & K \\ \hline
    \hat\ell_1 & 1 & 0 & 0 & 0 & 0 & -2 & 1 & 0 & 3 & -3 \\
    \hat\ell_2 & 0 & 0 & 1 & 1 & 1 & 0 & 0 & -3 & 0 & 0 \\
    \hat\ell_3 & 0 & 0 & 1 & 0 & 1 & 1 & -3 & 0 & 0 & 0 \\
    \hat\ell_4 & 0 & 1 & -1 & 0 & -1 & 0 & 3 & 0 & -2 & 0 \\
    \hat\ell_5 & 0 & -1 & -2 & 0 & -2 & 0 & 0 & 3 & 2 & 0 \\
  \end{array}
  \label{eq:mori-ambient-M_Y30}
\end{equation}
The K\"ahler cone of $\cA_\Y$, which is the dual of the Mori cone, is
then given by
\begin{equation}
  \begin{split}
    \Kcone(\cA_\Y) =&\; \Big\{ \hat D \in \Cl(\cA_\Y) ~\Big|~ \hat
    D\cdot \hat\ell_i >0 ~, i=1,\dots,5 \Big\}
    \\
    =&\; \Span\Big\{ \hat D_1 ,~ \hat D_4 ,~ 2 \hat D_1 + \hat D_6 ,~
    3 \hat D_4 + \hat D_8 ,~ 3 \hat D_2 + 3 \hat D_4 + \hat D_8
    \Big\}\, .
  \end{split}
\end{equation}

There are a few intersections of coordinate hyperplanes that do intersect
in the ambient space, but not on the Calabi-Yau hypersurface \M. These
are the surfaces $S\in \cA_\Y$ with 
\begin{equation}
  S \in \Big\{ 
  \hat D_1\cap \hat D_7,
  \hat D_1\cap \hat D_8,
  \hat D_2\cap \hat D_8,
  \hat D_7\cap \hat D_9
  \Big\}\, .
\end{equation}
Furthermore, the following intersections $\M \cap C$ with curves $C\in
\cA_\Y$ are empty:
\begin{equation}
  \begin{split}
    C \in \Big\{ & \hat D_1\cap\hat D_3\cap \hat D_5, \hat D_1\cap
    \hat D_3 \cap \hat D_9, \hat D_1 \cap \hat D_5 \cap \hat D_9, \hat
    D_2 \cap\hat D_3\cap \hat D_5, \hat D_3\cap \hat D_5 \cap \hat D_9
    \Big\}
  \end{split}\, .
\end{equation}
From \eqref{eq:mori-ambient-M_Y30} it follows that neither
$\hat\ell_4$ nor $\hat\ell_5$ can be Mori cone generators of $\M$,
because they are contained in $\hat D_3\cap \hat D_5 \cap \hat D_9$
and $ \hat D_2 \cap\hat D_3\cap \hat D_5$,
respectively. Systematically eliminating the curves that do not
intersect the Calabi-Yau threefold $\M$ leaves us with the Mori cone
for $\M$,
\begin{equation}
  \renewcommand{\arraystretch}{1.3}
  \begin{array}{c|rrrrrrrrrr}
    \ell \cdot D 
    & x_1 & x_2 & x_3 & x_4 & x_5 & x_6 & x_7 & x_8 & x_9 & K \\ \hline
    \ell_1 = \hat\ell_1 & 1 & 0 & 0 & 0 & 0 & -2 & 1 & 0 & 3 & -3 \\
    \ell_2 = \hat\ell_2 & 0 & 0 & 1 & 1 & 1 & 0 & 0 & -3 & 0 & 0 \\
    \ell_3 = \hat\ell_3 & 0 & 0 & 1 & 0 & 1 & 1 & -3 & 0 & 0 & 0 \\
    \ell_4 = \hat\ell_3 + \hat\ell_4
    & 0 & 1 & 0 & 0 & 0 & 1 & 0 & 0 & -2 & 0 \\
    \ell_5 = \tfrac{1}{3}(\hat\ell_4+\hat\ell_5)
    & 0 & 0 & -1 & 0 & -1 & 0 & 1 & 1 & 0 & 0 \\
  \end{array}
\end{equation}

We next identify the curves generating the Mori cone
(corresponding to facets of the K\"ahler cone) of $\M$. We  do this
by comparing their Chow cycle with the Chow 1-cycles of the form
$D_i\cap D_j\cap \M$. One finds that
\begin{equation}
  \begin{split}
    \ell_1 =&\; D_3\cap D_6 \cap \M = D_5\cap D_6 \cap \M
    \\
    \ell_2 =&\; D_4\cap D_8 \cap \M
    \\
    \ell_3 =&\; D_6\cap D_7 \cap \M
    \\
    \ell_4 =&\; \tfrac{1}{3}\big( D_3\cap D_9 \cap \M \big) =
    \tfrac{1}{3}\big( D_5 \cap D_9 \cap \M \big)
    \\
    \ell_5 =&\; D_3 \cap D_5 \cap \M\, .
  \end{split}
\end{equation}

From the Mori cone, we construct the (dual) K\"ahler cone by finding
linear combinations of the generators $D_1,\ldots, D_9$, such that in
the coordinates given by the corresponding columns, we get the five
standard basis vectors of $\bZ^5$. One particular choice allows us to
parametrize the K\"ahler form $\omega(t)$ by $t_i=\int_{\ell_i}\omega
\geq 0$, as
\begin{equation}
  \omega =
  t_1 D_1 +
  t_2 D_4 +
  t_3 \left( 2 D_1 - D_2 + D_6 \right) +
  t_4 D_2 +
  t_5 \left( 3 D_4 + D_8\right)
  \, .
\end{equation}
Note in particular that $t_5$ measures the volume of the $\bP^1$ that
resolves the $Y^{3,0}$ to $dP^0\times dP^0$.  In these coordinates,
the volume of $\M$ is
\begin{multline}
  \smash{\int_\M} \omega^3 = 
  54 t_1^3 + 36 t_1^2 t_2 + 90 t_1^2 t_3 + 243 t_1^2 t_4 + 
  108 t_1^2 t_5 + 6 t_1 t_2^2 + 36 t_1 t_2 t_3 + 
  \\
  108 t_1 t_2 t_4 + 
  36 t_1 t_2 t_5 + 48 t_1 t_3^2 + 270 t_1 t_3 t_4 + 108 t_1 t_3 t_5 + 
  351 t_1 t_4^2 + 
  \\
  324 t_1 t_4 t_5 + 
  54 t_1 t_5^2 + 3 t_2^2 t_3 + 
  9 t_2^2 t_4 + 3 t_2^2 t_5 + 9 t_2 t_3^2 + 54 t_2 t_3 t_4 + 
  \\
  18 t_2 t_3 t_5 + 81 t_2 t_4^2 + 54 t_2 t_4 t_5 + 9 t_2 t_5^2 + 
  8 t_3^3 + 72 t_3^2 t_4 + 
  27 t_3^2 t_5 + 
  \\
  198 t_3 t_4^2 + 
  162 t_3 t_4 t_5 + 27 t_3 t_5^2 + 168 t_4^3 + 
  243 t_4^2 t_5 + 81 t_4 t_5^2 + 9 t_5^3
  \, .
\end{multline}
The volumes $\Vol(D_i) = \tfrac{1}{3} \partial_{t_i} \Vol(\Y)$ of the
dual $4$-cycles are
\begin{equation}
  \begin{split}
    \tau_1 =&\; 
    54 t_1^2 + 24 t_1 t_2 + 60 t_1 t_3 + 162 t_1 t_4 + 
    72 t_1 t_5 + 2 t_2^2 + 12 t_2 t_3 + 36 t_2 t_4 + 
    \\ &\;
    12 t_2 t_5 + 16 t_3^2 + 90 t_3 t_4 + 
    36 t_3 t_5 + 117 t_4^2 + 108 t_4 t_5 + 18 t_5^2
    \\
    \tau_2 =&\; 
    (2 t_1 + t_3 + 3 t_4 + t_5)   
    (6 t_1 + 2 t_2 + 3 t_3 + 9 t_4 + 3 t_5)
    \\
    \tau_3 =&\;
    30 t_1^2 + 12 t_1 t_2 + 32 t_1 t_3 + 90 t_1 t_4 + 
    36 t_1 t_5 + t_2^2 + 6 t_2 t_3 + 18 t_2 t_4 + 
    \\ &\;
    6 t_2 t_5 + 8 t_3^2 + 48 t_3 t_4 + 18 t_3 t_5 + 
    66 t_4^2 + 54 t_4 t_5 + 9 t_5^2
    \\
    \tau_4 =&\;
    3 (3 t_1 + t_2 + 2 t_3 + 4 t_4 + 3 t_5)   
    (9 t_1 + t_2 + 4 t_3 + 14 t_4 + 3 t_5)
    \\
    \tau_5 =&\;
    (6 t_1 + t_2 + 3 t_3 + 9 t_4 + 3 t_5)^2\, .
  \end{split}
\end{equation}
We note that all of the nef divisors corresponding to the rays of the
K\"ahler cone have nonzero volume away from the origin of the K\"ahler
cone. In addition, by construction, the divisors $D_7$ and $D_8$ do
shrink on two distinct walls of the K\"ahler cone. Their volumes are
\begin{equation}
  \Vol\big( D_7 \big) = 
  \int_{D_7} \omega^2 = 
  t_3^2
  ,\qquad
  \Vol\big( D_8 \big) = 
  \int_{D_8} \omega^2 = 
  t_2^2
  \, .
\end{equation}
Hence the volumes of the divisors $D_7$ and $D_8$, respectively,
vanish when we shrink the two-cycles dual to $D_2$ and $D_3$, and the
two resulting singularities collide when we set $t_5=0$. Nevertheless,
the volume of the total Calabi-Yau  stays finite as long as $t_1>0$.

\paragraph{The Two Shrinking Divisors.}

We want to better understand the geometry of the divisors $D_7$ and
$D_8$ on $\M$. Since $\M\subset \cA_\Y$ is cut out by a section
of the anticanonical bundle, we have to 
\begin{enumerate}
\item Identify the corresponding divisors $\hat D_7$, $\hat D_8
  \subset \cA_\Y$ of the ambient space as 3-dimensional toric
  varieties.
\item Pull back the anticanonical bundle, $K_\nabla$, on $\cA_\Y$ to $\hat D_7$,
  $\hat D_8$.
\item Identify the divisors $D_7$, $D_8 \subset \M$ as a generic
  zero section of the pulled-back anticanonical bundle.
\end{enumerate}
The divisor $x_8=0$ is particularly simple to describe, so we 
start with this case. From the fan eq.~\eqref{eq:Y30fan} one can
easily identify $\hat D_8 = \CP^2\times\CP^1 \subset \cA_\Y$ on the
ambient space. In order to pull-back the anticanonical bundle, we note
that 
\begin{equation}
  \label{eq:aKnabla}
  -K_\nabla = 
  \sum_{i=1}^9 \Osheaf\big(\cA_\Y, \hat D_i \big)
  \sim
  2 \Osheaf\big(\cA_\Y, \hat D_2 \big) + 
  \Osheaf\big(\cA_\Y, \hat D_9 \big) 
\end{equation}
Using this choice of linear equivalence class, one easily finds the
pull-back
\begin{math}
  (-K_\nabla)\big|_{\hat D_8} 
  =
  \Osheaf_{\CP^2\times\CP^1}(0,1)
  .
\end{math}
A generic section of $\Osheaf_{\CP^1}(1)$ is a single point, therefore 
\begin{equation}
  D_8 = 
  \Big\{
  z=0
  ~\Big|~
  z \in H^0\Big(\hat D_8, (-K_\nabla)\big|_{\hat D_8} \Big)
  \Big\}
  =
  \CP^2\, .
\end{equation}

Identifying the other divisor $x_7=0$ is somewhat more technical
because $\hat D_7$ is not just a product of projective
spaces. However, as with any toric divisor, it is again a toric
variety. Here, it turns out to be the variety corresponding to the
face  of the lattice polytope
\begin{equation}
  \nabla_7 = 
  \conv\big\{
  y_1,~\dots,~y_5
  \big\}
  =
  \conv\big\{
  ( 0,  0,  1),~
  ( 0,  1,  0),~
  (-3,  2,  0),~
  ( 3, -2, -1),~
  (-2,  1,  0)
  \big\}
  .
\end{equation}
$\hat D_7$ is a toric fibration over $\CP^1$ with generic fiber
$\CP^2$, but its total space is a singular variety. Using
eq.~\eqref{eq:aKnabla}, one then identifies\footnote{We denote the
  divisors $y_i=0$ in $\hat D_7$ by $\tilde D_i$.}
\begin{equation}
  (-K_\nabla)\big|_{\hat D_7} 
  =
  \Osheaf\big( \tilde D_5)\, .
\end{equation}
It is now easy to identify the toric divisor $V(y_5)$. Again, one
finds a smooth projective plane, that is
\begin{equation}
  D_7 = 
  \hat D_7 \cap \Y 
  =
\tilde D_5
  = 
  \CP^2
  .
\end{equation}

\subsection{$\dPt\to \MPt$}
\Label{sec:M30}

\begin{figure}
  \begin{center}
    \includegraphics[width=0.3\textwidth]{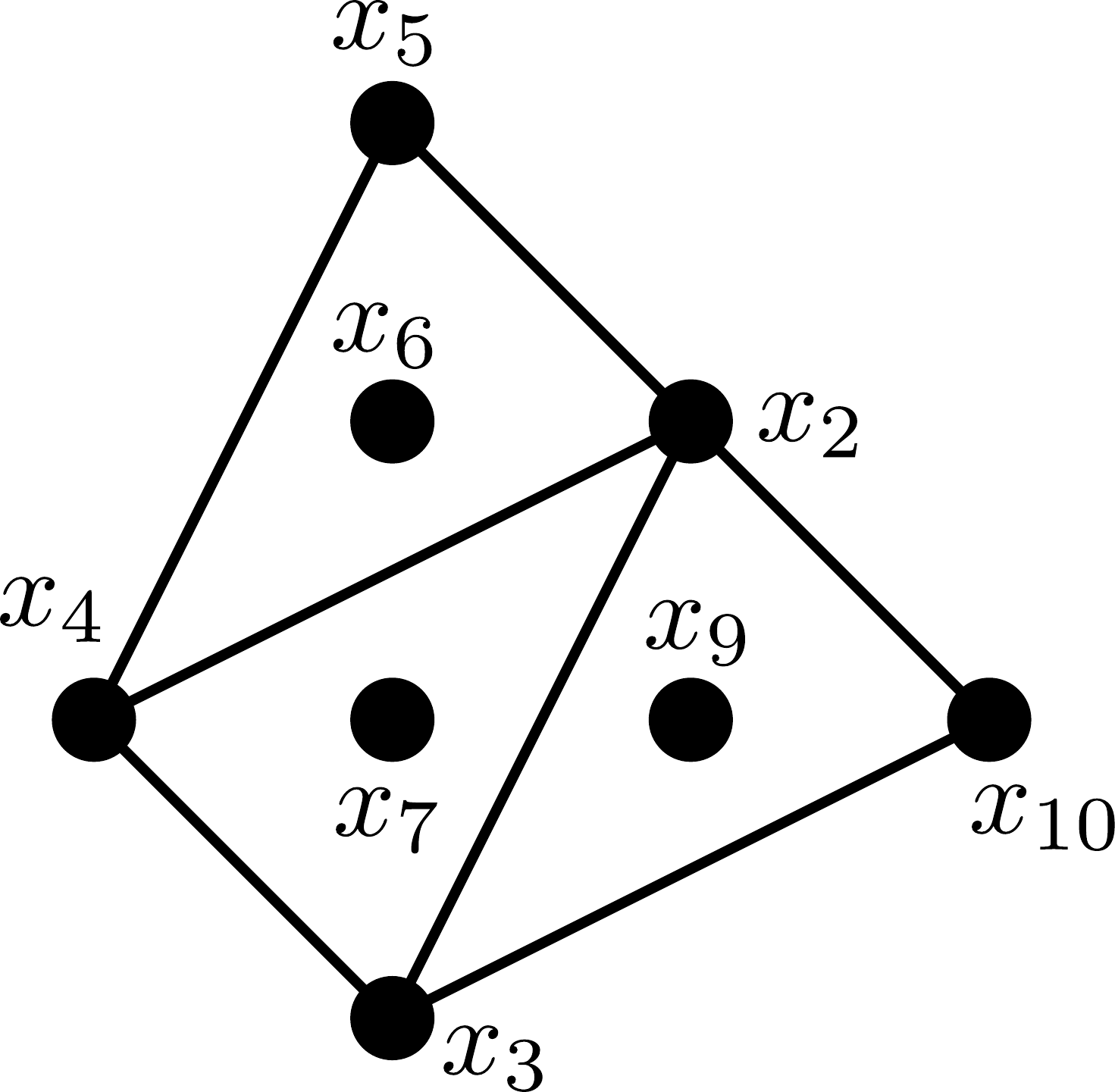}
  \end{center}

  \caption{Local model for \dPt.}
  \label{fig:dPt}
\end{figure}
In order to embed a $\bC^3/\bZ_3$ singularity such that the local
sector is placed away from the orientifold plane\footnote{The
  orientifold will be discussed in section~\ref{sec:semi-realistic}
  below.} requires \emph{three} copies of the $\bC^3/\bZ_3$, or
$\dPt$. The local singularity is described by the toric diagram in
\autoref{fig:dPt}. A simple ambient toric variety $\cA_\dPt$ giving
rise to the desired local singularity can be constructed by the
relatively simple reflexive polytope with integral points
\begin{align}
  \begin{array}{rrrrrrrrrrrr}
    x_0 & x_1 & x_2 & x_3 & x_4 & x_5 & x_6 & x_7 & x_8 & x_9 &
    x_{10} & x_{11}\\
    \hline
    0 & 0 & 1 & 0 & -1 & 0 & 0 & 0 & 0 &
    1 & 2 & 1 \\
    0 & 0 & 1 & -1 & 0 & 2 & 1 & 0 & 1 &
    0 & 0 & 0 \\
    -1 & 0 & 2 & 2 & 2 & 2 & 2 & 2 & 1 &
    2 & 2 & 1 \\
    0 & -1 & 3 & 3 & 3 & 3 & 3 & 3 & 1 &
    3 & 3 & 1
  \end{array}
  \label{eq:pMSSM-rays}
\end{align}
or equivalently, the following GLSM data:
\begin{align}
  \label{eq:pMSSM-GLSM}
  \begin{array}{r|rrrrrrrrrrrr}
      & x_0 & x_1 & x_2 & x_3 & x_4 & x_5 & x_6 & x_7 & x_8 & x_9 &
      x_{10} & x_{11}\\
      \hline
      \bC^*_1 & 1 & 0 & 0 & 0 & 0 & 0 & 0 & 0 & 0 &
      1 & -2 & 3 \\
      \bC^*_2 & 0 & 1 & 0 & 0 & 0 & 0 & 0 & 0 & 0 &
      0 & 1 & -2 \\
      \bC^*_3 & 0 & 0 & 1 & 0 & 0 & 0 & 0 & 0 & -1 &
      0 & -1 & 1 \\
      \bC^*_4 &  0 & 0 & 0 & 1 & 0 & 0 & 0 & 0 & 1 &
      -3 & 2 & -1 \\
      \bC^*_5 & 0 & 0 & 0 & 0 & 1 & 0 & 0 & 0 & 0 &
      -3 & 2 & 0 \\
      \bC^*_6 & 0 & 0 & 0 & 0 & 0 & 1 & 0 & 0 & -2 &
      0 & -1 & 2 \\
      \bC^*_7 & 0 & 0 & 0 & 0 & 0 & 0 & 1 & 0 & -1 &
      -1 & 0 & 1 \\
      \bC^*_8 & 0 & 0 & 0 & 0 & 0 & 0 & 0 & 1 & 0 &
      -2 & 1 & 0
\end{array}
\end{align}
The Calabi-Yau hypersurface $M_\dPt$ in this ambient toric variety has
Hodge numbers $(h^{11}, h^{21})=(8,158)$.

This polytope admits 236 different fine star triangulations. Out of
these, 30 are compatible with the partial resolution we wish to
perform, and out of these 30 we obtain 4 compatible with the $\bZ_2$
permutation involution, which we  choose in
section~\ref{sec:semi-realistic}. One of these triangulations is
described by the fan below:
\begin{align}
  \begin{split}
    \cF(\cA_\dPt) = \smash{\Big<}&\left\langle x_{0} x_{1} x_{2} x_{8}
    \right\rangle, \left\langle x_{0} x_{1} x_{2} x_{11}
    \right\rangle, \left\langle x_{0} x_{1} x_{3} x_{4} \right\rangle,
    \left\langle x_{0} x_{1} x_{3} x_{11}
    \right\rangle, \left\langle  x_{0} x_{1} x_{4} x_{8}  \right\rangle,\\
    & \left\langle x_{0} x_{2} x_{3} x_{7} \right\rangle, \left\langle
      x_{0} x_{2} x_{3} x_{9} \right\rangle, \left\langle x_{0} x_{2}
      x_{4} x_{6} \right\rangle, \left\langle x_{0} x_{2} x_{4} x_{7}
    \right\rangle, \left\langle  x_{0} x_{2} x_{5} x_{6}  \right\rangle,\\
    &\left\langle x_{0} x_{2} x_{5} x_{8} \right\rangle, \left\langle
      x_{0} x_{2} x_{9} x_{10} \right\rangle, \left\langle x_{0} x_{2}
      x_{10} x_{11} \right\rangle, \left\langle x_{0} x_{3} x_{4}
      x_{7}
    \right\rangle, \left\langle  x_{0} x_{3} x_{9} x_{10}  \right\rangle,\\
    &\left\langle x_{0} x_{3} x_{10} x_{11} \right\rangle,
    \left\langle x_{0} x_{4} x_{5} x_{6} \right\rangle, \left\langle
      x_{0} x_{4} x_{5} x_{8} \right\rangle, \left\langle x_{1} x_{2}
      x_{3} x_{7}
    \right\rangle, \left\langle  x_{1} x_{2} x_{3} x_{9}  \right\rangle,\\
    &\left\langle x_{1} x_{2} x_{4} x_{6} \right\rangle, \left\langle
      x_{1} x_{2} x_{4} x_{7} \right\rangle, \left\langle x_{1} x_{2}
      x_{6} x_{8} \right\rangle, \left\langle x_{1} x_{2} x_{9} x_{11}
    \right\rangle, \left\langle  x_{1} x_{3} x_{4} x_{7}  \right\rangle,\\
    &\left\langle x_{1} x_{3} x_{9} x_{11} \right\rangle, \left\langle
      x_{1} x_{4} x_{6} x_{8} \right\rangle, \left\langle x_{2} x_{5}
      x_{6} x_{8} \right\rangle, \left\langle x_{2} x_{9} x_{10}
      x_{11}
    \right\rangle, \left\langle  x_{3} x_{9} x_{10} x_{11}  \right\rangle,\\
    &\left\langle x_{4} x_{5} x_{6} x_{8} \right\rangle\smash{\Big>}
  \end{split}
\end{align}
corresponding to the following Stanley-Reisner ideal:
\begin{align}
  \label{eq:pMSSM-SRI}
  \begin{split}
    SR(\cA_\dPt) = \big< & x_{1} x_{5}, x_{1} x_{10}, x_{3} x_{8},
    x_{7} x_{8}, x_{8} x_{9}, x_{8} x_{10}, x_{8} x_{11}, x_{4}
    x_{11}, x_{5} x_{11}, \\ &x_{6} x_{11}, x_{7} x_{11}, x_{3} x_{5},
    x_{3} x_{6}, x_{4} x_{9}, x_{4} x_{10}, x_{5} x_{7}, x_{6}
    x_{7},\\& x_{7} x_{9}, x_{7} x_{10}, x_{5} x_{9}, x_{6} x_{9},
    x_{6} x_{10}, x_{5} x_{10}, x_{0} x_{1} x_{6}, x_{0} x_{1}
    x_{7},\\ & x_{0} x_{1} x_{9}, x_{0} x_{6} x_{8}, x_{2} x_{4}
    x_{8}, x_{0} x_{9} x_{11}, x_{2} x_{3} x_{11}, x_{2} x_{3} x_{4},
    \\ & x_{2} x_{3} x_{10}, x_{2} x_{4} x_{5}, x_{0} x_{1} x_{2}
    x_{3}, x_{0} x_{1} x_{2} x_{4}\big>.
  \end{split}
\end{align}

The Mori cone of the ambient space is then given by:
\begin{equation}
  \label{eq:pMSSM-ambient-MC}
  \renewcommand{\arraystretch}{1.1}
  \begin{array}{c|rrrrrrrrrrrrr}
    & x_0 & x_1 & x_2 & x_3 & x_4 & x_5 & x_6 & x_7 & x_8 & x_9 &
    x_{10} & x_{11} & K \\
    \hline
\hat\ell_1 & 0 & 0 & 1 & 1 & 1 & 0 & 0 & -3 & 0 &
0 & 0 & 0 & 0 \\
\hat\ell_2 & 0 & 0 & -1 & 0 & -1 & 0 & 1 & 1 & 0
& 0 & 0 & 0 & 0 \\
\hat\ell_3 & 1 & 0 & 0 & 0 & 0 & -2 & 1 & 0 & 3 &
0 & 0 & 0 & -3 \\
\hat\ell_4 & 0 & -1 & 1 & 1 & 0 & 0 & 0 & 0 & 0 &
-3 & 0 & 2 & 0 \\
\hat\ell_5 & 0 & 1 & 0 & 0 & 0 & 0 & 0 & 0 & 0 &
0 & 1 & -2 & 0 \\
\hat\ell_6 & 0 & 0 & -1 & -1 & 0 & 0 & 0 & 1 & 0
& 1 & 0 & 0 & 0 \\
\hat\ell_7 & 0 & 1 & 0 & 0 & 0 & 1 & 0 & 0 & -2 &
0 & 0 & 0 & 0 \\
\hat\ell_8 & 0 & -1 & 1 & 0 & 1 & 0 & -3 & 0 & 2
& 0 & 0 & 0 & 0 \\
\hat\ell_9 & 1 & 0 & 0 & 0 & 0 & 0 & 0 & 0 & 0 &
1 & -2 & 3 & -3
  \end{array}
\end{equation}
Notice that, while the Mori cone itself is 8-dimensional, we need 9
generators in order to span the whole cone of effective curves. In
other words, the Mori cone is not a simplicial cone.

Some of the curves in the Mori cone for $\cA_\dPt$ are not contained
in the Calabi-Yau hypersurface $M_\dPt$, and in order to obtain the
Mori cone for the Calabi-Yau hypersurface we need to eliminate these
curves. The curves actually contained in $M_\dPt$ then generate a
smaller cone, which often equals the Mori cone of the hypersurface. An
observation that is useful in order to do this systematically is that:
\begin{align}
  -K_\nabla \sim 3 \cO(\cA_\dPt,\hat D_0).
\end{align}
From this fact we can easily obtain which curves are not in the
Calabi-Yau hypersurface. For instance, since $\hat\ell_4$ is contained
in $\hat D_1\cap \hat D_9$, and $x_0 x_1 x_9$ is in the
Stanley-Reisner ideal~\eqref{eq:pMSSM-SRI}, $\hat\ell_4$ is not
contained in the Calabi-Yau hypersurface. Proceeding systematically,
one finds 
\begin{equation}
  \renewcommand{\arraystretch}{1.1}
  \begin{array}{r@{=}c|rrrrrrrrrrrrr}
    \multicolumn{2}{c}{\ell\cdot D}    & 
    x_0 & x_1 & x_2 & x_3 & x_4 & x_5 & x_6 & x_7 & 
    x_8 & x_9 & x_{10} & x_{11} & K \\
    \hline
    \ell_1 &\hat\ell_1  & 
    0 & 0 & 1 & 1 & 1 & 0 & 0 &-3 & 0 & 0 & 0 & 0 & 0 \\ 
    \ell_2 &\hat\ell_2  & 
    0 & 0 &-1 & 0 &-1 & 0 & 1 & 1 & 0 & 0 & 0 & 0 & 0 \\ 
    \ell_3 &\hat\ell_3  & 
    1 & 0 & 0 & 0 & 0 &-2 & 1 & 0 & 3 & 0 & 0 & 0 &-3 \\ 
    \ell_4 &\hat\ell_4+\hat\ell_5  & 
    0 & 0 & 1 & 1 & 0 & 0 & 0 & 0 & 0 &-3 & 1 & 0 & 0 \\ 
    \ell_5 &\hat\ell_5  & 
    0 & 1 & 0 & 0 & 0 & 0 & 0 & 0 & 0 & 0 & 1 &-2 & 0 \\ 
    \ell_6 &\hat\ell_6  & 
    0 & 0 &-1 &-1 & 0 & 0 & 0 & 1 & 0 & 1 & 0 & 0 & 0 \\ 
    \ell_7 &\hat\ell_7  & 
    0 & 1 & 0 & 0 & 0 & 1 & 0 & 0 &-2 & 0 & 0 & 0 & 0 \\ 
    \ell_8 &\hat\ell_7+\hat\ell_8  & 
    0 & 0 & 1 & 0 & 1 & 1 &-3 & 0 & 0 & 0 & 0 & 0 & 0 \\ 
    \ell_9 &\hat\ell_9  & 
    1 & 0 & 0 & 0 & 0 & 0 & 0 & 0 & 0 & 1 &-2 & 3 &-3 \\ 
    \ell_{10} &\tfrac{1}{2}(\hat\ell_4+\hat\ell_8+3 \hat\ell_6+3 \hat\ell_2+2 \hat\ell_1) & 
    0 &-1 &-1 & 0 & 0 & 0 & 0 & 0 & 1 & 0 & 0 & 1 & 0 \\ 
  \end{array}
\end{equation}
The $8$-dimensional Mori cone of the hypersurface has $10$ rays, so it
is even less simplicial than the Mori cone of the ambient space. Its
dual, the K\"ahler cone of the Calabi-Yau threefold $M_\dPt$, is an
$8$-dimensional cone generated by $29$ rays!

To make the discussion more manageable, we will restrict ourselves to
the $5$-dimensional subspace
\begin{equation}
  \mathop{\mathrm{span}}\big\{ D_0,~D_1,~ D_2,~ D_3+D_4,~ D_7\big\}
\end{equation}
of the $8$-dimensional K\"ahler moduli space. In
section~\ref{sec:semi-realistic}, this will turn out to be the
$h^{11}_+(M_\dPt)=5$-dimensional subspace of the orientifold-invariant
K\"ahler moduli, but for the purposes of this section we can just take
it to be a simplifying assumption. When intersected with this
subspace, the K\"ahler cone is generated by the 8 divisor classes
\begin{equation}
  \begin{split}
    \omega \in \mathop{\mathrm{span}}{}_{\bR_\geq} \big\{&
    D_0,~ 
    D_1-D_2+(D_3+D_4),~
    D_1-D_2+2 (D_3+D_4)+D_7,~
    \\&
    3 D_1-3 D_2+3 (D_3+D_4)+D_7,~
    -3 D_2+3 (D_3+D_4)+D_7,~ 
    \\&
    -D_2+(D_3+D_4), ~
    -D_2+2 (D_3+D_4)+D_7
    \big\}
  \end{split}
\end{equation}
A particular patch of the (non-simplicial) K\"ahler cone that will be
useful in the following can be parametrized as
\begin{equation}
  \begin{split}
    \omega =&\;
    t_0 D_0 + 
    t_1 \big(D_1-D_2+(D_3+D_4)\big) + 
    t_2 \big(D_1-D_2+2 (D_3+D_4)+D_7\big) +
    \\&
    t_3 \big(3 D_1-3 D_2+3 (D_3+D_4)+D_7\big) +
    t_4 \big(-3 D_2+3 (D_3+D_4)+D_7\big)
  \end{split}
\end{equation}
With respect to this K\"ahler class, the volume of the invariant
divisors $D_7$ and $D_6+D_9\simeq D_0-3 (D_3+D_4)-2 D7$ is 
\begin{align}
    \Vol(D_7) =  \int_{D_7}\omega^2 = t_1^2
    ,\qquad
    \Vol(D_6+D_9) = \int_{D_6+D_9}\omega^2 = 2 t_2^2
    ,
\end{align}
and the volume of the Calabi-Yau is
\begin{equation}
  \begin{split}
    \Vol\big(M_\dPt\big)  = & \int_{\MPt}\omega^3 \\
    = &\; 36 t_0^3 + 198
    t_0^2 t_1 + 330 t_0 t_1^2 + 183 t_1^3 + 234 t_0^2 t_2 + 792 t_0
    t_1 t_2 + 660 t_1^2 t_2 \\ & + 474 t_0 t_2^2 + 792 t_1 t_2^2 +
    316 t_2^3 + 594 t_0^2 t_3 + 1980 t_0 t_1 t_3 + 1650 t_1^2 t_3 \\
    & + 2376 t_0 t_2 t_3 + 3960 t_1 t_2 t_3 + 2376 t_2^2 t_3 + 2970
    t_0 t_3^2 + 4950 t_1 t_3^2 \\ & + 5940 t_2 t_3^2 + 4950 t_3^3 +
    108 t_0^2 t_4 + 360 t_0 t_1 t_4 + 300 t_1^2 t_4 + 432 t_0 t_2
    t_4 \\ & + 720 t_1 t_2 t_4 + 432 t_2^2 t_4 + 1080 t_0 t_3 t_4 +
    1800 t_1 t_3 t_4 + 2160 t_2 t_3 t_4 \\ & + 2700 t_3^2 t_4 + 54
    t_0 t_4^2 + 90 t_1 t_4^2 + 108 t_2 t_4^2 + 270 t_3 t_4^2 + 9
    t_4^3 .
  \end{split}
\end{equation}
Hence, at $t_1=0$ the $\dP$ surface $D_7$ shrinks and at $t_2=0$ the
two $\dP$ surfaces $D_6$ and $D_9$ shrink simultaneously, see
\autoref{fig:dPt}. Furthermore, the toric curves $D_2\cap D_3$ and
$D_2\cap D_4$ separate $D_7$ from $D_9$ and $D_7$ from $D_6$,
respectively. Their volume is
\begin{equation}
  \frac{1}{2} \Vol\big(D_2 \cap (D_3+D_4) \big) = t_3+t_4.
\end{equation}
Therefore, if we set $t_1=t_2=t_3=t_4=0$, the three $\bC^3/\bZ_3$
singularities, obtained by shrinking $D_6$, $D_7$, and $D_9$, respectively, collide and
produce an enhanced (non-orbifold) singularity. As long as $t_0$ stays
finite, the Calabi-Yau threefold is of finite volume in these limits.

\section{Adding D-branes to the model}
\Label{sec:branes}

So far we have only discussed the closed string sector, but realistic
models also require the specification of an open string sector. There
are two main ingredients in the open string sector of type IIB
compactifications: D-branes giving the gauge dynamics, and
orientifolds canceling the tadpoles. We will postpone the discussion
of orientifolds to section~\ref{sec:orientifold}, and deal with the
systematic incorporation of D-branes in this section.

We will assume that we have a consistent (i.e., tadpole-free) local
model of branes at singularities, described in terms of a quiver gauge
theory. In order to view this local model as part of a global model we
need to give a description of the fractional branes and flavor branes
in terms of the geometry of the global model. In order to keep our
discussion concrete, we will focus on a particular choice of branes on the
$\bC^3/\bZ_3$ singularity.

\begin{figure}
  \begin{center}
    
\ifpdf
  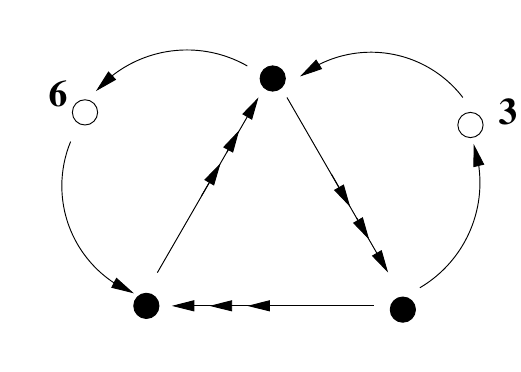
  \else
  \input{dp0mssm.pstex_t}
\fi
  \end{center}

  \caption{A toy model for the MSSM, from \cite{Aldazabal:2000sa}. The
    filled dark dots denote gauge groups, while the white dots denote
    global symmetry groups, coming from non-compact $D7$ branes. The
    labels on the arrows denote with which MSSM field they should be
    identified.}

  \label{fig:dp0mssm}
\end{figure}

\begin{figure}
  \begin{center}
    \includegraphics[width=0.3\textwidth]{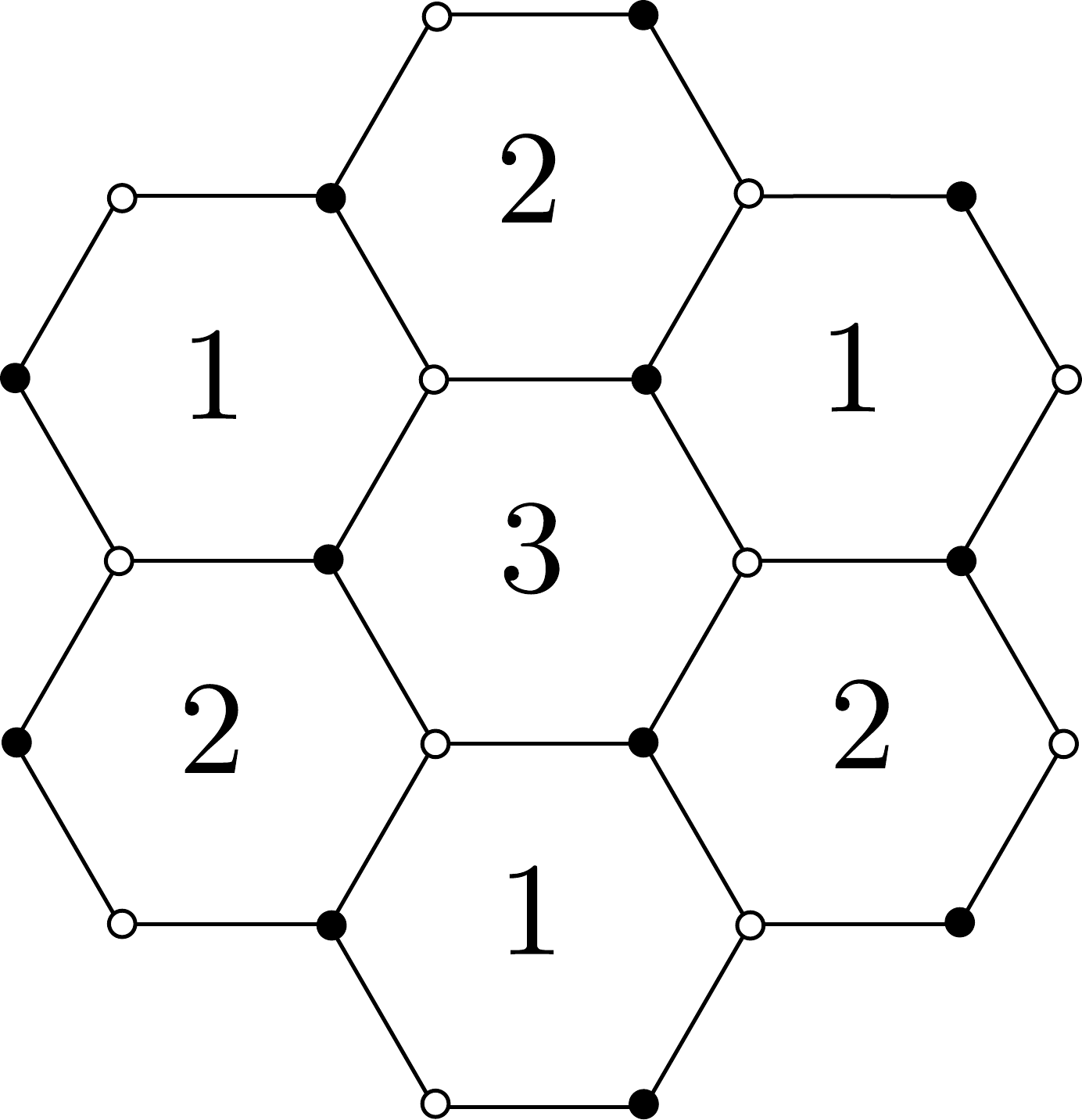}
  \end{center}

  \caption{Dimer model for branes at the $\bC^3/\bZ_3$ singularity,
    given by the periodic honeycomb lattice (we have only shown a few
    cells). The labels on the faces of the dimer model indicate which
    gauge factor in \autoref{fig:dp0mssm} they correspond to.}

  \label{fig:honeycomb}
\end{figure}

\paragraph{The $\bC^3/\bZ_3$ MSSM.}
The model that we will use to illustrate our discussion was originally
introduced in \cite{Aldazabal:2000sa} as a toy model for the MSSM. We
reproduce it in \autoref{fig:dp0mssm}. Notice that in this quiver we
have both gauge groups and flavor groups. The gauge groups are
obtained by introducing fractional branes on the $\bC^3/\bZ_3$
singularity. In order to analyze the physics of branes at this
singularity, it is convenient to have the dimer model for
$\bC^3/\bZ_3$. This is given by the honeycomb periodic lattice, shown
in \autoref{fig:honeycomb}.

\medskip

Despite focusing on the previous example --- which is all we need for
our chosen toy models $(dP_0)^n$ --- we will formulate the discussion
in general terms, and it will be applicable to any singularity with at
least one contracting 4-cycle. Models with no contracting 4-cycles
($L^{a,b,a}$ geometries, or $\bC^3/(\bZ_2\times\bZ_2)$) can also be
interesting for model building purposes, see for example
\cite{Argurio:2006ny,Argurio:2007qk}. They cannot support
standard-model-like structures, though, so we omit their detailed
description, although it should be possible to give a description
quite similar to the one below. In the context of the Toric Lego such
supersymmetry breaking sectors can be incorporated into a Gauge Mediated Supersymmetry Breaking (GMSB) model
\cite{Balasubramanian:2009tv}, and the total model can then be
analyzed using the techniques below.

\medskip

We review the large volume description of gauge branes in
section~\ref{sec:fractional-branes}. This construction is well known
in the literature
\cite{Diaconescu:1999dt,Douglas:2000ah,Douglas:2000qw,Cachazo:2001sg,Wijnholt:2002qz,Aspinwall:2004jr,Herzog:2004qw,Aspinwall:2004vm,Herzog:2005sy,Hanany:2006nm}. We
will follow the convention of describing the branes by elements in the
derived category of coherent sheaves. This description differs from
the more conventional one in physics (a brane with a vector bundle on
top) by some subtleties that will not matter much in our analysis,
except for a factor of $\sqrt{K_S^\vee}$ relating the sheaf and the
bundle \cite{Freed:1999vc,Katz:2002gh}, with $K_S$ the canonical class
of the divisor wrapped by the brane. Accordingly, we will distinguish
the sheaf $\cE$ from its associated vector bundle $F_\cE=\cE\otimes
\sqrt{K_S^\vee}$.\footnote{When $S$ is not $Spin$ we have that
  $\cF_\cE$ is not an honest bundle. Nevertheless all the formulas for
  physical quantities are well defined, as it is manifest from the
  sheaf description of the brane. This is one reason why we will
  prefer the ``sheafy'' description.}

To our knowledge a similar systematic dictionary between the quiver
and large volume languages has not been given for the kind of flavor
branes and singularities that we are interested in. We give the first
steps in this direction in section~\ref{sec:flavor-branes}, by
obtaining some of the charges of the flavor branes by imposing the
right quiver structure. It would certainly be interesting to give a
general and complete dictionary for flavor branes at the same level as
that for gauge branes, but we do not attempt to do so in this paper.

\subsection{Blowing up the fractional branes}
\label{sec:fractional-branes}

Space-time filling D-branes on a IIB compactification on a smooth
Calabi-Yau $X$ are described by objects in the bounded derived
category of coherent sheaves on $X$, generally denoted by $D^b(X)$. We
do not review this here, and instead refer the reader to
\cite{Aspinwall:2004jr} for a nice review. This category can be
thought of as the set of branes in the B-model. As such, it does not
depend on where we are in K\"ahler moduli space, and in particular it
can also describe branes at the singular quiver point.

Such a description is not necessarily the most suitable one at every
point in moduli space. Close to orbifold points, for example, a more
convenient description of D-brane states is  given in terms of
quiver representations, and in the toric case dimer models. The
information describing the D-brane is in this case encoded in the
ranks of the gauge nodes, and the vevs of the
bifundamentals. Nevertheless, as mentioned above, the categorical
description of branes is valid everywhere in moduli space, and thus
one expects an equivalence between the respective categories. In
particular cases such a correspondence can be proven, one of the most
celebrated such results being the one by Bridgeland, King and Reid
\cite{BKR}, which states that:
\begin{align}
  D_G(M) = D^b(\widetilde{M/G})\, ,
\end{align}
with $M$ typically $\bC^3$ in physical applications, and $G$ a finite
subgroup of $SU(d)$, $d=\dim(M)$. The term on the left represents the
branes in $\bC^3$ together with an action under $G$ (an equivariant
structure), i.e., the ordinary description of branes on orbifolds,
while the term on the right represents the category of coherent
sheaves on the resolution of the orbifold $M/G$.

We are interested in continuing the quiver description to large
volume in cases where the singularity is not of orbifold type, and not
necessarily a contracting del Pezzo surface either. In this case there
are only partial results, the most useful for us being the method
proposed in \cite{Hanany:2006nm}, which we proceed to review now.

\subsubsection{The $\Psi$ map}

Let us start by describing in terms of the global geometry the
fractional branes giving the gauge group. That is, we want to obtain
the exceptional collection that describes the fractional branes at
large volume. The basic technology we need for doing this was
described in \cite{Hanany:2006nm}. Notice that the exceptional
collection for a local $\bC^3/\bZ_3$ singularity is well-known, and
has been obtained by other methods
\cite{Diaconescu:1999dt,Douglas:2000qw,Cachazo:2001sg,Wijnholt:2002qz,Aspinwall:2004jr}. In
particular it is an orbifold, and as such falls into the class studied
by \cite{BKR}, who give the explicit functor between the large volume
and quiver categories. We have chosen to compute the basis of
fractional branes using the $\Psi$ map \cite{Hanany:2006nm} since this
method works for any toric singularity with a compact 4-cycle.

\begin{figure}
  \begin{center}
    \includegraphics[width=\textwidth]{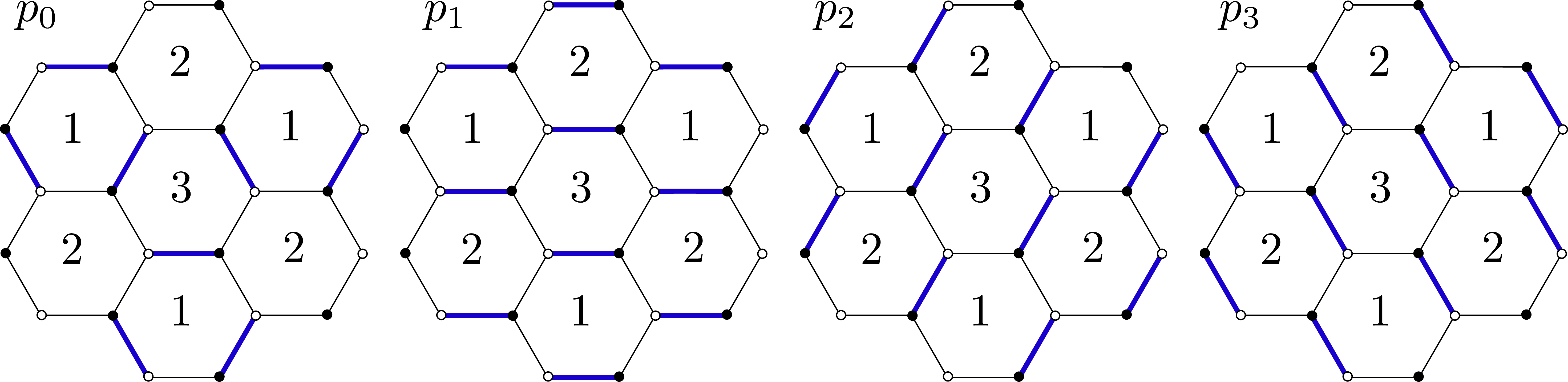}
  \end{center}

  \caption{Relevant perfect matchings for the $\bC^3/\bZ_3$ dimer
    model. We have omitted one of the internal perfect matchings.}

  \label{fig:dp0-matchings}
\end{figure}

The first step in constructing the $\Psi$ map consists of determining
the perfect matchings of the dimer model, and their interpretation in
terms of divisors of the local geometry. There are 5 perfect matchings
of the honeycomb lattice in \autoref{fig:honeycomb}, of which 2 are
reference matchings --- associated with the interior point of the
toric diagram, or equivalently with the compact $\bP^2$ --- and the
other three correspond to the exterior points of the toric diagram. We
show this structure in \autoref{fig:dp0-matchings}.

From the perfect matchings in \autoref{fig:dp0-matchings} we 
reconstruct the geometry, as explained in
\cite{Hanany:2005ve,Franco:2005rj,Franco:2006gc}. In particular, to
each external perfect matching we  associate a non-compact divisor
of the $\bC^3/\bZ_3$ geometry, i.e., the $D_1,D_2,D_3$ divisors in
\autoref{fig:dP0-toric-diagram}. This is done as follows: notice that
each edge in the dimer has a natural orientation (going from the white
node to the black node, for example). Given this orientation, it makes
sense to consider $p_i-p_0$ as a closed oriented cycle in the
dimer. The cycle is obtained by superposing the edges belonging to
$p_i$ with the edges belonging to $p_0$, with the understanding that
edges belonging to both perfect matchings ``annihilate'' each
other. This operation defines a $(p,q)$ cycle on the $T^2$ where the
dimer model is defined. Taking this $(p,q)$ as a point in the integer
lattice, the convex hull of the resulting set of points turns out to
be the diagram for the toric geometry giving rise to the dimer. For
our particular example, we obtain the following winding numbers:
\begin{align}
  p_0 - p_0 &= (0,0) & p_1 - p_0 &= (1,1) & p_2 - p_0 &= (-1,0) & p_3
  - p_0 &= (0,-1)\, .
\end{align}
By comparison with \autoref{fig:dP0-toric-diagram} we thus obtain the
following identification between perfect matchings and external
divisors:
\begin{align}
  \label{eq:matchings-to-divisors}
  (p_0,p_1,p_2,p_3) \sim (D_6, D_3, D_2, D_1)\, .
\end{align}

\begin{figure}
  \begin{center}
    \includegraphics[width=0.3\textwidth]{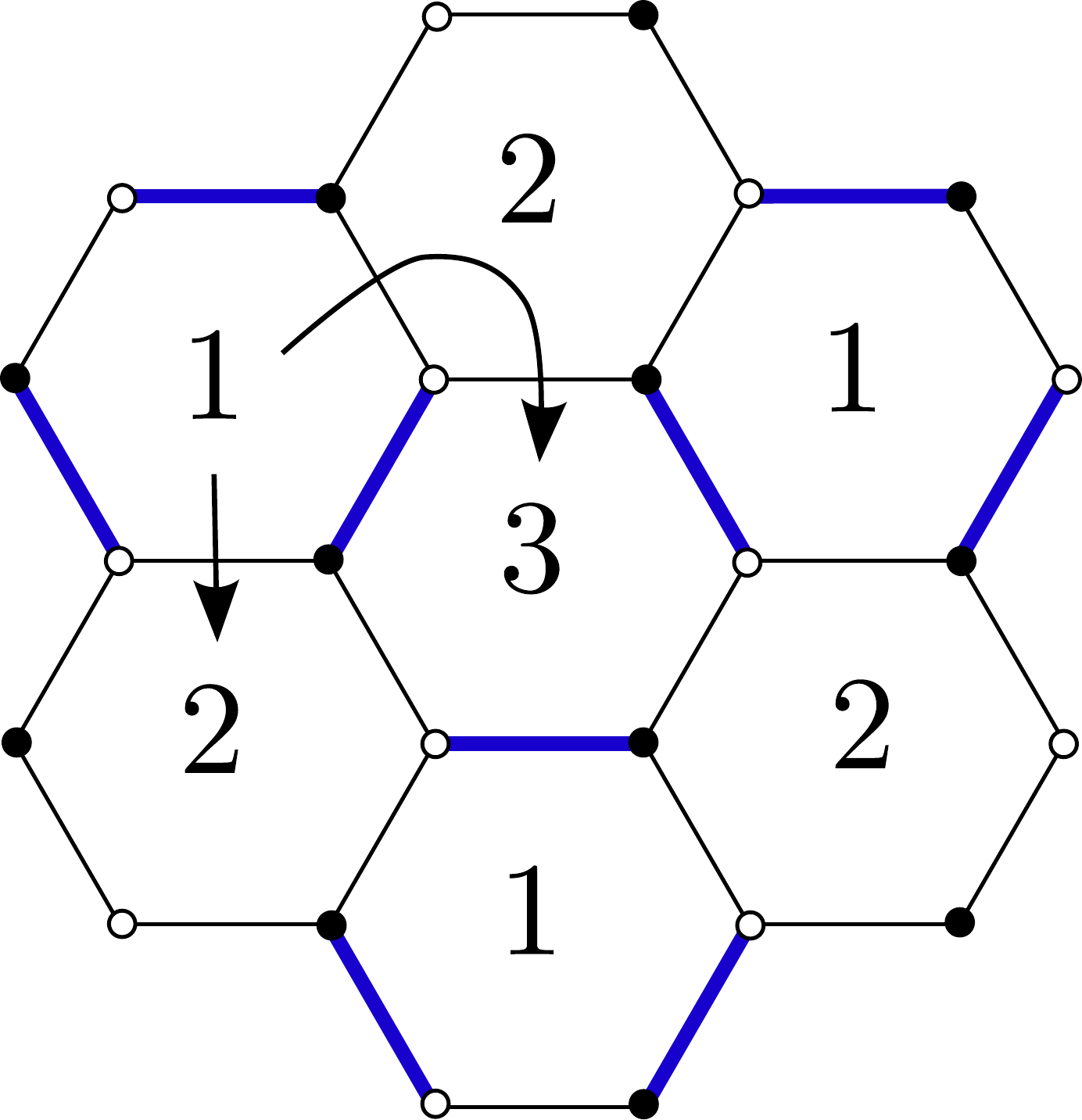}
  \end{center}
  
  \caption{Allowed open paths giving the elements of the exceptional
    collection associated to the Beilinson quiver for $\bC^3/\bZ_3$,
    taking $p_0$ as a reference matching.}

  \label{fig:dp0-beilinson}
\end{figure}

We  now proceed to describe the exceptional collection obtained
from the $\Psi$ map in our particular example. Let us take as our
reference fractional brane the gauge factor denoted by ``1'' in
\autoref{fig:honeycomb}. In order to obtain the exceptional collection
we construct an open (\emph{allowed} \cite{Hanany:2006nm}) path in the
dimer from 1 to 2 and another from 1 to 3. The paths we have chosen
are depicted in \autoref{fig:dp0-beilinson}. The exceptional
collection giving the Beilinson quiver is then determined by the
crossing of the open paths with the perfect matchings. In our
conventions, and taking into account the
relation~\eqref{eq:matchings-to-divisors} between perfect matchings
and divisors we obtain:\footnote{Our notation departs slightly from
  the conventions in
  \cite{Herzog:2004qw,Herzog:2005sy,Hanany:2006nm}. We denote the
  collection~\eqref{eq:projective-basis} $P=\{P_i\}$ since it is a
  collection of \emph{projective} objects in the del Pezzo case
  \cite{Aspinwall:2004vm}, and we will reserve the symbol $\cE_i$ for
  the fractional branes in $\bC^3/\bZ_3$ themselves. $\cE_i^\vee$ will
  then denote the dual of the sheaf $\cE_i$ in the ordinary sense (for
  example $c_i(\cE^\vee)=(-1)^i c_i(\cE)$).}
\begin{align}
  \label{eq:projective-basis}
  P = (\cO_{\bP^2}, \cO_{\bP^2}(D_3), \cO_{\bP^2}(D_3+D_1))\, .
\end{align}
Taking into account linear equivalence of divisors on the $\bP^2$, and
simplifying notation a bit, we end up with:
\begin{align}
  \label{eq:beilinson-collection}
  P = (\cO, \cO(1), \cO(2))\, ,
\end{align}
which is a well-known projective basis for the derived category of branes on
$\bP^2$.

In order to identify the physical fractional branes from the
collection~\eqref{eq:beilinson-collection} we still need to work a
little bit more, and dualize $P$.\footnote{Technically, what we are
  doing here is constructing the fractional branes in terms of
  projective objects by mutation, and taking Chern characters. We
  refer the reader interested in the details to the nice exposition in
  \cite{Aspinwall:2004vm}.} That is, the collection that describes the
physical fractional branes is not~\eqref{eq:beilinson-collection}, but
rather the following related collection:
\begin{align}
  \cE = (\cE_1, \cE_2, \cE_3)\, .
\end{align}
Constructing the objects of the collection $\cE$ is non-trivial, but
thankfully their Chern character is much more easily computed. We have
the basic relation:
\begin{align}
  \label{eq:dual-chern-character}
  \ch(\cE_i) = (S^{-1})_{ji} \ch(P_j)\, .
\end{align}
The matrix $S$ is defined as follows:
\begin{align}
  S_{ij} = \dim\Hom(P_i,P_j) = H^0(\bP^2,P_i^\vee\otimes P_j)\, .
\end{align}
Since $P_i$ are line bundles on a toric variety, it is easy to
compute $S$ just by counting sections:
\begin{align}
  S = \begin{pmatrix}
    1 & 3 & 6\\
    0 & 1 & 3\\
    0 & 0 & 1
  \end{pmatrix}\, .
\end{align}
The inverse matrix $S^{-1}$ is thus given by:
\begin{align}
  S^{-1} = \begin{pmatrix}
    1 & -3 & 3\\
    0 & 1 & -3\\
    0 & 0 & 1
  \end{pmatrix}\, .
\end{align}
Plugging this matrix into~\eqref{eq:dual-chern-character}, we
obtain:
\begin{align}
  \label{eq:P2-sheaves-characters-wrong-sign}
  \begin{split}
    \ch (\cE_1) &= 1\\
    \ch (\cE_2) &= -2 + H + \frac{1}{2}H^2\\
    \ch (\cE_3) &= 1 - H + \frac{1}{2}H^2
  \end{split}
\end{align}
with $H=c_1(\cO_{\bP^2}(1))$ the hyperplane class of $\bP^2$.
A simple check of this expression is that if we add up the Chern
characters of the three D-branes we get $H^2$. Due to the Chern-Simons
coupling on the worldvolume of the branes, we have that this induces
precisely one unit of $D3$-brane charge (there are also well known
curvature contributions to the charge, but they do not contribute
here). This agrees with the expected result that the quiver with all
ranks equal represents the theory on a single $D3$ brane probing the
singularity.

\medskip

In our particular case we want to introduce fractional branes of
different ranks, in particular we want to consider the object:
\begin{align}
  \cE_{MSSM} = \cE_1 + 2\cE_2 + 3\cE_3\,.
\end{align}
The Chern-Simons coupling for a $D7$-brane described by a sheaf $\cE$
wrapping a divisor $S$ is given by:
\begin{align}
  \label{eq:$D7$-CS-coupling}
  S_{C.S.} = \int_{S\times\bR^{3,1}} C^{(RR)}\wedge
  e^{-B}\wedge\ch(\cE)\wedge \sqrt{\frac{\Td(T_S)}{\Td(N_S)}}\, ,
\end{align}
with $C^{(RR)}=C_0 + C_2 + \ldots$ the formal sum of $RR$ fields;
$T_S$, $N_S$ refer to the tangent and normal bundles of $S$ in the
Calabi-Yau $X$, respectively, and $B$ is the NSNS 2-form, which we
will set to 0 in what follows.\footnote{In the particular case of the
  $\bC^3/\bZ_3$ orbifold on which we focus in this paper, the quiver
  point is located at $B=J=0$
  \cite{Morrison:1994fr,Aspinwall:2004jr}.} The Todd class $\Td(T_S)$
of $S$ is defined as follows:
\begin{align}
  \Td(T_S) = 1 + \frac{1}{2} c_1(T_S) + \frac{c_1(T_S)^2 + c_2(T_S)}{12}\,.
\end{align}
For example, in the particular case of $\bP^2$, we have:
\begin{align}
  \Td(T_{\bP^2}) = 1 + \frac{3}{2} H + H^2\, .
\end{align}
with $H$ the hyperplane class.

We  encode this charge information into the charge vector:
\begin{align}
  \label{eq:charge-vector}
  \Gamma_\cE = [S]\ch(\cE)\sqrt{\frac{\Td(T_S)}{\Td(N_S)}}\, ,
\end{align}
with $[S]$ the class Poincare dual to $S$. In what follows we will
often forget the distinction between cohomology classes and divisors,
in order to lighten the notation, and we will omit the space-time part
of the charge.

Computing the induced $D5$ and $D3$ charges
from~\eqref{eq:charge-vector},
\eqref{eq:P2-sheaves-characters-wrong-sign} and the facts that $N_S|_S
= D_S|_S = \cO(-3H)$, $c(T_S) = (1+H)^3$:
\begin{align}
  \begin{split}
    \Gamma_{\cE_1} & = [S]\wedge \left(1+\frac{3}{2} H+\frac{5}{4}H^2\right)\\
    \Gamma_{\cE_2} & = [S]\wedge \left(-2-2H-\frac{1}{2}H^2\right)\\
    \Gamma_{\cE_3} & = [S]\wedge
    \left(1+\frac{1}{2}H+\frac{1}{4}H^2\right)\, ,
  \end{split}
\end{align}
so we get the total charge:
\begin{align}
  \begin{split}
    \Gamma_{\cE_{MSSM}} & = \Gamma_{\cE_1} + 2\Gamma_{\cE_2} +
    3\Gamma_{\cE_3}\\
    & = [S]\wedge (-H+H^2)\, .
  \end{split}
\end{align}
We see that we have a total non-vanishing induced $D_5$ charge, and
also the expected induced $D3$ charge. The induced $D_5$ charge
signals a tadpole, since it is supported on a 2-cycle with a compact
dual cycle given in this case by the $\bP^2$ itself.  We need to
cancel this tadpole by introducing extra ingredients into our
configuration in the form of flavor $D7$ branes. We will do this in
section~\ref{sec:flavor-branes}.

\paragraph{A subtlety in the definition of the $D3$ charge.}

In the discussion above we have taken the usual conventions in the
exceptional collection literature. Unfortunately there is a subtlety
that is important in physics applications of these formulas: what we
ordinarily call a $D3$ brane (defined as the object mutually
supersymmetric with respect to a large volume $D7$ brane) has
\begin{align}
  \label{eq:D3-charge}
  \Gamma_{D3} = -[pt]\, ,
\end{align}
with $[pt]$ the class of a point in the Calabi-Yau, instead of
$\Gamma_{D3}=[pt]$, as one may guess at first. An easy way to show
this is by noticing that the expression for the central charge of a $D7$
with support on $S$ at large volume, given by:
\begin{align}
  Z(D7) = \int_X e^{-(B+iJ)}\wedge\Gamma \sim - \frac{1}{2}\int_S J\wedge J\, , 
\end{align}
is a large negative number. Taking the sign as in~\eqref{eq:D3-charge}
gives a central charge with the same sign:
\begin{align}
  Z(D3) = \int_X e^{-(B+iJ)}\wedge(-[pt]) = -1\, ,
\end{align}
and thus both objects preserve the same supersymmetry.\footnote{The
  negative sign also follows from duality and a familiar fact in the
  heterotic string: in order to cancel the heterotic tadpole in a
  $K3\times T^2$ compactification without switching on instantons on the
  gauge bundle, one needs to introduce 24 mobile $NS5$s. Dualizing to
  F-theory, these branes appear as 24 mobile $D3$s. The $D3$ tadpole
  created by these branes is canceled by the $D3$ charge induced by
  the curvature couplings of the 24 7-branes wrapping the base $K3$ in
  F-theory. Choosing the (standard) Chern-Simons coupling for $D7$s as
  in \eqref{eq:$D7$-CS-coupling} then forces us to set
  $\Gamma_{D3}=-[pt]$.}

The simplest way of taking this issue into account is simply to
multiply by $-1$ the Chern characters that we found above for the
fractional branes. This does not change the quiver, but now the
charges of the fractional branes add up to minus the class of a
point. The Chern characters for the objects in the exceptional
collection are thus given by:
\begin{align}
  \label{eq:P2-sheaves-characters}
  \begin{split}
    \ch (\cE_1) &= -1\\
    \ch (\cE_2) &= 2 - H - \frac{1}{2}H^2\\
    \ch (\cE_3) &= -1 + H - \frac{1}{2}H^2\,.
  \end{split}
\end{align}
and the corresponding charges by:
\begin{align}
  \label{eq:fractional-brane-charges}
  \begin{split}
    \Gamma_{\cE_1} & = [S]\wedge \left(-1-\frac{3}{2} H-\frac{5}{4}H^2\right)\\
    \Gamma_{\cE_2} & = [S]\wedge \left(2+2H+\frac{1}{2}H^2\right)\\
    \Gamma_{\cE_3} & = [S]\wedge
    \left(-1-\frac{1}{2}H-\frac{1}{4}H^2\right)\, .
  \end{split}
\end{align}

\subsubsection{Computing the spectrum}

Given two branes, described by sheaves $\cE,\cF$ on the divisors $S$
and $T$, embedded on the Calabi-Yau by the maps $i:S\hookrightarrow
X$, $j: T\hookrightarrow X$ respectively, we have that the spectrum of
strings between them is expected to be counted by \cite{Katz:2002gh}:
\begin{align}
  \Ext^\bullet(i_*\cE, j_*\cF)\, .
\end{align}
$\Ext$ groups are typically hard to compute, but luckily in the cases
that we are dealing with the calculation simplifies. For two $D7$ branes
wrapping the same divisor, one has that \cite{Aspinwall:2004vm}:
\begin{align}
  \Ext_X^i(i_*\cE, j_*\cF) = \Ext_S^i(\cE, \cF) \oplus \Ext_S^{3-i}(\cF,
  \cE)\, .
\end{align}
If one is just interested in computing indices the calculation
simplifies further:
\begin{align}
  \begin{split}
    \sum (-1)^i \Ext_X^i(i_*\cE, j_*\cF) & = \sum (-1)^i \Ext_S(\cE, \cF)
    - \sum (-1)^i \Ext_S(\cF, \cE)\\
    & = \chi(\cE,\cF) - \chi(\cF, \cE)\, .
  \end{split}
\end{align}
where we have used Hirzebruch-Riemann-Roch in the last line, and
$\chi(\cE,\cF)$ is defined as:
\begin{align}
  \label{eq:euler-index}
  \chi(\cE,\cF) = \int_S \ch(\cE^\vee) \ch(\cF) \Td(T_S)\, .
\end{align}

In the case in which we have a couple of $D7$ branes intersecting
transversely over a curve $C$, the calculation of the $\Ext$ groups
also simplifies \cite{Katz:2002gh}:
\begin{align}
  \Ext^{i+1}(i_*\cE, j_*\cF) = H^i(S, F_\cE^\vee\otimes F_\cF \otimes
  \sqrt{K_\cC})
\end{align}
for $i<2$, and 0 otherwise. If we are interested in purely computing
indices, the result can again be expressed in terms of integrals of
characteristic classes:
\begin{align}
  \sum (-1)^i \Ext^i(i_*\cE, j_*\cF) = \int_C \ch(F_\cF^\vee\otimes
  F_\cE)\, .
\end{align}

It is illuminating and useful to rewrite the formulas above as
follows. Define the antisymmetric Dirac-Schwinger-Zwanziger (DSZ)
product as:
\begin{align}
  \label{eq:dsz}
  \dsz{\Gamma_\cE}{\Gamma_\cF} = \sum_{n=0}^3 \int_X (-1)^n\,
  \Gamma_\cE^{(2n)}\wedge \Gamma_\cF^{(6-2n)}\, ,
\end{align}
where $\Gamma^{(k)}$ denotes the part of the form $\Gamma$ of degree
$k$. Then it is an easy exercise to check, given the index formulas
above, that
\begin{align}
  \sum (-1)^i\Ext^i(i_*\cE, j_*\cF) = \dsz{\Gamma_\cE}{\Gamma_\cF}\, .
\end{align}
Notice that in our context the 6-form part of $\Gamma$ plays no role,
since its magnetic dual, the 0-form, is always absent. We will thus
often ignore the 6-form part without further notice in any computation
of chiral quantities.

As an illustration, the quiver for our example is now 
reconstructed easily using ~\eqref{eq:dsz} and the charges
in~\eqref{eq:fractional-brane-charges}:
\begin{align}
  \dsz{\Gamma_{\cE_1}}{\Gamma_{\cE_2}} & = \dsz{[S]\wedge
    \left(-1-\frac{3}{2} H\right)}{[S]\wedge \left(2+2H\right)}\\
  & = \int_S \left((-\frac{3}{2} H) \wedge 2 [S] - 2H\wedge (-[S])\right)\\
  & = 6  \left(\frac{3}{2} - 1\right) \int_S H^2 \\
  & = 3\, ,
\end{align}
and similarly $\dsz{\Gamma_{\cE_1}}{\Gamma_{\cE_3}}=-3$,
$\dsz{\Gamma_{\cE_2}}{\Gamma_{\cE_3}}=3$.

\subsection{Flavor $D7$ branes}
\Label{sec:flavor-branes}

Looking to the MSSM quiver in \autoref{fig:dp0mssm}, we see that there
are three basic flavor $D7$ branes we can consider, classified by 
which fractional branes they couple to. In particular, in
\autoref{fig:dp0mssm} we have a rank 6 stack coupling to $\cE_3$ and
$\cE_2$ (but not $\cE_1$), a rank 3 stack coupling only to $\cE_3$ and
$\cE_2$, and finally a rank 0 stack (which we have not drawn),
coupling only to $\cE_2$ and $\cE_1$. We will denote them respectively
as $\cF_6$, $\cF_3$ and $\cF_0$. Our task is to promote $\cF_{3i}$ to
geometrical objects in the Calabi-Yau background.

We  obtain a first piece of information by imposing that the right
bifundamentals exist between the gauge and flavor branes. Given that
we are computing intersection numbers, in this way we will only obtain 
a restricted amount of information about $\Gamma_{\cF_i}$. Let us make
this explicit by parametrizing:
\begin{align}
  \Gamma_{\cF_i} = a_i [D_H] + b_i [D_H]^2\, ,
\end{align}
where the only thing we need to know about $[D_H]$ is that
$[D_H]|_{\bP^2}=H$. Looking to figure~\ref{fig:dp0mssm}, we have to
impose that:
\begin{align}
  \dsz{\Gamma_{\cE_1}}{\Gamma_{\cF_3}} = 1\, ,
\end{align}
which implies
\begin{align}
  \dsz{[S]\wedge \left(-1-\frac{3}{2} H\right)}{[D_H]\wedge (a_3+b_3[D_H])}
  = \left(-\frac{3}{2}a_3 + b_3 \right) \int_S H^2 = 1\, .
\end{align}
Similarly, $\dsz{\Gamma_{\cE_2}}{\Gamma_{\cF_3}} = 0$ implies:
\begin{align}
  \dsz{[S]\wedge \left(2+2H\right)}{[D_H]\wedge (a_3+b_3[D_H])} = 2
  \left(a_3 - b_3\right) = 0\, .
\end{align}
These two equations together imply that:
\begin{align}
  \label{eq:F3-charge}
  \Gamma_{\cF_3} = [D_H]\wedge (-2-2[D_H])\, .
\end{align}
The third condition $\dsz{\Gamma_{\cE_3}}{\Gamma_{\cF_3}}=-1$ is now
automatic, since $\sum_i \Gamma_{\cE_i}=0$ (up to a 6-form), and the
DSZ product is linear.
We  proceed similarly for $\Gamma_{\cF_6}$ and $\Gamma_{\cF_0}$,
obtaining:
\begin{align}
  \Gamma_{\cF_6} & = [D_H]\wedge \left(1+\frac{3}{2}[D_H]\right)\\
  \Gamma_{\cF_0} & = [D_H]\wedge
  \left(1+\frac{1}{2}[D_H]\right) \label{eq:F6-charge} 
\end{align}

\medskip

We now lift this local information to global
information. As an illustration, we will embed our local system into
the compact Calabi-Yau in section~\ref{sec:MP}, namely an elliptic
fibration over $\bP^2$. In doing so, global tadpoles with charge in
the non-compact divisors will not be canceled. We  ignore
this effect momentarily and will come back to it later. Recall from
section~\ref{sec:MP} that $D_6$ described our local $\bP^2$. It is
also easy to see that $D_1|_{D_6}=H$, for example by computing the
triple intersection number $D_1\cdot D_1 \cdot D_6=1$ in the
Calabi-Yau. In order to emphasize the fibration structure of the
threefold, we  relabel our coordinates
\begin{align}
  (s, t, u, x, y, z) \equiv (x_1, x_2, x_3, x_4, x_5, x_6)\, .
\end{align}
From table~\ref{table:P^2-fibration}, we see that $(s,t,u)$ are the
coordinates of the $\bP^2$, located at $z=0\cap M_{dP_0}$. Similarly, the
elliptic curve is the hypersurface in  $\bP^2_{(2,3,1)}$ described by
$(x,y,z)$, as in  section~\ref{sec:MP}.

A generic $D7$ brane $\cF$ on our global embedding is described by:
\begin{align}
  \Gamma_\cF = [aD_s + bD_z] \wedge (r + cD_s + dD_z)\wedge
  \left(1-\frac{1}{2}(aD_s+bD_z)\right)\, ,
\end{align}
with $a,b,c,d,r\in\bZ$, and the last term comes from expanding
$\sqrt{\Td(T_D)/\Td(N_D)}$ (here $D\equiv aD_s+bD_z$). We have ignored
6-form terms, as usual. Notice that the divisor $aD_s+bD_z$ will in
general have more two-forms than those induced from the ambient space,
so we could add further 4-form terms (an example is $D_s$
itself). Nevertheless, since the DSZ product is taken in $X$, it will
not see these, so we have set them to 0. We want the flavor branes to
look like ordinary $D7$ branes away from the singularity, so we impose
$r=a=1$:
\begin{align}
  \label{eq:flavor-brane-charge}
  \Gamma_\cF = [D_s + bD_z] \wedge \left(1+ \left(c-\frac{1}{2}\right)D_s +
    \left(d-\frac{b}{2}\right)D_z \right)\, .
\end{align}
Using the restrictions $D_s|_{D_z}=H$, $D_z|_{D_z}=-3H$:
\begin{align}
  \label{eq:restricted-flavor-charge}
  \Gamma_\cF|_{D_z} = (1-3b)H\wedge \left(1 +
    \left(c-3d-\frac{1-3b}{2}\right)H\right)\, .
\end{align}
From here it is easy to read the global information given by the
quiver. We find that:
\begin{align}
  \label{eq:flavor-charges-MP}
  \begin{split}
    \Gamma_{\cF_6} & = [D_s] \wedge \left(1 + (3d_6+2)[D_s] + d_6[D_z]\right)\wedge\sqrt{\frac{T_{D_s}}{N_{D_s}}}\\
    \Gamma_{\cF_3} & = [D_s+D_z]\wedge \left(1 + 3d_3[D_s] + d_3[D_z]\right)\wedge\sqrt{\frac{T_{D_s+D_z}}{N_{D_s+D_z}}}\\
    \Gamma_{\cF_0} & = [D_s] \wedge \left(1 + (3d_0+1)[D_s] +
      d_0[D_z]\right)\wedge\sqrt{\frac{T_{D_s}}{N_{D_s}}}\, .
  \end{split}
\end{align}
Notice that there is certain ambiguity, and with what we have said so
far one can only fix $c-3d$, but not $c$ and $d$ individually. We have
reflected this ambiguity in the unknown coefficients $d_i\in\bZ$.

\subsection{$D5$-brane tadpole}

Our original motivation for introducing flavor branes was that the $D5$
brane charge was not canceled with the desired assignment of ranks for
the gauge groups, and this induced a tadpole. As a consistency check,
let us now verify that the local tadpole cancels once we introduce the
flavor branes.

We denote the curves that are Poincare dual to $D_s$ and $D_z$ by $\ell_s$ and
$\ell_z$, respectively. These curves are defined in homology by
$\ell_s\cdot D_s=\ell_z\cdot D_z = 1$ and $\ell_s\cdot D_z =
\ell_z\cdot D_s = 0$. By comparing with curves of the form $D_i\cdot
D_j$, we find that
\begin{align}
  \label{eq:P2-Mori-curves}
  \begin{split}
    \ell_z & = D_s\cdot D_s\\
    \ell_s & = D_s\cdot(3D_s + D_z) \, .
  \end{split}
\end{align}

$D5$  tadpole cancellation requires that all $D5$ brane charge is
supported on a 2-cycle that does not intersect any compact
4-cycle. Since the compact 4-cycle in the non-compact geometry is
$D_z$, we need all of our $D5$ charge to be supported on $\ell_s$.
Recall from~\eqref{eq:flavor-brane-charge} that a general flavor $D7$
brane $\cF$ has a charge vector:
\begin{align}
  \Gamma_\cF = [D_s + bD_z]\wedge
  \left(1+\left(c-\frac{1}{2}\right)D_s +
    \left(d-\frac{b}{2}\right)D_z\right)\, .
  \tag{\ref{eq:flavor-brane-charge}}
\end{align}
The $D5$ charge is given by the 4-form part
of~\eqref{eq:flavor-brane-charge}:
\begin{align}
  \begin{split}
    \Gamma_\cF^{(D5)} & = \left(c-\frac{1}{2}\right)D_s\cdot D_s +
    \left[\left(c-\frac{1}{2}\right)b +
      \left(d-\frac{b}{2}\right)\right]D_s\cdot D_z +
    b\left(d-\frac{b}{2}\right)D_z\cdot D_z\\
    & = \left[d + b\left(c - 3d - 1 + \frac{3b}{2}\right)\right]\ell_s + (1-3b)\left(c-3d-\frac{1-3b}{2}\right)\ell_z
  \end{split}
\end{align}
where in the second line we have used~\eqref{eq:P2-Mori-curves}, and
the fact that $D_z\cdot D_z=9\ell_z-3\ell_s$. As one may have
expected, the $\ell_z$ charge is precisely the 4-form term in the
restriction of the flavor brane charge to $D_z$
(recall~\eqref{eq:restricted-flavor-charge}).

Using this last observation, and (\ref{eq:F3-charge}-\ref{eq:F6-charge}) it
is easy to compute the contribution of the flavor branes to the
compact $D5$ tadpole; in our case:
\begin{align}
  \label{eq:flavor-local-D5-charge}
  \begin{split}
    \Gamma_\cF^{(D5)} & = 6\,\Gamma_{\cF_6}^{(D5)} + 3\, \Gamma_{\cF_3}^{(D5)}\\
    & = \biggl(6d_6 + 3d_3 + \frac{3}{2}\biggr)\ell_s + 3\ell_z\, .
  \end{split}
\end{align}

The fractional branes are dealt with similarly. A general
fractional brane is written as:
\begin{align}
  \Gamma_\cE = D_z \wedge (a+b D_s)\, .
\end{align}
The $D5$ brane charge is thus given by $bD_s\cdot D_z = b(\ell_s -
3\ell_z)$. Using the charges given
in~\eqref{eq:fractional-brane-charges}, we then find
that:\footnote{There is a slight clash of notation here. Notice that
  $[S]$ in \eqref{eq:fractional-brane-charges} refers to $D_z$, not
  $D_s$.}
\begin{align}
  \Gamma_\cE^{(D5)} = 3\, \Gamma_{\cE_3}^{(D5)} + 2\,
  \Gamma_{\cE_2}^{(D5)} + \Gamma_{\cE_1}^{(D5)} = \ell_s - 3\ell_z\, .
\end{align}
As we see, this cancels the $\ell_z$ part
of~\eqref{eq:flavor-local-D5-charge}, leaving:
\begin{align}
  \label{eq:total-D5-charge}
  \Gamma_\cF^{(D5)} + \Gamma_\cE^{(D5)} = \biggl(6d_6 + 3d_3 +
  \frac{5}{2}\biggr)\ell_s\, .
\end{align}

\medskip

Let us point out that global tadpole cancellation requires \emph{all}
$D5$ charge to vanish, once we embed the local geometry into a compact
model. This is clearly not possible in~\eqref{eq:total-D5-charge},
since $d_i\in\bZ$, but this is a shortcoming of this particular
(oversimplified) embedding.  In general cases where the $D7$ sector
admits a consistent embedding we expect the equation analogous
to~\eqref{eq:total-D5-charge} to admit solutions.

\section{Type IIB orientifolds}
\Label{sec:orientifold}

Next we turn to the study of various orientifold involutions,
anticipating the D-brane/open string sector required for the gauge
theory/matter, and how to uplift this to F-theory. We will mostly
follow the generalization of Sen's original work connecting type IIB
orientifolds and F-theory \cite{Sen:1996vd,Sen:1997kw,Sen:1997gv}, as
can be found in \cite{Collinucci:2008zs, Collinucci:2009uh,
  Blumenhagen:2009up}.

\subsection{Generalities}
\Label{sec:orientifolding-generalities}

With the inclusion of the open string sector, and the associated
D-brane charge, we need to construct the orientifold model from the
Calabi-Yau manifold in which the local geometry was embedded, in order
to satisfy the various tadpole conditions. The choice of the
orientifold involution $\sigma$ has to be made such that the fixed
point set left invariant under $\sigma$ gives a divisor which when
wrapped by the orientifold plane leads to tadpole cancellation.  For
the $D7$-brane charge this corresponds to
\begin{align}
  \label{eq:tadpole-condition}
  8 D_{O7^-} = D_{D7}
\end{align}
in conventions of the double cover, i.e., counting each $D7$ and its
image separately.

One possible type of orientifold involution that will be of particular
interest to us is a permutation symmetry acting on the coordinates of
the ambient space,
\begin{align}
  \label{eq:permutation-action-general}
  \sigma:(x_0,\ldots,x_{n})\leftrightarrow (x_{\sigma(0)},\ldots,x_{\sigma(n)})\, ,
\end{align}
where $x_{\sigma(i)}$ refers to the coordinate which $x_i$ is mapped
to under the orientifold involution $\sigma$.  The points in the fixed
locus are defined by:
\begin{align}
  \label{eq:pMSSM-invariance-general}
  (x_0,\ldots,x_{n}) =
  \mathfrak{g}[(x_{\sigma(0)},\ldots,x_{\sigma(n)})]\, ,
\end{align}
where $\mathfrak{g}$ is a gauge transformation of the underlying
gauged sigma model.

\medskip

The reader may wonder whether~\eqref{eq:permutation-action-general} is
a well defined action on the toric ambient space, since it is not
obvious that it commutes with the gauge action. That this is the case
is argued in general as follows. Start from the fact that the
generators of the Mori cone give a basis for the $U(1)$ generators of
the GLSM, with charges of the divisors given by the intersection
numbers between the divisor and the Mori cone generator. Now assume
that we have a permutation involution $\sigma$,
\eqref{eq:permutation-action-general}, of the toric fan. In terms of
the GLSM, this induces a $\bZ_2$ permutation of the columns. We now
argue that this permutation of the columns of the GLSM is undone
by a permutation of the rows.

Calling $\ell_i$ the Mori cone generators, and $D_j$ the toric
divisors, one has that $\ell_i\cdot D_j = \sigma(\ell_i)\cdot
\sigma(D_j)$. Since this is an involution of the fan we have that
$\sigma(\ell_i) = \ell_k$, i.e. one of the original charges. Another way of
saying this is that the permutation involution is a permutation of the
generators of the Mori cone, if we choose the involution to be an
involution of the fan. Thus the permutation of the columns can be
undone by (i.e., it is equivalent to, since we are in $\bZ_2$) a
permutation of the rows.

From here it is easy to see that the permutation orbifold is well
defined. We need to show that, for any gauge transformation
$\mathfrak{g}_1$ of the GLSM, $\mathfrak{g}_2 \equiv \sigma \mathfrak{g}_1
\sigma$ is another gauge transformation of the GLSM. Looking to the
coordinates, it is easy to convince oneself that $\sigma \mathfrak{g}_1 \sigma$
induces the action on the Mori cone given above, and thus an action on
the charge vector which amounts to a relabeling.

\medskip

Given any $\bZ_2$ action on the coordinates of the ambient space,
\eqref{eq:permutation-action-general} or otherwise, we  construct
the quotient space by constructing the set of coordinates invariant
under the orientifold action, and impose any relations that follow
tautologically from the definition of the invariant coordinates in
terms of the original coordinates. There will be no constraints
between the invariant coordinates in our examples below, but they 
appear in general, as the following example shows. Consider the
$\bC^2/\bZ_2$ orbifold, with orbifold action on the $x_1,x_2$
coordinates of $\bC$ given by:
\begin{align}
  (x_1,x_2)\to (-x_1,-x_2)\, .
\end{align}
Invariant coordinates are given by $a=x_1^2$, $b=x_2^2$,
$c=x_1x_2$. It is easy to convince oneself that any invariant
polynomial can be constructed out of these coordinates, so $a,b,c$
generate the ring of functions on the quotient. Due to their
definition, the constraint between the coordinates is given by:
\begin{align}
  \label{eq:orbifold-hypersurface}
  ab = c^2\, .
\end{align}
We thus reproduce the well-known fact that $\bC^2/\bZ_2$ is
alternatively described as the
hypersurface~\eqref{eq:orbifold-hypersurface} in $\bC^3$.

\subsection{Sign orientifolds: $\MP$}

In section~\ref{sec:branes} we have described in large volume terms
the flavor and color branes in our local model. As a result of this
analysis, we found that the total $D7$ brane charge in the embedding
of $\bC^3/\bZ_3$ into $\MP$ is given by
(recall~\eqref{eq:flavor-charges-MP}):
\begin{align}
  \label{eq:P2-total-$D7$-charge}
  Q_{D7} = 9D_s + 3D_z \sim D_y\, .
\end{align}
Notice that this charge is concentrated on a divisor which does not
intersect $D_z$, consistent with local tadpole cancellation. In order
to cancel the resulting global tadpole while keeping a supersymmetric
model, we  need to introduce a $O7^-$ orientifold. The $\MP$
embedding is too simple, so cannot introduce the
orientifold without also perturbing  the local model, but it serves as
a good stepping stone to the more realistic model in
section~\ref{sec:semi-realistic}.

In view of the relations~\eqref{eq:tadpole-condition} and
\eqref{eq:P2-total-$D7$-charge}, we  cancel the tadpoles by wrapping an
orientifold on $D_y$, and adding extra branes on top of
$D_y$. This is easy to achieve by quotienting the space by the
involution:
\begin{align}
  \label{eq:P2-fibration-involution}
  (s,t,u,x,y,z) \to (s,t,u,x,-y,z)\, .
\end{align}
This involution also acts on the equation defining the Calabi-Yau
hypersurface, and we should make sure that the involution and the
Calabi-Yau manifold are compatible. The most general Calabi-Yau hypersurface in
the ambient space is given by (ignoring the $\bZ_2$ action for a
moment):
\begin{align}
  \label{eq:P2-tate-form}
  y^2 + a_1 xyz + a_3 y z^3 = x^3 + a_2 x^2 z^2 + a_4 xz^4 + a_6 z^6\, ,
\end{align}
where $a_n$ is a homogeneous polynomial of order $3n$ on the $s,t,u$
variables. If we now impose invariance
under~\eqref{eq:P2-fibration-involution}, the fixed locus is indeed at
$D_y$. It is easy to see that this involution is compatible with the
equation~\eqref{eq:P2-tate-form} defining the Calabi-Yau hypersurface if
we set $a_1=a_3=0$. The gauge invariance of the background forces us
to consider also the following $\bZ_2$ actions:
\begin{align}
  (s,t,u,x,y,z) &\to (s,t,u,x,y,-z)\\
  (s,t,u,x,y,z) &\to (-s,-t,-u,x,y,z)\, .
\end{align}
The second involution has no fixed points, since $\{s=t=u=0\}$ is in
the Stanley-Reisner ideal~\eqref{eq:elliptic-P^2-SRI}, but the first
involution has fixed points at $\{z=0\}\sim D_z$. In order to cancel
the charge coming from this component of the orientifold, according
to~\eqref{eq:tadpole-condition}, we need to wrap extra $D7$ branes on
$D_z$, the cycle supporting the gauge sector. This will render a
consistent model, but not the one that we wanted to embed.

\subsection{Permutation orientifolds: $\MPt$}
\Label{sec:semi-realistic}

With hindsight, the fact that the model in the previous section
required important modifications to the open string sector is not too
surprising. The orientifold that we chose leaves all divisors classes
invariant, and thus typically will project down some unitary gauge
factors to symplectic or orthogonal subgroups, in addition to
identifying nodes in the quiver. What we want is an involution that
exchanges the gauge sector with a copy somewhere else in the
Calabi-Yau, without the orientifold intersecting the gauge
sector.\footnote{For a related use of the orientifold permutation
  involution, see~\cite{Diaconescu:2005pc}.}

\begin{figure}
  \begin{center}
    \includegraphics[width=0.3\textwidth]{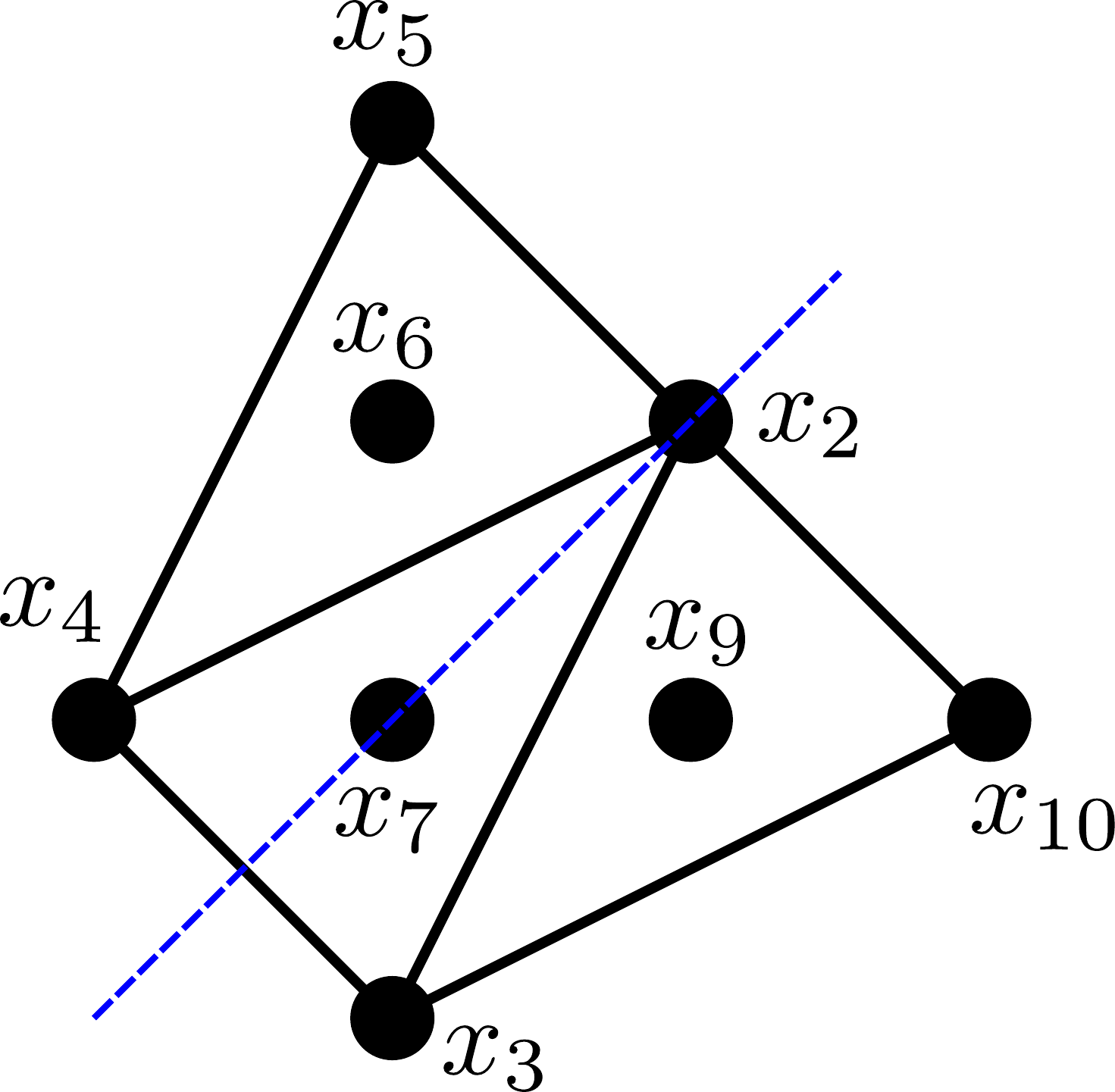}
  \end{center}
  
  \caption{Local geometry for the model in this section. The
    orientifold involution acts by reflection on the dashed blue
    line. We have partially resolved the singularity into three
    separated $\bC^3/\bZ_3$ singularities. We have also indicated the
    position of the coplanar divisors in~\eqref{eq:pMSSM-rays}.}

  \label{fig:permutation-MSSM}
\end{figure}

In this section we discuss a simple extension of the model of the
previous section that achieves this. In fact, we have already
encountered a suitable geometry: it is the $\MPt$ geometry we
discussed in section~\ref{sec:M30}.\footnote{The hyperconifold in
  section~\ref{sec:hyperconifold} also has an involution of the local
  geometry. Unfortunately this involution does not extend to the
  global embedding $M_{Y^{3,0}}$ that we chose. $\MPt$ also has the
  advantage of having a divisor mapped to itself under the orientifold
  involution, and which could thus be interesting from the point of
  view of generating non-perturbative superpotentials.} We show the
action of the involution on the local geometry
in~\autoref{fig:permutation-MSSM}. The induced $\bZ_2$ action on
$\MPt$ is given by:
\begin{align}
  \label{eq:permutation-action}
  (x_0,x_1,\ldots,x_{10},x_{11})\leftrightarrow (x_0,x_1,x_2,x_4,x_3,x_{10},x_9,x_7,x_{11},x_{6},x_{5},x_{8})\, .
\end{align}
According to the discussion in
section~\ref{sec:orientifolding-generalities}, this action can
equivalently be seen as an action on the Mori cone
generators~\eqref{eq:pMSSM-ambient-MC}:
\begin{align}
 \label{eq:pMSSM-MC-involution}
 (\hat\ell_3, \hat\ell_5, \hat\ell_6, \hat\ell_8) \leftrightarrow
 (\hat\ell_9, \hat\ell_7, \hat\ell_2, \hat\ell_4)
\end{align}
with $\hat\ell_1$ fixed.

\medskip

Identifying the monomials invariant under
involution~\eqref{eq:permutation-action} we obtain the quotient map
\begin{align}
  \label{eq:involution-map}
  \begin{split}
    (x_0,\cdots,x_{11})\to & \phantom{=\,}(z_0,\cdots,z_{8}) \\  & = (x_0, x_1, x_2, x_3x_4,
    x_5x_{10}, x_6 x_9, x_8 x_{11}, x_7,
    x_3x_9x_{10}^2x_{11}+x_4x_5^2x_6x_8)\, .
  \end{split}
\end{align}
Notice in particular that the contracting $dP_0$ on which we want to
put our gauge sector is now given by $z_5=0$.

The points in the fixed locus satisfy:
\begin{align}
  \label{eq:pMSSM-invariance}
  (x_0,x_1,\ldots,x_{10},x_{11}) =
  \mathfrak{g}[(x_0,x_1,x_2,x_4,x_3,x_{10},x_9,x_7,x_{11},x_{6},x_{5},x_{8})]\, ,
\end{align}
where $\mathfrak{g}$ is a gauge transformation of the
GLSM~\eqref{eq:pMSSM-GLSM}, in other words an element of $(\bC^*)^8$.
Condition \eqref{eq:pMSSM-invariance} can be seen to be equivalent to:
\begin{align}
  \label{eq:pMSSM-orientifold-locus}
  x_8x_5^2x_4x_6 = x_3x_{10}^2x_9x_{11}\, ,
\end{align}
and thus the fixed locus is in the divisor class
\begin{align}
D_{O7} = -D_2+D_3+D_4\, .
\end{align}
Notice that, as one may guess by looking to
figure~\ref{fig:permutation-MSSM}, the orientifold locus does not
intersect the cycle on which we are wrapping our branes, even in the
ambient space. Take for example the $dP_0$ coming from the $x_6=0$
divisor. If this cycle intersected the orientifold locus,
from~\eqref{eq:pMSSM-orientifold-locus} this would imply that
$x_3x_{10}^2x_9x_{11}=0$, but notice that all of $x_6x_3$,
$x_6x_{10}$, $x_6x_9$ and $x_6x_{11}$ are in the Stanley-Reisner
ideal~\eqref{eq:pMSSM-SRI}, and thus $x_3x_{10}^2x_9x_{11}\neq 0$ when
$x_6=0$.

One interesting subtlety that appears in this example, and not in the
examples studied in
\cite{Collinucci:2008zs,Collinucci:2009uh,Blumenhagen:2009up}, is the
following. Notice that we constructed 9 invariant coordinates, but the
dimension of the Mori cone before the involution was 8. We also
constructed the action on the Mori cone generators
in~\eqref{eq:pMSSM-MC-involution}, and we saw that it permutes 8
generators in pairs. It is thus a little bit puzzling how one would
get the $9-4=5$ Mori cone generators that one would need in order to
construct the ambient fourfold where the base lives as a
hypersurface. Furthermore, it is easy to convince oneself that the
$z_i$ coordinates do not satisfy any identity.

This is resolved as follows. Notice that because the Mori cone is not
simplicial, in order to generate it we need 9 effective curves,
rather than 8 which the dimension counting may suggest. There is a linear
relation between the Mori cone generators
in~\eqref{eq:pMSSM-ambient-MC} (this is easy to verify for instance by
computing the rank of~\eqref{eq:pMSSM-ambient-MC} seen as a matrix),
but interestingly, once we quotient by the orientifold action, the
linear relation no longer holds, and we generate a new Mori cone
generator, for a total of 5, as one expects. The ambient space
$\cA_\dPt/\bZ_2$ of the quotient is then described by the set of Mori
cone generators invariant under~\eqref{eq:pMSSM-MC-involution}:
\begin{align}
  \renewcommand{\arraystretch}{1.3}
  \begin{array}{r|rrrrrrrrrr}
    & z_0 & z_1 & z_2 & z_3 & z_4 & z_5 & z_6 & z_7 & z_8\\
    \hline
    \hat\ell_1 \sim \check\ell_0 & 0 & 0 & 1 & 2 & 0 & 0 & 0 & -3 & 1\\
    \hat\ell_2\sim \hat\ell_6\sim\check\ell_1 & 0 & 0 & -1 & -1 & 0 & 1 &
    0 & 1 & 0\\
    \hat\ell_3\sim \hat\ell_9\sim\check\ell_2 & 1 & 0 & 0 & 0 & -2 & 1
    &3 & 0 & 0\\
    \hat\ell_4\sim \hat\ell_8\sim\check\ell_3 & 0 & -1 & 1 & 1 & 0 &
    -3 & 2 & 0 & 0\\
    \hat\ell_5\sim \hat\ell_7\sim\check\ell_4 & 0 & 1 & 0 & 0 & 1 & 0
    & -2 & 0 & 0
    \label{eq:pMSSM-MC-after-involution}
  \end{array}
\end{align}
Here we have denoted the Mori cone generators by $\check\ell_i$, and
we have also indicated where they come from in the double
quotient. The resulting space is a toric variety, given by the
polytope:
\begin{align}
  \begin{array}{rrrrrrrrr}
      z_0 & z_1 & z_2 & z_3 & z_4 & z_5 & z_6 & z_7 & z_8\\
      \hline
      1 & 0 & 0 & 0 & 6 & 2 & 3 & -2 & -6 \\
      0 & 1 & 0 & 0 & 9 & 3 & 5 & -3 & -9 \\
      0 & 0 & 1 & 0 & 2 & 1 & 1 & 0 & -1 \\
      0 & 0 & 0 & 1 & 2 & 1 & 1 & 0 & -2
    \end{array}
\end{align}
and Stanley-Reisner ideal:
\begin{align}
  \label{eq:dPt-quotient-SRI}
  \begin{split}
    SR(\cA_{\dPt/\bZ_2}) & = \bigl\langle z_{1} z_{4}, z_{6} z_{7}, z_{6} z_{8},
      z_{4} z_{8}, z_{5} z_{8}, z_{5} z_{7}, z_{4} z_{7}, z_{0} z_{1} z_{5}, \\
      & \phantom{=\bigl\langle\quad} 
      z_{0} z_{1} z_{7}, z_{0} z_{5} z_{6}, z_{2} z_{3} z_{6},
      z_{2} z_{3} z_{8}, z_{2} z_{3} z_{4}, z_{0} z_{1} z_{2} z_{3}
    \bigr\rangle\, .
  \end{split}
\end{align}

After the involution~\eqref{eq:pMSSM-MC-involution}, the original
Calabi-Yau equation $P(x_i)=0$ with degrees
\begin{align}
  \deg(P) = (0,0,3,0,0,0,0,0,3)
\end{align}
(see the $K$ column in~\eqref{eq:pMSSM-ambient-MC}) becomes an
equation $\check P(z_i)=0$ of degree
\begin{align}
  \deg(\check P) = (0,0,3,0,0).
\end{align}
On the other hand, the degrees of the anti-canonical class of
$\cA_\dPt/\bZ_2$ are given by:
\begin{align}
\deg(\ov K_{\cA_\dPt/\bZ_2}) = (1,0,3,0,0)\, ,
\end{align}
and from here we easily deduce the anti-canonical class of the
hypersurface $\cB_3=\{\check P(z_i) = 0\}$:
\begin{align}
  \label{eq:base-degree}
  \deg(\ov K_{\cB_3}) = \deg(\ov K_{\cA_\dPt/\bZ_2}) - \deg(\check P)
  = (1,0,0,0,0) \, .
\end{align}
We  verify in section~\ref{sec:MPt-F} that this agrees beautifully
with what one  expects from Sen's limit, similarly to the
discussion in \cite{Collinucci:2009uh}.

\medskip

Rather than adding the flavor branes directly in the type IIB
orientifold setting we will find it more convenient to uplift the above
orientifold picture to F-theory, to which we  turn next.

\section{Hybrid embeddings}
\Label{sec:hybrid}

Let us present a systematic method for embedding the previous
configurations in F-theory. Although F-theory adequately captures
non-perturbative $g_s$ effects, naively it is ill-suited for
describing branes at singularities, which require non-perturbative
$\alpha'$ effects for a full description.  (Alternatively, if we want
to describe fractional branes at large volume we have to deal with
anti-$D7$ branes in F-theory.)

The basic idea we use for overcoming this obstruction is the
following: notice that by local tadpole cancellation, the total
$D7$-brane charge of the local configuration wraps a cycle that does
\emph{not} intersect the collapsing cycle. Thus, at the level of
cohomology (what F-theory describes most naturally), the contracting
cycle is generically far away from the discriminant! From the point of
view of F-theory, our quiver configurations are then described by
ordinary Calabi-Yau 4-fold compactifications at a very non-generic
point in their moduli space, where the discriminant intersects a
singular point in the geometry.  This observation, while simple, is
clearly very general. Let us illustrate how it works for our working
example $\MPt$.

\subsection{\MPt}
\Label{sec:MPt-F}

We follow the techniques introduced in
\cite{Collinucci:2008zs,Collinucci:2009uh,Blumenhagen:2009up} to
uplift IIB configurations to F-theory. One starts with the threefold
base $\cB_3$ of the fourfold obtained by taking the $\bZ_2$ quotient
of $\MPt$, as constructed in section~\ref{sec:semi-realistic}. In
general, one can present a Calabi-Yau fourfold elliptically fibered
over a base $\cB_3$ using a Weierstrass form:\footnote{As it is
  conventional when writing elliptic fibrations, we denote the
  coordinates on the fiber by $x,y,z$. We hope the reader will not get
  confused by the unrelated elliptic fibration and corresponding
  $x,y,z$ coordinates studied in section~\ref{sec:flavor-branes}.}
\begin{align}
  y^2 = x^3 + f x z^4 + g z^6\, ,
\end{align}
with $f$ a section of $K_{\cB_3}^{-4}$ and $g$ a section of
$K_{\cB_3}^{-6}$. Taking Sen's limit, one has that
\begin{align}
  \label{eq:Sen-limit-f}
  f\sim -3h^2 + \ldots
\end{align}
with $h=0$ the orientifold locus. In \ref{sec:semi-realistic} we found
that the orientifold locus in the Calabi-Yau threefold is located at
\begin{align}
\sqrt{h}=x_3x_9x_{10}^2x_{11}-x_4x_5^2x_6x_8=0
\end{align}
with the square root encoding the fact that the Calabi-Yau threefold
is the double cover of $\cB_3$, where $h$ is most naturally
defined. We thus have that:
\begin{align}
  \begin{split}
    h  & = \left(x_3x_9x_{10}^2x_{11}-x_4x_5^2x_6x_8\right)^2\\
    & = \left(x_3x_9x_{10}^2x_{11}+x_4x_5^2x_6x_8\right)^2 - 4x_3x_4
    x_6 x_9 (x_5x_{10})^2x_8x_{11}\\
    & = z_8^2 - 4 z_3 z_4^2 z_5 z_6\, ,
  \end{split}
\end{align}
which has degree $(2,0,0,0,0)=\deg(\ov K_{\cB_3}^2)$. Taking into
account~\eqref{eq:Sen-limit-f} and \eqref{eq:base-degree} this gives
the expected degree $\deg(\ov K_{\cB_3}^4)$ for $f$.

\medskip

In addition to having constructed the Calabi-Yau fourfold itself, we
want to go to a point in moduli space where the flavor and gauge $D7$
branes have the right structure close to the singularity. From the
local analysis in section~\ref{sec:flavor-branes} (see in particular
eqs.~\eqref{eq:flavor-charges-MP}) we want the discriminant to
degenerate as a $U(3)$ stack on top of the collapsing cycle, and 9
non-compact branes to intersect $dP_0$ on its hyperplane class,
locally splitting into a stack of $6$ $D7$ branes times a stack of $3$
$D7$ branes. Due to $\alpha'$ corrections the stack on top of the
collapsing cycle will recombine with the non-compact $U(3)$ stack to
give the $\cF_3$ flavor branes, while the 6 non-compact branes will
give rise to the $\cF_6$ branes.

It is not hard to see that there are indeed locations in the moduli
space of the elliptic fibration such that the discriminant has this
structure. In order to explicitly obtain these loci it is easiest to
work with the elliptic fibration in its Tate form \cite{Tate}:
\begin{align}
  \label{eq:tate-form}
  y^2 + a_1 xyz + a_3 y z^3 = x^3 + a_2 x^2 z^2 + a_4 xz^4 + a_6 z^6\, ,
\end{align}
where the $a_i$ are sections of $K_{\cB_3}^{-i}$. In order to have a
$U(3)$ stack on $z_5=0$ (the locus of the contracting cycle, recall
the map~\eqref{eq:involution-map}), we impose the following degrees of
vanishing \cite{Bershadsky:1996nh}:
\begin{align}
  \label{eq:U(3)-vanishing-degrees}
  \begin{split}
    \deg(a_1)=0 \qquad ; \qquad \deg(a_2)&=1 \qquad ; \qquad
    \deg(a_3)=1\\
    \deg(a_4)=2 \qquad &; \qquad \deg(a_6)=3\, ,
  \end{split}
\end{align}
with the notation meaning that close to the $z_5=0$ locus, $a_i$
vanishes as $z_5^{\deg(a_i)}(\ldots)$, with the quantity in
parenthesis generically non-vanishing. 
We find the following space of
solutions:
\begin{align}
  \label{eq:Tate-coeffs}
  \begin{split}
    a_1 & = c_{11} z_{2} z_{4}^{2} z_{5} z_{6} + c_{12} z_{8}\\
    a_2 & = z_5\left(c_{21} z_{2}^{2} z_{4}^{4} z_{5} z_{6}^{2} +
      c_{22} z_{2} z_{4}^{2} z_{6} z_{8} + c_{23} z_{3} z_{4}^{2}
      z_{6} \right)\\
    a_3 & = z_5\bigl(c_{31} z_{2}^{3} z_{4}^{6} z_{5}^{2} z_{6}^{3} +
    c_{32} z_{2} z_{4}^{2} z_{6} z_{8}^{2} + c_{33} z_{3} z_{4}^{2}
    z_{6}
    z_{8}\\
    &\phantom{=z_5\bigl(+} +c_{34} z_{2} z_{3} z_{4}^{4} z_{5}
    z_{6}^{2} +
    c_{35} z_{2}^{2} z_{4}^{4} z_{5} z_{6}^{2} z_{8} \bigr)\\
    a_4 & = z_5^{2}\bigl(c_{41} z_{2}^{4} z_{4}^{8} z_{5}^{2}
    z_{6}^{4} + c_{42} z_{2}^{2} z_{4}^{4} z_{6}^{2} z_{8}^{2} +
    c_{43} z_{3}^{2} z_{4}^{4} z_{6}^{2} \\
    & \phantom{= z_5^{2}\bigl(+} + c_{44} z_{2} z_{3} z_{4}^{4}
    z_{6}^{2} z_{8} + c_{45} z_{2}^{3} z_{4}^{6} z_{5} z_{6}^{3} z_{8}
    + c_{46} z_{2}^{2} z_{3} z_{4}^{6} z_{5} z_{6}^{3}
    \bigr)\\
    a_6 & = z_5^{3}\bigl(c_{61} z_{2}^{6} z_{4}^{12} z_{5}^{3}
    z_{6}^{6} + c_{62} z_{2}^{3} z_{4}^{6} z_{6}^{3} z_{8}^{3} +
    c_{63} z_{3}^{3} z_{4}^{6} z_{6}^{3}\\
    & \phantom{= z_5^{3}\bigl(+} + c_{64} z_{2}^{2} z_{3} z_{4}^{6}
    z_{6}^{3} z_{8}^{2} + c_{65} z_{2} z_{3}^{2} z_{4}^{6} z_{6}^{3}
    z_{8} + c_{66} z_{2}^{4} z_{4}^{8} z_{5} z_{6}^{4}
    z_{8}^{2}\\
    & \phantom{= z_5^{2}\bigl(+} + c_{67} z_{2}^{3} z_{3} z_{4}^{8}
    z_{5} z_{6}^{4} z_{8} + c_{68} z_{2}^{2} z_{3}^{2} z_{4}^{8} z_{5}
    z_{6}^{4} + c_{69} z_{2}^{5} z_{4}^{10} z_{5}^{2} z_{6}^{5}
    z_{8}\\
    & \phantom{= z_5^{2}\bigl(+} + c_{6,10} z_{2}^{4} z_{3} z_{4}^{10}
    z_{5}^{2} z_{6}^{5} \bigr)\, ,
  \end{split}
\end{align}
with the $c_{ij}$ arbitrary coefficients parametrizing the complex
structure moduli space. It is a straightforward calculation to verify
that the discriminant restricted to $z_5=0$ has the local form:
\begin{align}
  \label{eq:flavor-discriminant}
\frac{\Delta}{z_5^3}\Bigl|_{z_5=0}= z_4^6z_6^3f(z_0,
z_1,z_2,z_3,z_7,z_8)\, ,
\end{align}
with $f$ a section of $\cO(3)$ (here, and it what follows, the line
bundles are over $\bP^2=\{z_5=0\}$). In order to see this, notice
that, when restricted to $z_5=0$, the divisors $D_i=[\{z_i=0\}]$
become:
\begin{align}
  \label{eq:divisor-restrictions}
  (D_0,D_1,D_2,D_3,D_4,D_5,D_6,D_7,D_8)|_{z_5=0} = (\cO, \cO, \cO(1),
  \cO(1), \cO(1), \cO(-3), \cO, \cO, \cO)\, .
\end{align}
Note in particular that~\eqref{eq:base-degree} and
\ref{eq:pMSSM-MC-after-involution}, imply that $\ov K_{\cB_3}=D_8$,
and thus $\ov K_{\cB_3}|_{z_5=0}=\cO$, or in other words the geometry
is locally Calabi-Yau, as we expected. Since $\Delta$ is a section of
$\ov K_{\cB_3}^{12}$, and thus trivial when restricted to $z_5=0$, and
we have split off a factor of $z_5^3z_4^6z_6^3$, which is a locally
section of $\cO(3\cdot(-3) + 6\cdot 1 + 3\cdot 0) = \cO(-3)$, we
conclude that $f$ is a section of $\cO(3)$, or in other words a cubic
polynomial in the local coordinates of the contracting $\bP^2$.

It is natural to associate the $U(6)$ flavor stack with $z_4=0$. One
may be tempted to interpret the $z_6^3$ factor
in~\eqref{eq:flavor-discriminant} as the non-compact part of the
$U(3)$ stack. This is not correct, though: notice
from~\eqref{eq:divisor-restrictions} that $D_6$ becomes trivial when
restricted to $z_5=0$, so $z_6$ is effectively a non-zero
constant. Another way of seeing this is by recalling that the
hypersurface defining $\cB_3$ is in the same class as $z_0^3=0$, and
$z_0z_5z_6$ is in the Stanley-Reisner
ideal~\eqref{eq:dPt-quotient-SRI} of $\cA_{\dPt/\bZ_2}$.

The non-compact part of the $U(3)$ stack must then come from $f$. Thus, we learn that generically the $U(3)$ part of the flavor symmetry is
broken in this embedding, since $f$ does not generically vanish to
cubic order. This could be desirable for model building purposes, but
one can also tune coefficients to locally recover the symmetry. One
way of achieving this is to impose the vanishing
degrees~\eqref{eq:U(3)-vanishing-degrees} for the coordinate $z_2$
(so, in particular, the discriminant behaves as
$\Delta=z_2^3\bigl(\ldots\bigr)$). In this way we have a $U(3)$
singularity at $z_2=0$, which by~\eqref{eq:divisor-restrictions} will
intersect the contracting locus at a $\bP^1$. We  do this by
choosing coefficients in~\eqref{eq:Tate-coeffs} as follows:
\begin{align}
  c_{23} = c_{33} = c_{43} = c_{44} = c_{64} = c_{65} = c_{68} = 0
\end{align}
keeping the rest of the $c_{ij}$ arbitrary.

\section{Mapping the landscape of singularities}
\Label{sec:stats}

So far we have described how to analyze each embedding
individually. In order to find actual examples, one needs to scan over
a large class of toric ambient spaces, comparing their two dimensional
faces to our desired target geometry. We will focus our efforts on the
set of Calabi-Yau threefolds constructed by Kreuzer and Skarke
\cite{Kreuzer:2000xy}, containing 473,800,776 reflexive polytopes. We
refer to this set of Calabi-Yau manifolds as the \emph{KS landscape}
in what follows. Doing an exhaustive scan of such a big dataset
requires the systematic use of computers; in
section~\ref{sec:implementation} we give a short overview of the
computer tools we developed for performing this task.

The following set of questions, though by no means exhaustive, gives a
flavor of how singular the landscape is. Summarizing the results, it
is very simple to find singular Calabi-Yau spaces (including very
singular Calabi-Yau spaces).  Even relatively complicated Toric Lego
models is found in rather large numbers.

\subsection{How singular is the KS landscape?}

One of the most important questions one may try to answer concerns the
degree to which the landscape is singular. In our framework we address
this question by counting the number of interior points of the two
dimensional faces of the 4d polytopes. Locally, each of this interior
points corresponds to a zero-size 4-cycle. Naturally, whether this
4-cycle has zero size or not depends on where we are in the K\"ahler
cone of the Calabi-Yau hypersurface, and a priori it may be the case
that we cannot contract the 4-cycle without sending the whole
Calabi-Yau to zero size. Nevertheless, as we have seen in the examples
in section~\ref{sec:globalembed}, it is possible to contract many four
cycles even in relatively complex examples.

\begin{figure}
  \begin{center}
    \includegraphics[width=0.8\textwidth]{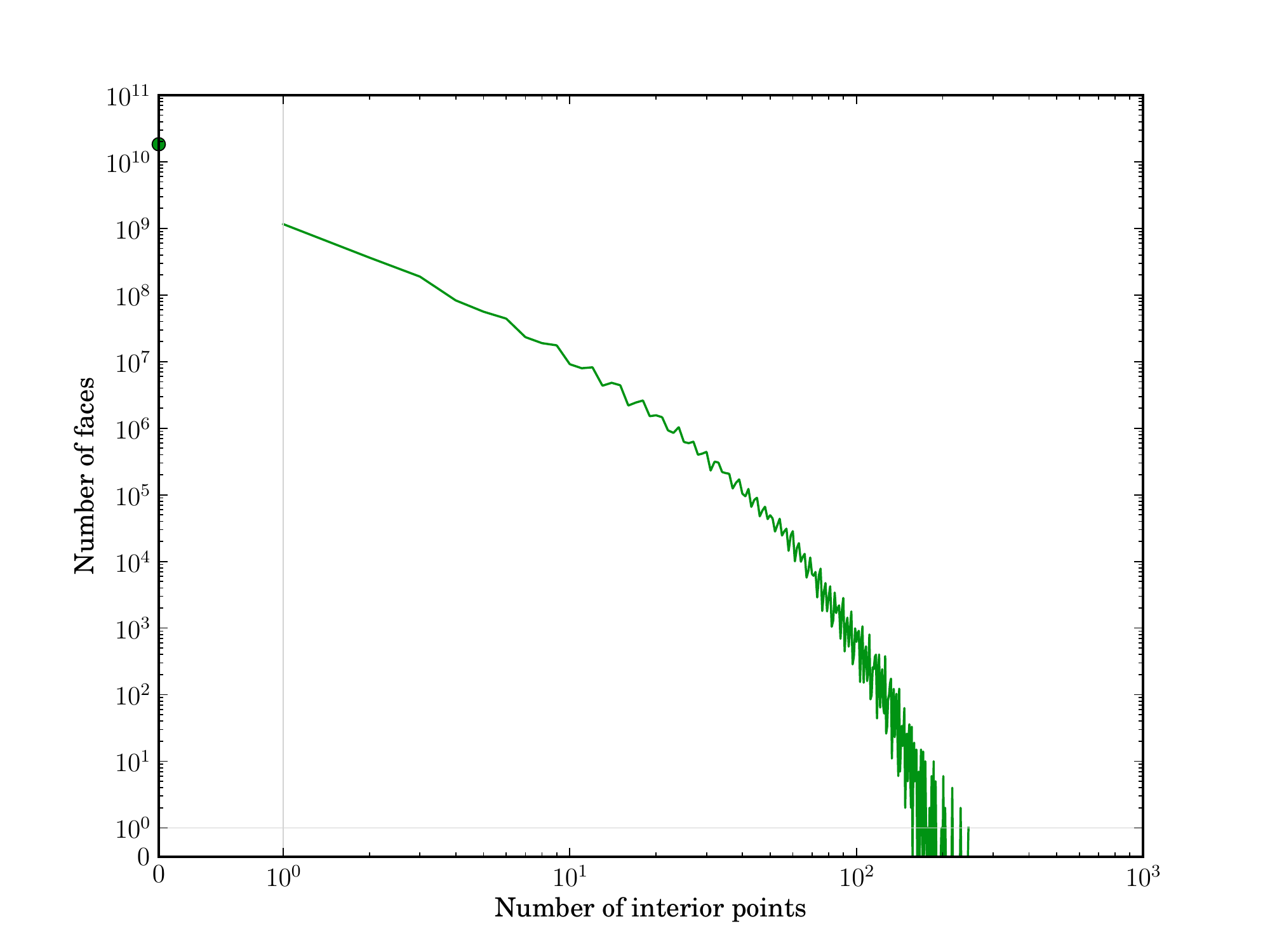}
  \end{center}

  \caption{Number of two dimensional faces in the KS landscape having
    the given number of interior points. The isolated dot shows the
    number of faces with no interior points.}
  \label{fig:IPH}
\end{figure}

According to this philosophy, we scanned the KS landscape counting the
number of internal points in each of the two-dimensional faces. The
result is displayed in figure~\ref{fig:IPH}.

One particularly remarkable feature of figure~\ref{fig:IPH} is its
smoothness. Despite the finite sample size, introducing some noise as
we go towards low frequencies, it does definitely look like the
singularity structure follows a well defined probability
distribution. A preliminary fit gives a power law distribution
$N_{faces}\sim N_{int}^{-0.4}$ up to $N_{int}\approx 50$, and from
then on an approximate exponential behavior: $N_{faces}\sim e^{-0.6
  N_{int}}$. It would be interesting to see whether this structure
persists as we consider larger datasets (natural generalizations are
higher dimensional reflexive polytopes). We leave this as an empirical
observation for the moment.

The distribution of interior points peaks at 0 interior points, but it
is rather common having just a few interior points. For reference, in
\autoref{table:interior-points} we include the number of two
dimensional faces with less than 10 interior points.
\begin{table}
  \centering
  \begin{tabular}{c|c}
    Interior points & Number of faces\\
    \hline
    0 & 18,348,252,546\\
    1 & 1,160,340,121\\
    2 & 364,176,255\\
    3 & 188,901,035\\
    4 & 82,981,171\\
    5 & 56,180,491\\
    6 & 44,224,288\\
    7 & 23,299,165\\
    8 & 18,939,629\\
    9 & 17,560,669
  \end{tabular}
  \caption{Number of two dimensional faces with less than 10 interior
    points.}
  \label{table:interior-points}
\end{table}

\subsection{Singularities vs.\ Hodge numbers}

From the combinatorial formula~\cite{1993alg.geom.10003B} for the
Hodge numbers of toric hypersurfaces
\begin{align}
  \begin{split}
    h^{11}(X) \;&= 
    \#(\nabla)
    - 4 - 1
    - \sum_{{\mathop{\mathrm{codim}}(\nu)=1}}\mathrm{Int}(\nu) 
    + \sum_{{\mathop{\mathrm{codim}}(\nu)=2}}\mathrm{Int}(\nu)\mathrm{Int}(\nu^*)
    \\[1ex]
    h^{21}(X) \;&= 
    \#(\Delta)
    - 4 - 1
    - \sum_{{\mathop{\mathrm{codim}}(\delta)=1}}\mathrm{Int}(\delta) 
    + \sum_{{\mathop{\mathrm{codim}}(\delta)=2}}\mathrm{Int}(\delta)\mathrm{Int}(\delta^*)
  \end{split}
\end{align}
\label{eq:Hodge-from-lattice}
\begin{figure}
  \centering
  \includegraphics{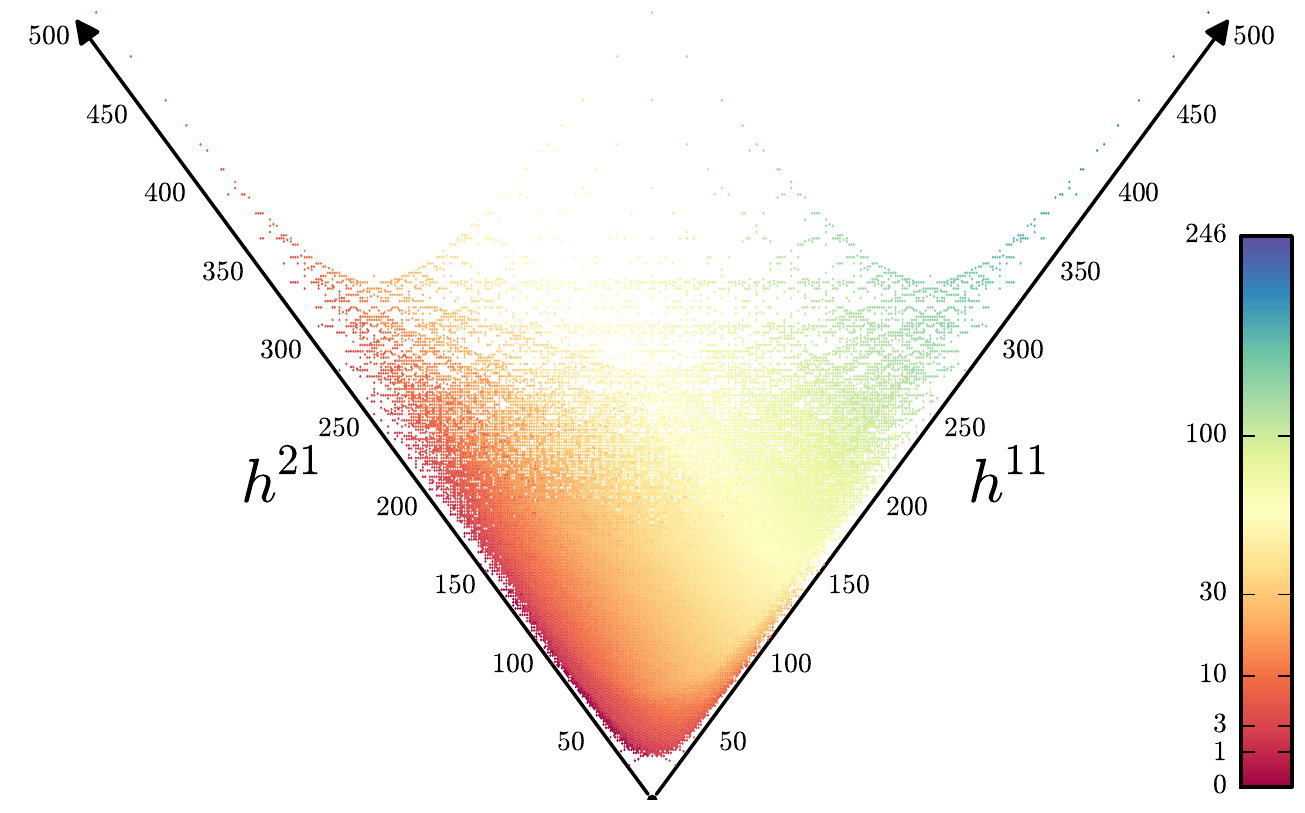}
  \caption{Maximal number of integral points on a single 2-face for each Hodge number pair.}
  \label{fig:max_2face}
\end{figure}
it is clear that the number of internal points in various faces of the
polytope is related to the Hodge numbers of the Calabi-Yau
hypersurface. In \autoref{fig:max_2face}, we plot the number of
internal points by Hodge numbers. The plot shows no mirror
symmetry. In fact, the number of internal points is mostly related to
$h^{11}$ only.

\subsection{The most singular polytope}

A natural question is how singular we can make the space. This has
been partially answered above, by studying the interior point
structure of the KS landscape. In this section we focus explicitly on
the most singular polytope that we found. Here we define \emph{most
  singular} as the polytope containing the two-dimensional face with
the largest number of interior points. The most singular manifold
defined in such a way is in fact the rightmost manifold in
\autoref{fig:max_2face}, with Hodge numbers $(h^{11}, h^{21})=(491,
11)$. Its polytope $\Sigma$ is defined by the following vertices:
\begin{equation}
  \begin{array}{rrrrr}
    v_1 & v_2 & v_3 & v_4 & v_5 \\
    \hline
    1 & -1 & 1 & 1 & 1\\
    1 & 1 & -2 & 1 & 1\\
    1 & 1 & 1 & -6 & 1\\
    1 & 1 & 1 & 1 & -83
  \end{array}
\end{equation}
In particular, its most singular two-dimensional face is defined by
$v_1$, $v_4$ and $v_5$, and it has 246 integral points. Locally, it
defines a $\bC^3/(\bZ_{84}\times\bZ_{7})$ singularity. In fact, all 10
of the two dimensional faces of this polytope are of the form
$\bC^3/(\bZ_p\times\bZ_q)$, with
\begin{equation}
  \left\{(p,q)\right\} = \left\{(84,3), (84,7), (7,3), (7,3), (84,2),
      (3,2), (3,2), (7,2), (7,2), (1,1)\right\} \,.
\end{equation}

Let us remark in passing that the Calabi-Yau manifold $X$ obtained
from $\Sigma$ is both elliptically and K3 fibered, and thus gives an
interesting background for studying heterotic/F-theory duality.  It
was in fact studied in \cite{Candelas:1997eh}, where it was found that
compactifying F-theory on $X$ gives rise to a 6d theory with gauge
group $G=E_8^{17}\times F_4^{16}\times G_2^{32}\times SU(2)^{32}$ and
$n_T=193$.\footnote{We would like to thank W. Taylor for a remark
  about this point.}

\subsection{Toric del Pezzo singularities}

Calabi-Yau spaces with del Pezzo singularities are of particular
interest to model building
\cite{Aldazabal:2000sa,Wijnholt:2002qz,Verlinde:2005jr,Dolan:2011qu,Donagi:2008ca,Beasley:2008dc,Hayashi:2008ba,Beasley:2008kw,Donagi:2008kj}. The
results of a scan for two dimensional faces being the toric diagrams
of del Pezzo singularities is displayed in \autoref{table:del-Pezzos}.
\begin{table}
  \centering
  \begin{tabular}{c|c}
    Singularity & Number of faces\\
    \hline
    $dP_0$ & 6,438,735\\
    $dP_1$ & 33,073,205\\
    $dP_2$ & 60,732,256\\
    $dP_3$ & 17,085,648 
  \end{tabular}

  \caption{Local del Pezzo singularities.}
  \label{table:del-Pezzos}
\end{table}

Note that some care is required in interpreting the results in
table~\ref{table:del-Pezzos}. While the results denote two-dimensional
faces which correspond precisely to the toric diagrams for del Pezzo
surfaces, there is a much larger number of faces that \emph{contain}
the del Pezzo diagrams. If one has one such face containing a del
Pezzo singularity, a series of small resolutions of the singularity
may leave behind precisely the desired del Pezzo. A similar caveat
applies to the rest of the discussion in this section.

\subsection{$Y^{(p,q)}$ and $L^{(a,b,a)}$ cones}

\begin{figure}[htp]
  \centering
  \includegraphics[width=0.8\textwidth]{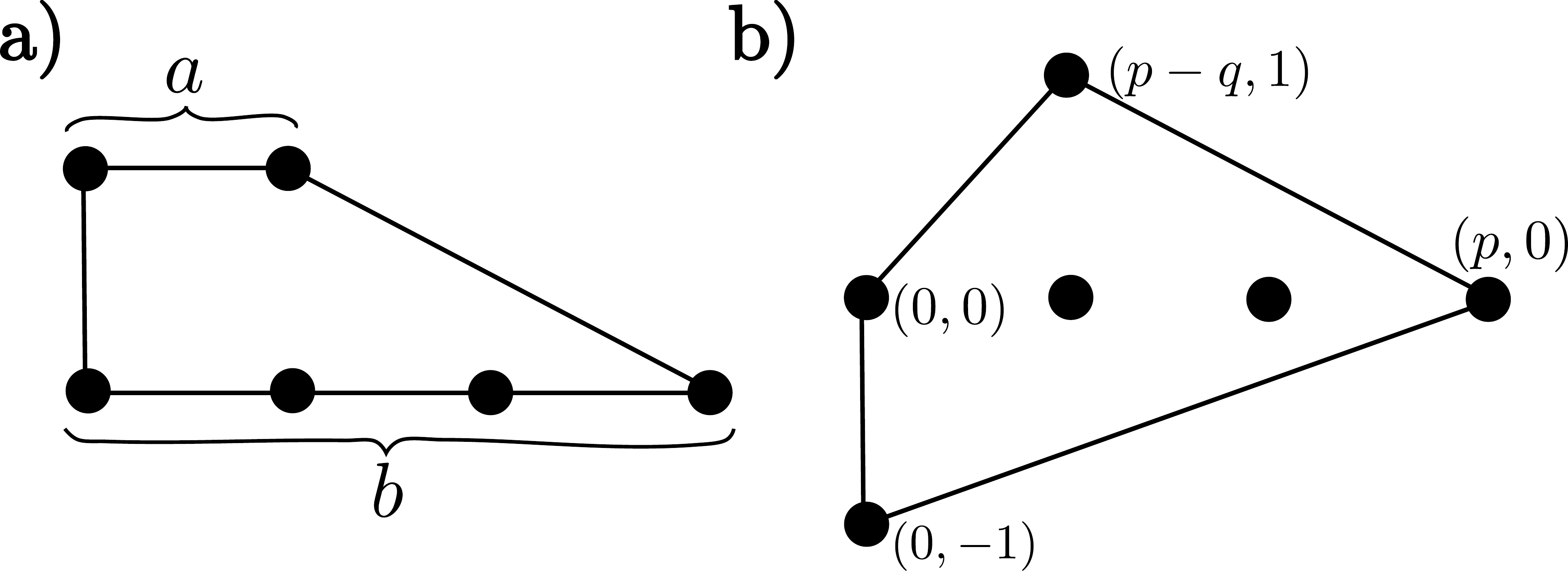}

  \caption{a) Toric diagram for a $L^{(a,b,a)}$ singularity. The
    particular example shown is $L^{(2,4,2)}$. b) Toric diagram for
    $Y^{(p,q)}$. We have shown $Y^{(3,2)}$.}
  \label{fig:Laba-Ypq-diagrams}
\end{figure}

The $L^{(a,b,a)}$ family of local toric singularities generalizes the
conifold and the suspended pinch point (SPP) singularities (these geometries belong to the
general $L^{(a,b,c)}$ family \cite{Cvetic:2005ft,Martelli:2005wy},
which also comprises the $Y^{(p,q)}$ geometries below). The generic
toric diagram for one such singularity is displayed in
\autoref{fig:Laba-Ypq-diagrams}a, it has vertices at $(0,0)$, $(0,1)$,
$(a,1)$ and $(b,0)$. We choose the convention $a\leq b$.

\begin{figure}[htp]
  \centering
  \includegraphics[width=0.7\textwidth]{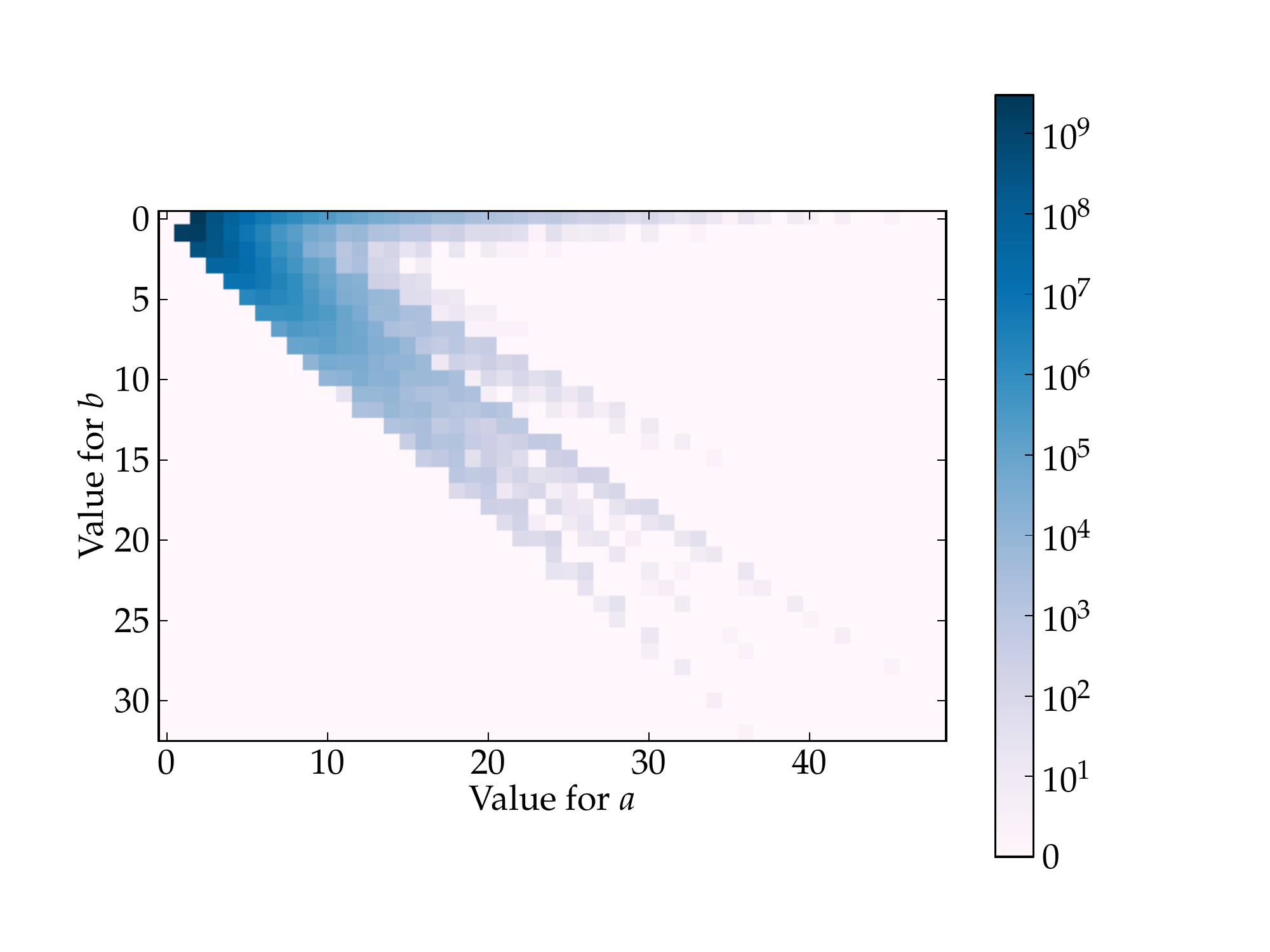}

  \caption{Plot of the number of local $L^{(a,b,a)}$ geometries in the
    KS landscape. $L^{(0,1,0)}$ is just a copy of flat space, so we
    have omitted it from the plot.}
  \label{fig:Laba-plot}
\end{figure}

These singularities find a particularly nice model building use in the
context of metastable supersymmetry breaking, see for example
\cite{Argurio:2006ny,Argurio:2007qk}. We have performed an exhaustive
scan over the KS landscape for these singularities, with the results
shown in \autoref{fig:Laba-plot}.
Notice that very singular $L^{(a,b,a)}$ singularities tend to have
either $a\approx b$, i.e. they are roughly orbifolds of the conifold,
or $b\approx 0$, in which case are geometries of the form
$\bC^2/\bZ_n\times\bC$.

A related large class of local singularities that we studied are the
$Y^{(p,q)}$ cones
\cite{Gauntlett:2004zh,Gauntlett:2004yd,Martelli:2004wu}. They can
also play an important role in the local model building approach (see
for example~\cite{GarciaEtxebarria:2007vh} for applications to
metastable supersymmetry breaking) and also in $\cN=1$ AdS/CFT. We
show the corresponding toric diagram in
\autoref{fig:Laba-Ypq-diagrams}b. We have collected the results of the
exhaustive scan for these singularities in \autoref{table:Ypq}.
\begin{table}
\begin{tabular}{c|cccccccccccc}
  & $q=0$ & 1 & 2 & 3 & 4 & 5 & 6 & 7 & 8 & 9 & 10 & 11 \\
  \hline
  $Y^{2,q}$ & 12175355 & 33073205 & & & & & & & &  & & \\
  $Y^{3,q}$ &  203038 & 531568 & 311508 & & & & & & & & & \\
  $Y^{4,q}$ &  13123 & 14868 & 19772 & 10553 & & & & & &  & & \\
  $Y^{5,q}$ &  200 & 672 & 357 & 632 & 351 & & & &  & & &  \\
  $Y^{6,q}$ &  114 & 129 & 171 & 130 & 167 & 129 & & & &  & &  \\
  $Y^{7,q}$ &  20 & 38 & 35 & 38 & 35 & 38 & 35 & & & & &  \\
  $Y^{8,q}$ &  14 & 19 & 20 & 19 & 20 & 19 & 20 & 19 & & & & \\
  $Y^{9,q}$ &  3 & 5 & 5 & 5 & 5 & 5 & 5 & 5 & 5 & & & \\
  $Y^{10,q}$ &  2 & 3 & 3 & 3 & 3 & 3 & 3 & 3 & 3 & 3 & & \\
  $Y^{11,q}$ &  0 & 0 & 0 & 0 & 0 & 0 & 0 & 0 & 0 & 0 & 0 & \\
  $Y^{12,q}$ &  1 & 1 & 1 & 1 & 1 & 1 & 1 & 1 & 1 & 1 & 1 & 1
\end{tabular}
\caption{Number of local $Y^{p,q}$ cones. No $Y^{p,q}$ spaces with
  $p>12$ were found.}
\label{table:Ypq}
\end{table}

\subsection{Toric Lego models}

\begin{figure}[htp]
  \centering
  \includegraphics[width=0.3\textwidth]{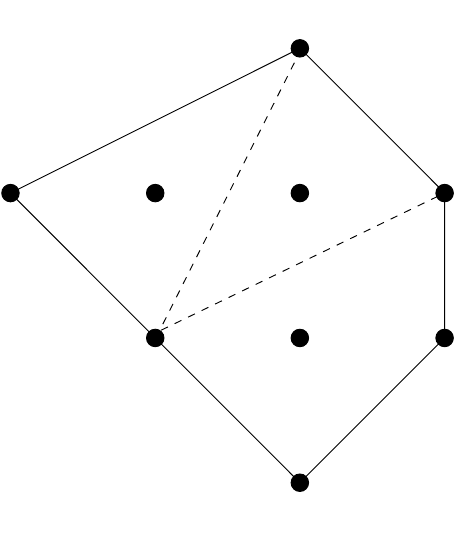}

  \caption{Three sector ($dP_0+dP_0+dP_1$) model considered in
    \cite{Balasubramanian:2009tv}.}
  \label{fig:three-sector-lego}
\end{figure}

Our original motivation in approaching the problem of finding global
embeddings was to see whether there was any obstruction to finding
global embeddings for Toric Lego singularities. As representative
examples, we have performed exhaustive scans for a model consisting of
three $dP_0$ sectors (given by the toric diagram in
figure~\ref{fig:permutation-MSSM}), and one three sector model
($dP_0+dP_0+dP_1$) considered previously in
\cite{Balasubramanian:2009tv}, which we reproduce in
figure~\ref{fig:three-sector-lego} for convenience.
For the model in figure~\ref{fig:permutation-MSSM} we found 41,799
polytopes having that toric diagram as one of their two-dimensional
faces, and 292,691 polytopes has figure~\ref{fig:three-sector-lego} as
one of their two dimensional faces.

\subsection{Computer implementation}
\Label{sec:implementation}

The scans performed above were done on a 2.5 GHZ Intel Core 2 Duo,
with 4 GB of 667 MHz DDR2 SDRAM (MacBook Pro), using PALP
\cite{Kreuzer:2002uu} to generate the polytopes and some custom C code
to analyze the polytopes for particular singularities. The code is
available at the address
\begin{center}
  \url{http://cern.ch/inaki/scan.tar.gz}
\end{center}
together with the results of the scan. The whole
scan was completed in 4 days, with the two cores running in parallel.

In order to analyze the resulting set of polytopes, we used Sage
\cite{sage47}, and in particular its toric varieties package
\cite{toricsage}.

\section{Conclusions and generalizations}
\Label{sec:conclusions}

In this paper we have provided a prescription for how to find global
realizations of local models of gauge and matter content in which
D-branes are placed at toric singularities. In particular, we focus on
hypersurfaces in four dimensional toric varieties, using the complete
classification of such a class of ambient
spaces~\cite{Kreuzer:2000xy}. The global models are then either given
in terms of type IIB orientifolds on compact Calabi-Yau manifolds or
by considering a Calabi-Yau four-fold elliptically fibered over a
compact three-dimensional base.  Our construction provides a large
class of models in which we can explore further important issues such
as moduli stabilization in flux compactifications on type IIB
orientifolds. In this way, our results link the rich literature on
models at local toric singularities (see e.g.
\cite{Aldazabal:2000sk,Berenstein:2001nk,Cascales:2003wn,Dolan:2011qu})
with the wealth of results on moduli stabilization (see
\cite{Witten:1996bn,Kachru:2003aw,Balasubramanian:2005zx,Kallosh:2005gs,Bergshoeff:2005yp}
for some of the classical results, and
\cite{Blumenhagen:2007sm,Uranga:2008nh,Blumenhagen:2009gk,Bobkov:2010rf,Grimm:2011dj,Cicoli:2011it,Cicoli:2011qg}
for some interesting recent developments).  The models studied in the present paper also present
a way of embedding local singularities into the very large class of
backgrounds appearing in F-theory. As opposed to the usual F-theory
construction of GUTs, in our construction the MSSM sector comes from a
very singular region, which cannot be described in F-theory. This, of
course, does not preclude the bulk F-theory to have some interesting
dynamics of its own.
  
In our discussion of the landscape, we found that singularities in the
Calabi-Yau are \emph{generic}. One can of course blow up to a smooth
manifold, but since this breaks part of the symmetry once we have
branes around, one expects that configurations with branes at
singularities are dynamically preferred \cite{Kofman:2004yc}.  It is
then an interesting and natural question to explore what happens in
very singular manifolds. Our work provides some tools for analyzing
this question in the context of toric geometries.

Finally, we have focused on the tools and the generic discussion. We
hope to come back to the construction of realistic models using the
techniques presented in this paper.

\subsection{More general F-theory bases}
\Label{sec:conclusions-F}

While in this paper we have proceeded mostly along the traditional IIB
route, it is clear that one can directly construct many bases of
elliptically fibered fourfolds having the required structure for the
analysis in this paper to apply. Let us assume that we construct
the basis $\cB_3$ as a hypersurface in an ambient toric space
$\cA_\nabla$,\footnote{The extension to complete intersections
  proceeds very similarly to that in
  section~\ref{sec:conclusions-CI}.} described by a polytope $\nabla$
(this setup has been recently analyzed in
\cite{Blumenhagen:2009yv,Grimm:2009yu,Chen:2010ts,Knapp:2011wk,Knapp:2011ip}). Then we require that:
\begin{enumerate}
  \item $\nabla$ has a two-dimensional face describing the singularity
    that we want to embed.
  \item The restriction of $K_{\cB_3}$, the canonical class of
    $\cB_3$, to the contracting locus is trivial.
\end{enumerate}
The first condition follows from the same arguments as in
section~\ref{sec:globalembed-general}. The second condition is
slightly more subtle, and encodes the fact that in the neighborhood of
the singularity there are no orientifolds. This is the case in many of
the examples in the literature. Condition 2 then ensures that close to
the singularity the space is Calabi-Yau, as in
section~\ref{sec:MPt-F}. This last condition can easily be modified if
one wants local models with orientifolds. One just has to compute the
canonical class of the base after the involution, similar to the
computation that was done in section~\ref{sec:semi-realistic}.

\subsection{Complete intersections and toroidal orbifolds}
\Label{sec:conclusions-CI}

It would be interesting to extend our searches to the more general
class of complete intersection Calabi-Yau manifolds (CICYs) in a toric
variety. Unfortunately, the complete classification of higher
dimensional toric ambient spaces ($n>4$, with $n$ the complex
dimension) is unknown, though partial results are known for $n=5$ and
$n=6$~\cite{Kreuzer:2001fu}. While this implies that it is at present
not possible to carry out a complete search for the most general class
of Calabi-Yau manifolds, our construction, and in particular the
general discussion in section~\ref{sec:globalembed-general}, carries
over straightforwardly to the case of CICYs, mutatis mutandis.

Consider a $n$-dimensional reflexive polytope $\nabla$, together with
some nef partition, giving a three-dimensional Calabi-Yau space $M_X$
as a complete intersection on $\cA_\nabla$, the toric space associated
to $\nabla$. Assume that we find that the toric diagram for the local
toric Calabi-Yau $X$ in which we are interested appears as one of the
two-dimensional faces of $\nabla$, or equivalently, the toric fan for
$X$ is embedded as a three-dimensional cone in $\nabla$. Then
$\cA_\nabla$ has a local patch of the form $X\times (\bC^*)^{n-3}$,
and the set of hypersurfaces defining the complete intersection will
generically give rise to a copy of $X$ embedded in $M_X$.

\medskip

This observation allows one to study toroidal orbifolds using our
methods. The only requirement is being able to describe the toroidal
orbifold of interest as a complete intersection in an ambient toric
variety. We refer the reader to the nice recent paper
\cite{Blaszczyk:2011hs} for a careful analysis of how, and in which
cases, it is possible to realize toroidal orbifolds as CICYs.

\acknowledgments 

We would like to thank Mirjam Cveti\v{c}, Andr\'es Collinucci and
\'Angel Uranga for enlightening discussions. P.B. acknowledges the
hospitality of the theory group at CERN, where this project was
initiated and completed, as well as the Dublin Institute for Advanced
Studies, the University of Pennsylvania, the KITPC, Beijing and the
Berkeley Center for Theoretical Physics. V.B.\ acknowledges the
hospitality of the KITP (which is supported in part by the National
Science Foundation under Grant No.\ NSF PHY05-51164) during the final
stage of this project.  The other V.B. was supported in part by  DOE grant DE-FG02-95ER40893.
I.G.-E. acknowledges the hospitality of the
HKUST Institute for Advanced Study and the University of New
Hampshire, where parts of this work were
completed. I.G.\nobreakdash-E. would also like to thank N. Hasegawa
for kind encouragement and support. The work of P.B. is supported by
the NSF CAREER grant PHY-0645686 and by the University of New
Hampshire through its Faculty Scholars Award Program.

\bibliographystyle{JHEP}
\bibliography{refs}

\providecommand{\href}[2]{#2}\begingroup\raggedright\begin{thebibliography}{10}

\bibitem{Douglas:1996sw}
M.~R. Douglas and G.~W. Moore, {\it {D-branes, quivers, and ALE instantons}},
  \href{http://xxx.lanl.gov/abs/hep-th/9603167}{{\tt hep-th/9603167}}.

\bibitem{Klebanov:1998hh}
I.~R. Klebanov and E.~Witten, {\it {Superconformal field theory on three-branes
  at a Calabi-Yau singularity}},  {\em Nucl.Phys.} {\bf B536} (1998) 199--218,
  [\href{http://xxx.lanl.gov/abs/hep-th/9807080}{{\tt hep-th/9807080}}].

\bibitem{Morrison:1998cs}
D.~R. Morrison and M.~Plesser, {\it {Nonspherical horizons. 1.}},  {\em
  Adv.Theor.Math.Phys.} {\bf 3} (1999) 1--81,
  [\href{http://xxx.lanl.gov/abs/hep-th/9810201}{{\tt hep-th/9810201}}].
  Revised.

\bibitem{Aldazabal:2000sa}
G.~Aldazabal, L.~E. Ibanez, F.~Quevedo, and A.~Uranga, {\it {D-branes at
  singularities: A Bottom up approach to the string embedding of the standard
  model}},  {\em JHEP} {\bf 0008} (2000) 002,
  [\href{http://xxx.lanl.gov/abs/hep-th/0005067}{{\tt hep-th/0005067}}].

\bibitem{Berenstein:2001nk}
D.~Berenstein, V.~Jejjala, and R.~G. Leigh, {\it {The Standard model on a
  D-brane}},  {\em Phys.Rev.Lett.} {\bf 88} (2002) 071602,
  [\href{http://xxx.lanl.gov/abs/hep-ph/0105042}{{\tt hep-ph/0105042}}].

\bibitem{Verlinde:2005jr}
H.~Verlinde and M.~Wijnholt, {\it {Building the Standard Model on a D3-brane}},
   {\em JHEP} {\bf 01} (2007) 106,
  [\href{http://xxx.lanl.gov/abs/hep-th/0508089}{{\tt hep-th/0508089}}].

\bibitem{Balasubramanian:2009tv}
V.~Balasubramanian, P.~Berglund, and I.~Garc\'ia-Etxebarria, {\it {Toric Lego:
  A Method for modular model building}},  {\em JHEP} {\bf 1001} (2010) 076,
  [\href{http://xxx.lanl.gov/abs/0910.3616}{{\tt arXiv:0910.3616}}].

\bibitem{GarciaEtxebarria:2006aq}
I.~Garc\'ia-Etxebarria, F.~Saad, and A.~M. Uranga, {\it {Quiver gauge theories
  at resolved and deformed singularities using dimers}},  {\em JHEP} {\bf 06}
  (2006) 055, [\href{http://xxx.lanl.gov/abs/hep-th/0603108}{{\tt
  hep-th/0603108}}].

\bibitem{GarciaEtxebarria:2006rw}
I.~Garc\'ia-Etxebarria, F.~Saad, and A.~M. Uranga, {\it {Local models of gauge
  mediated supersymmetry breaking in string theory}},  {\em JHEP} {\bf 08}
  (2006) 069, [\href{http://xxx.lanl.gov/abs/hep-th/0605166}{{\tt
  hep-th/0605166}}].

\bibitem{Kreuzer:2000xy}
M.~Kreuzer and H.~Skarke, {\it {Complete classification of reflexive polyhedra
  in four dimensions}},  {\em Adv. Theor. Math. Phys.} {\bf 4} (2002)
  1209--1230, [\href{http://xxx.lanl.gov/abs/hep-th/0002240}{{\tt
  hep-th/0002240}}].

\bibitem{Donagi:2008ca}
R.~Donagi and M.~Wijnholt, {\it {Model Building with F-Theory}},
  \href{http://xxx.lanl.gov/abs/0802.2969}{{\tt arXiv:0802.2969}}.

\bibitem{Beasley:2008dc}
C.~Beasley, J.~J. Heckman, and C.~Vafa, {\it {GUTs and Exceptional Branes in
  F-theory - I}},  {\em JHEP} {\bf 01} (2009) 058,
  [\href{http://xxx.lanl.gov/abs/0802.3391}{{\tt arXiv:0802.3391}}].

\bibitem{Hayashi:2008ba}
H.~Hayashi, R.~Tatar, Y.~Toda, T.~Watari, and M.~Yamazaki, {\it {New Aspects of
  Heterotic--F Theory Duality}},  {\em Nucl. Phys.} {\bf B806} (2009) 224--299,
  [\href{http://xxx.lanl.gov/abs/0805.1057}{{\tt arXiv:0805.1057}}].

\bibitem{Beasley:2008kw}
C.~Beasley, J.~J. Heckman, and C.~Vafa, {\it {GUTs and Exceptional Branes in
  F-theory - II: Experimental Predictions}},  {\em JHEP} {\bf 01} (2009) 059,
  [\href{http://xxx.lanl.gov/abs/0806.0102}{{\tt arXiv:0806.0102}}].

\bibitem{Donagi:2008kj}
R.~Donagi and M.~Wijnholt, {\it {Breaking GUT Groups in F-Theory}},
  \href{http://xxx.lanl.gov/abs/0808.2223}{{\tt arXiv:0808.2223}}.

\bibitem{Blumenhagen:2008zz}
R.~Blumenhagen, V.~Braun, T.~W. Grimm, and T.~Weigand, {\it {GUTs in Type IIB
  Orientifold Compactifications}},  {\em Nucl.Phys.} {\bf B815} (2009) 1--94,
  [\href{http://xxx.lanl.gov/abs/0811.2936}{{\tt arXiv:0811.2936}}].

\bibitem{Diaconescu:2005pc}
D.-E. Diaconescu, B.~Florea, S.~Kachru, and P.~Svrcek, {\it {Gauge - mediated
  supersymmetry breaking in string compactifications}},  {\em JHEP} {\bf 02}
  (2006) 020, [\href{http://xxx.lanl.gov/abs/hep-th/0512170}{{\tt
  hep-th/0512170}}].

\bibitem{Buican:2006sn}
M.~Buican, D.~Malyshev, D.~R. Morrison, H.~Verlinde, and M.~Wijnholt, {\it
  {D-branes at singularities, compactification, and hypercharge}},  {\em JHEP}
  {\bf 01} (2007) 107, [\href{http://xxx.lanl.gov/abs/hep-th/0610007}{{\tt
  hep-th/0610007}}].

\bibitem{Dolan:2011qu}
M.~J. Dolan, S.~Krippendorf, and F.~Quevedo, {\it {Towards a Systematic
  Construction of Realistic D-brane Models on a del Pezzo Singularity}},  {\em
  JHEP} {\bf 1110} (2011) 024, [\href{http://xxx.lanl.gov/abs/1106.6039}{{\tt
  arXiv:1106.6039}}].

\bibitem{Blumenhagen:2005mu}
R.~Blumenhagen, M.~Cveti\v{c}, P.~Langacker, and G.~Shiu, {\it {Toward
  realistic intersecting D-brane models}},  {\em Ann.Rev.Nucl.Part.Sci.} {\bf
  55} (2005) 71--139, [\href{http://xxx.lanl.gov/abs/hep-th/0502005}{{\tt
  hep-th/0502005}}].

\bibitem{Kumar:2005hf}
J.~Kumar and J.~D. Wells, {\it {Surveying standard model flux vacua on T**6 /
  Z(2) x Z(2)}},  {\em JHEP} {\bf 0509} (2005) 067,
  [\href{http://xxx.lanl.gov/abs/hep-th/0506252}{{\tt hep-th/0506252}}].

\bibitem{Blumenhagen:2004xx}
R.~Blumenhagen, F.~Gmeiner, G.~Honecker, D.~Lust, and T.~Weigand, {\it {The
  Statistics of supersymmetric D-brane models}},  {\em Nucl.Phys.} {\bf B713}
  (2005) 83--135, [\href{http://xxx.lanl.gov/abs/hep-th/0411173}{{\tt
  hep-th/0411173}}].

\bibitem{Gmeiner:2005vz}
F.~Gmeiner, R.~Blumenhagen, G.~Honecker, D.~Lust, and T.~Weigand, {\it {One in
  a billion: MSSM-like D-brane statistics}},  {\em JHEP} {\bf 0601} (2006) 004,
  [\href{http://xxx.lanl.gov/abs/hep-th/0510170}{{\tt hep-th/0510170}}].

\bibitem{Douglas:2006xy}
M.~R. Douglas and W.~Taylor, {\it {The Landscape of intersecting brane
  models}},  {\em JHEP} {\bf 0701} (2007) 031,
  [\href{http://xxx.lanl.gov/abs/hep-th/0606109}{{\tt hep-th/0606109}}].

\bibitem{Gmeiner:2007zz}
F.~Gmeiner and G.~Honecker, {\it {Mapping an Island in the Landscape}},  {\em
  JHEP} {\bf 0709} (2007) 128, [\href{http://xxx.lanl.gov/abs/0708.2285}{{\tt
  arXiv:0708.2285}}].

\bibitem{Gmeiner:2008xq}
F.~Gmeiner and G.~Honecker, {\it {Millions of Standard Models on Z-prime(6)?}},
   {\em JHEP} {\bf 0807} (2008) 052,
  [\href{http://xxx.lanl.gov/abs/0806.3039}{{\tt arXiv:0806.3039}}].

\bibitem{Davies:2009ub}
R.~Davies, {\it {Quotients of the conifold in compact Calabi-Yau threefolds,
  and new topological transitions}},  {\em Adv.Theor.Math.Phys.} {\bf 14}
  (2010) 965--990, [\href{http://xxx.lanl.gov/abs/0911.0708}{{\tt
  arXiv:0911.0708}}].

\bibitem{Davies:2011is}
R.~Davies, {\it {Hyperconifold Transitions, Mirror Symmetry, and String
  Theory}},  {\em Nucl.Phys.} {\bf B850} (2011) 214--231,
  [\href{http://xxx.lanl.gov/abs/1102.1428}{{\tt arXiv:1102.1428}}].

\bibitem{fulton}
W.~Fulton, {\em Introduction to toric varieties}.
\newblock Annals of mathematics studies. Princeton University Press, 1993.

\bibitem{Hori:2003ic}
K.~Hori, S.~Katz, A.~Klemm, R.~Pandharipande, R.~Thomas, {\em et.~al.}, {\it
  {Mirror symmetry}}, .

\bibitem{Bouchard:2007ik}
V.~Bouchard, {\it {Lectures on complex geometry, Calabi-Yau manifolds and toric
  geometry}},  \href{http://xxx.lanl.gov/abs/hep-th/0702063}{{\tt
  hep-th/0702063}}. An older version of these notes was published in the
  Proceedings of the Modave Summer School in Mathematical Physics 2005.

\bibitem{CLS}
D.~A. Cox, J.~B. Little, and H.~K. Schenck, {\em Toric Varieties}.
\newblock Graduate Studies in Mathematics. AMS, 2011.

\bibitem{Hosono:1993qy}
S.~Hosono, A.~Klemm, S.~Theisen, and S.-T. Yau, {\it {Mirror symmetry, mirror
  map and applications to Calabi-Yau hypersurfaces}},  {\em Commun.Math.Phys.}
  {\bf 167} (1995) 301--350,
  [\href{http://xxx.lanl.gov/abs/hep-th/9308122}{{\tt hep-th/9308122}}].

\bibitem{Candelas:1994hw}
P.~Candelas, A.~Font, S.~H. Katz, and D.~R. Morrison, {\it {Mirror symmetry for
  two parameter models. 2.}},  {\em Nucl.Phys.} {\bf B429} (1994) 626--674,
  [\href{http://xxx.lanl.gov/abs/hep-th/9403187}{{\tt hep-th/9403187}}].

\bibitem{Morrison:1996pp}
D.~R. Morrison and C.~Vafa, {\it {Compactifications of F theory on Calabi-Yau
  threefolds. 2.}},  {\em Nucl.Phys.} {\bf B476} (1996) 437--469,
  [\href{http://xxx.lanl.gov/abs/hep-th/9603161}{{\tt hep-th/9603161}}].

\bibitem{Braun:2011ux}
V.~Braun, {\it {Toric Elliptic Fibrations and F-Theory Compactifications}},
  \href{http://xxx.lanl.gov/abs/1110.4883}{{\tt arXiv:1110.4883}}.

\bibitem{Witten:1993yc}
E.~Witten, {\it {Phases of N=2 theories in two-dimensions}},  {\em Nucl.Phys.}
  {\bf B403} (1993) 159--222,
  [\href{http://xxx.lanl.gov/abs/hep-th/9301042}{{\tt hep-th/9301042}}].

\bibitem{Aspinwall:1993yb}
P.~S. Aspinwall, B.~R. Greene, and D.~R. Morrison, {\it {Multiple mirror
  manifolds and topology change in string theory}},  {\em Phys.Lett.} {\bf
  B303} (1993) 249--259, [\href{http://xxx.lanl.gov/abs/hep-th/9301043}{{\tt
  hep-th/9301043}}].

\bibitem{Berglund:1995gd}
P.~Berglund, S.~H. Katz, and A.~Klemm, {\it {Mirror symmetry and the moduli
  space for generic hypersurfaces in toric varieties}},  {\em Nucl.Phys.} {\bf
  B456} (1995) 153--204, [\href{http://xxx.lanl.gov/abs/hep-th/9506091}{{\tt
  hep-th/9506091}}].

\bibitem{Balasubramanian:2005zx}
V.~Balasubramanian, P.~Berglund, J.~P. Conlon, and F.~Quevedo, {\it
  {Systematics of moduli stabilisation in Calabi-Yau flux compactifications}},
  {\em JHEP} {\bf 0503} (2005) 007,
  [\href{http://xxx.lanl.gov/abs/hep-th/0502058}{{\tt hep-th/0502058}}].

\bibitem{Argurio:2006ny}
R.~Argurio, M.~Bertolini, S.~Franco, and S.~Kachru, {\it {Gauge/gravity duality
  and meta-stable dynamical supersymmetry breaking}},  {\em JHEP} {\bf 01}
  (2007) 083, [\href{http://xxx.lanl.gov/abs/hep-th/0610212}{{\tt
  hep-th/0610212}}].

\bibitem{Argurio:2007qk}
R.~Argurio, M.~Bertolini, S.~Franco, and S.~Kachru, {\it {Metastable vacua and
  D-branes at the conifold}},  {\em JHEP} {\bf 06} (2007) 017,
  [\href{http://xxx.lanl.gov/abs/hep-th/0703236}{{\tt hep-th/0703236}}].

\bibitem{Diaconescu:1999dt}
D.-E. Diaconescu and J.~Gomis, {\it {Fractional branes and boundary states in
  orbifold theories}},  {\em JHEP} {\bf 0010} (2000) 001,
  [\href{http://xxx.lanl.gov/abs/hep-th/9906242}{{\tt hep-th/9906242}}].

\bibitem{Douglas:2000ah}
M.~R. Douglas, B.~Fiol, and C.~Romelsberger, {\it {Stability and BPS branes}},
  {\em JHEP} {\bf 0509} (2005) 006,
  [\href{http://xxx.lanl.gov/abs/hep-th/0002037}{{\tt hep-th/0002037}}].

\bibitem{Douglas:2000qw}
M.~R. Douglas, B.~Fiol, and C.~Romelsberger, {\it {The Spectrum of BPS branes
  on a noncompact Calabi-Yau}},  {\em JHEP} {\bf 0509} (2005) 057,
  [\href{http://xxx.lanl.gov/abs/hep-th/0003263}{{\tt hep-th/0003263}}].

\bibitem{Cachazo:2001sg}
F.~Cachazo, B.~Fiol, K.~A. Intriligator, S.~Katz, and C.~Vafa, {\it {A
  geometric unification of dualities}},  {\em Nucl. Phys.} {\bf B628} (2002)
  3--78, [\href{http://xxx.lanl.gov/abs/hep-th/0110028}{{\tt hep-th/0110028}}].

\bibitem{Wijnholt:2002qz}
M.~Wijnholt, {\it {Large volume perspective on branes at singularities}},  {\em
  Adv. Theor. Math. Phys.} {\bf 7} (2004) 1117--1153,
  [\href{http://xxx.lanl.gov/abs/hep-th/0212021}{{\tt hep-th/0212021}}].

\bibitem{Aspinwall:2004jr}
P.~S. Aspinwall, {\it {D-branes on Calabi-Yau manifolds}},
  \href{http://xxx.lanl.gov/abs/hep-th/0403166}{{\tt hep-th/0403166}}.

\bibitem{Herzog:2004qw}
C.~P. Herzog, {\it {Seiberg duality is an exceptional mutation}},  {\em JHEP}
  {\bf 08} (2004) 064, [\href{http://xxx.lanl.gov/abs/hep-th/0405118}{{\tt
  hep-th/0405118}}].

\bibitem{Aspinwall:2004vm}
P.~S. Aspinwall and I.~V. Melnikov, {\it {D-branes on vanishing del Pezzo
  surfaces}},  {\em JHEP} {\bf 0412} (2004) 042,
  [\href{http://xxx.lanl.gov/abs/hep-th/0405134}{{\tt hep-th/0405134}}].

\bibitem{Herzog:2005sy}
C.~P. Herzog and R.~L. Karp, {\it {Exceptional collections and D-branes probing
  toric singularities}},  {\em JHEP} {\bf 02} (2006) 061,
  [\href{http://xxx.lanl.gov/abs/hep-th/0507175}{{\tt hep-th/0507175}}].

\bibitem{Hanany:2006nm}
A.~Hanany, C.~P. Herzog, and D.~Vegh, {\it {Brane tilings and exceptional
  collections}},  {\em JHEP} {\bf 07} (2006) 001,
  [\href{http://xxx.lanl.gov/abs/hep-th/0602041}{{\tt hep-th/0602041}}].

\bibitem{Freed:1999vc}
D.~S. Freed and E.~Witten, {\it {Anomalies in string theory with D-branes}},
  {\em Asian J.Math} {\bf 3} (1999) 819,
  [\href{http://xxx.lanl.gov/abs/hep-th/9907189}{{\tt hep-th/9907189}}].

\bibitem{Katz:2002gh}
S.~H. Katz and E.~Sharpe, {\it {D-branes, open string vertex operators, and Ext
  groups}},  {\em Adv.Theor.Math.Phys.} {\bf 6} (2003) 979--1030,
  [\href{http://xxx.lanl.gov/abs/hep-th/0208104}{{\tt hep-th/0208104}}].

\bibitem{BKR}
T.~{Bridgeland}, A.~{King}, and M.~{Reid}, {\it {Mukai implies McKay: the McKay
  correspondence as an equivalence of derived categories}},  {\em ArXiv
  Mathematics e-prints} (Aug., 1999)
  [\href{http://xxx.lanl.gov/abs/math/9908}{{\tt math/9908}}].

\bibitem{Hanany:2005ve}
A.~Hanany and K.~D. Kennaway, {\it {Dimer models and toric diagrams}},
  \href{http://xxx.lanl.gov/abs/hep-th/0503149}{{\tt hep-th/0503149}}.

\bibitem{Franco:2005rj}
S.~Franco, A.~Hanany, K.~D. Kennaway, D.~Vegh, and B.~Wecht, {\it {Brane Dimers
  and Quiver Gauge Theories}},  {\em JHEP} {\bf 01} (2006) 096,
  [\href{http://xxx.lanl.gov/abs/hep-th/0504110}{{\tt hep-th/0504110}}].

\bibitem{Franco:2006gc}
S.~Franco and D.~Vegh, {\it {Moduli spaces of gauge theories from dimer models:
  Proof of the correspondence}},  {\em JHEP} {\bf 0611} (2006) 054,
  [\href{http://xxx.lanl.gov/abs/hep-th/0601063}{{\tt hep-th/0601063}}].

\bibitem{Morrison:1994fr}
D.~R. Morrison and M.~Plesser, {\it {Summing the instantons: Quantum cohomology
  and mirror symmetry in toric varieties}},  {\em Nucl.Phys.} {\bf B440} (1995)
  279--354, [\href{http://xxx.lanl.gov/abs/hep-th/9412236}{{\tt
  hep-th/9412236}}].

\bibitem{Sen:1996vd}
A.~Sen, {\it {F theory and orientifolds}},  {\em Nucl.Phys.} {\bf B475} (1996)
  562--578, [\href{http://xxx.lanl.gov/abs/hep-th/9605150}{{\tt
  hep-th/9605150}}].

\bibitem{Sen:1997kw}
A.~Sen, {\it {F theory and the Gimon-Polchinski orientifold}},  {\em
  Nucl.Phys.} {\bf B498} (1997) 135--155,
  [\href{http://xxx.lanl.gov/abs/hep-th/9702061}{{\tt hep-th/9702061}}].

\bibitem{Sen:1997gv}
A.~Sen, {\it {Orientifold limit of F theory vacua}},  {\em Phys.Rev.} {\bf D55}
  (1997) 7345--7349, [\href{http://xxx.lanl.gov/abs/hep-th/9702165}{{\tt
  hep-th/9702165}}].

\bibitem{Collinucci:2008zs}
A.~Collinucci, {\it {New F-theory lifts}},  {\em JHEP} {\bf 0908} (2009) 076,
  [\href{http://xxx.lanl.gov/abs/0812.0175}{{\tt arXiv:0812.0175}}].

\bibitem{Collinucci:2009uh}
A.~Collinucci, {\it {New F-theory lifts. II. Permutation orientifolds and
  enhanced singularities}},  {\em JHEP} {\bf 1004} (2010) 076,
  [\href{http://xxx.lanl.gov/abs/0906.0003}{{\tt arXiv:0906.0003}}].

\bibitem{Blumenhagen:2009up}
R.~Blumenhagen, T.~W. Grimm, B.~Jurke, and T.~Weigand, {\it {F-theory uplifts
  and GUTs}},  {\em JHEP} {\bf 0909} (2009) 053,
  [\href{http://xxx.lanl.gov/abs/0906.0013}{{\tt arXiv:0906.0013}}].

\bibitem{Tate}
J.~Tate, {\it Algorithm for determining the type of a singular fiber in an
  elliptic pencil},  in {\em Modular Functions of One Variable IV} (B.~Birch
  and W.~Kuyk, eds.), vol.~476 of {\em Lecture Notes in Mathematics},
  pp.~33--52.
\newblock Springer Berlin / Heidelberg, 1975.
\newblock 10.1007/BFb0097582.

\bibitem{Bershadsky:1996nh}
M.~Bershadsky, K.~A. Intriligator, S.~Kachru, D.~R. Morrison, V.~Sadov, {\em
  et.~al.}, {\it {Geometric singularities and enhanced gauge symmetries}},
  {\em Nucl.Phys.} {\bf B481} (1996) 215--252,
  [\href{http://xxx.lanl.gov/abs/hep-th/9605200}{{\tt hep-th/9605200}}].

\bibitem{1993alg.geom.10003B}
V.~V. {Batyrev}, {\it {Dual Polyhedra and Mirror Symmetry for Calabi-Yau
  Hypersurfaces in Toric Varieties}},  in {\em eprint arXiv:alg-geom/9310003},
  p.~10003, Oct., 1993.

\bibitem{Candelas:1997eh}
P.~Candelas, E.~Perevalov, and G.~Rajesh, {\it {Toric geometry and enhanced
  gauge symmetry of F theory / heterotic vacua}},  {\em Nucl.Phys.} {\bf B507}
  (1997) 445--474, [\href{http://xxx.lanl.gov/abs/hep-th/9704097}{{\tt
  hep-th/9704097}}].

\bibitem{Cvetic:2005ft}
M.~Cveti\v{c}, H.~Lu, D.~N. Page, and C.~N. Pope, {\it {New Einstein-Sasaki
  spaces in five and higher dimensions}},  {\em Phys. Rev. Lett.} {\bf 95}
  (2005) 071101, [\href{http://xxx.lanl.gov/abs/hep-th/0504225}{{\tt
  hep-th/0504225}}].

\bibitem{Martelli:2005wy}
D.~Martelli and J.~Sparks, {\it {Toric Sasaki-Einstein metrics on S**2 x
  S**3}},  {\em Phys. Lett.} {\bf B621} (2005) 208--212,
  [\href{http://xxx.lanl.gov/abs/hep-th/0505027}{{\tt hep-th/0505027}}].

\bibitem{Gauntlett:2004zh}
J.~P. Gauntlett, D.~Martelli, J.~Sparks, and D.~Waldram, {\it {Supersymmetric
  AdS(5) solutions of M theory}},  {\em Class.Quant.Grav.} {\bf 21} (2004)
  4335--4366, [\href{http://xxx.lanl.gov/abs/hep-th/0402153}{{\tt
  hep-th/0402153}}].

\bibitem{Gauntlett:2004yd}
J.~P. Gauntlett, D.~Martelli, J.~Sparks, and D.~Waldram, {\it {Sasaki-Einstein
  metrics on S**2 x S**3}},  {\em Adv.Theor.Math.Phys.} {\bf 8} (2004)
  711--734, [\href{http://xxx.lanl.gov/abs/hep-th/0403002}{{\tt
  hep-th/0403002}}].

\bibitem{Martelli:2004wu}
D.~Martelli and J.~Sparks, {\it {Toric geometry, Sasaki-Einstein manifolds and
  a new infinite class of AdS/CFT duals}},  {\em Commun.Math.Phys.} {\bf 262}
  (2006) 51--89, [\href{http://xxx.lanl.gov/abs/hep-th/0411238}{{\tt
  hep-th/0411238}}].

\bibitem{GarciaEtxebarria:2007vh}
I.~Garc\'ia-Etxebarria, F.~Saad, and A.~M. Uranga, {\it {Supersymmetry breaking
  metastable vacua in runaway quiver gauge theories}},  {\em JHEP} {\bf 05}
  (2007) 047, [\href{http://xxx.lanl.gov/abs/0704.0166}{{\tt
  arXiv:0704.0166}}].

\bibitem{Kreuzer:2002uu}
M.~Kreuzer and H.~Skarke, {\it {PALP: A Package for analyzing lattice polytopes
  with applications to toric geometry}},  {\em Comput.Phys.Commun.} {\bf 157}
  (2004) 87--106, [\href{http://xxx.lanl.gov/abs/math/0204356}{{\tt
  math/0204356}}].

\bibitem{sage47}
W.~Stein {\em et.~al.}, {\em {S}age {M}athematics {S}oftware ({V}ersion 4.7)}.
\newblock The Sage Development Team, 2011.
\newblock {\tt http://www.sagemath.org}.

\bibitem{toricsage}
V.~Braun and A.~Novoseltsev, ``{Toric Geometry in the Sage CAS}.'' \emph{in
  preparation}.

\bibitem{Aldazabal:2000sk}
G.~Aldazabal, L.~E. Ibanez, and F.~Quevedo, {\it {A $D^-$ brane alternative to
  the MSSM}},  {\em JHEP} {\bf 0002} (2000) 015,
  [\href{http://xxx.lanl.gov/abs/hep-ph/0001083}{{\tt hep-ph/0001083}}].

\bibitem{Cascales:2003wn}
J.~Cascales, M.~Garcia~del Moral, F.~Quevedo, and A.~Uranga, {\it {Realistic
  D-brane models on warped throats: Fluxes, hierarchies and moduli
  stabilization}},  {\em JHEP} {\bf 0402} (2004) 031,
  [\href{http://xxx.lanl.gov/abs/hep-th/0312051}{{\tt hep-th/0312051}}].

\bibitem{Witten:1996bn}
E.~Witten, {\it {Nonperturbative superpotentials in string theory}},  {\em
  Nucl.Phys.} {\bf B474} (1996) 343--360,
  [\href{http://xxx.lanl.gov/abs/hep-th/9604030}{{\tt hep-th/9604030}}].

\bibitem{Kachru:2003aw}
S.~Kachru, R.~Kallosh, A.~D. Linde, and S.~P. Trivedi, {\it {De Sitter vacua in
  string theory}},  {\em Phys.Rev.} {\bf D68} (2003) 046005,
  [\href{http://xxx.lanl.gov/abs/hep-th/0301240}{{\tt hep-th/0301240}}].

\bibitem{Kallosh:2005gs}
R.~Kallosh, A.-K. Kashani-Poor, and A.~Tomasiello, {\it {Counting fermionic
  zero modes on M5 with fluxes}},  {\em JHEP} {\bf 0506} (2005) 069,
  [\href{http://xxx.lanl.gov/abs/hep-th/0503138}{{\tt hep-th/0503138}}].

\bibitem{Bergshoeff:2005yp}
E.~Bergshoeff, R.~Kallosh, A.-K. Kashani-Poor, D.~Sorokin, and A.~Tomasiello,
  {\it {An Index for the Dirac operator on D3 branes with background fluxes}},
  {\em JHEP} {\bf 0510} (2005) 102,
  [\href{http://xxx.lanl.gov/abs/hep-th/0507069}{{\tt hep-th/0507069}}].

\bibitem{Blumenhagen:2007sm}
R.~Blumenhagen, S.~Moster, and E.~Plauschinn, {\it {Moduli Stabilisation versus
  Chirality for MSSM like Type IIB Orientifolds}},  {\em JHEP} {\bf 0801}
  (2008) 058, [\href{http://xxx.lanl.gov/abs/0711.3389}{{\tt
  arXiv:0711.3389}}].

\bibitem{Uranga:2008nh}
A.~M. Uranga, {\it {D-brane instantons and the effective field theory of flux
  compactifications}},  {\em JHEP} {\bf 0901} (2009) 048,
  [\href{http://xxx.lanl.gov/abs/0808.2918}{{\tt arXiv:0808.2918}}].

\bibitem{Blumenhagen:2009gk}
R.~Blumenhagen, J.~Conlon, S.~Krippendorf, S.~Moster, and F.~Quevedo, {\it
  {SUSY Breaking in Local String/F-Theory Models}},  {\em JHEP} {\bf 0909}
  (2009) 007, [\href{http://xxx.lanl.gov/abs/0906.3297}{{\tt
  arXiv:0906.3297}}].

\bibitem{Bobkov:2010rf}
K.~Bobkov, V.~Braun, P.~Kumar, and S.~Raby, {\it {Stabilizing All Kahler Moduli
  in Type IIB Orientifolds}},  {\em JHEP} {\bf 1012} (2010) 056,
  [\href{http://xxx.lanl.gov/abs/1003.1982}{{\tt arXiv:1003.1982}}].

\bibitem{Grimm:2011dj}
T.~W. Grimm, M.~Kerstan, E.~Palti, and T.~Weigand, {\it {On Fluxed Instantons
  and Moduli Stabilisation in IIB Orientifolds and F-theory}},  {\em Phys.Rev.}
  {\bf D84} (2011) 066001, [\href{http://xxx.lanl.gov/abs/1105.3193}{{\tt
  arXiv:1105.3193}}].

\bibitem{Cicoli:2011it}
M.~Cicoli, M.~Kreuzer, and C.~Mayrhofer, {\it {Toric K3-Fibred Calabi-Yau
  Manifolds with del Pezzo Divisors for String Compactifications}},
  \href{http://xxx.lanl.gov/abs/1107.0383}{{\tt arXiv:1107.0383}}.

\bibitem{Cicoli:2011qg}
M.~Cicoli, C.~Mayrhofer, and R.~Valandro, {\it {Moduli Stabilisation for Chiral
  Global Models}},  \href{http://xxx.lanl.gov/abs/1110.3333}{{\tt
  arXiv:1110.3333}}.

\bibitem{Kofman:2004yc}
L.~Kofman, A.~D. Linde, X.~Liu, A.~Maloney, L.~McAllister, {\em et.~al.}, {\it
  {Beauty is attractive: Moduli trapping at enhanced symmetry points}},  {\em
  JHEP} {\bf 0405} (2004) 030,
  [\href{http://xxx.lanl.gov/abs/hep-th/0403001}{{\tt hep-th/0403001}}].

\bibitem{Blumenhagen:2009yv}
R.~Blumenhagen, T.~W. Grimm, B.~Jurke, and T.~Weigand, {\it {Global F-theory
  GUTs}},  {\em Nucl.Phys.} {\bf B829} (2010) 325--369,
  [\href{http://xxx.lanl.gov/abs/0908.1784}{{\tt arXiv:0908.1784}}].

\bibitem{Grimm:2009yu}
T.~W. Grimm, S.~Krause, and T.~Weigand, {\it {F-Theory GUT Vacua on Compact
  Calabi-Yau Fourfolds}},  {\em JHEP} {\bf 1007} (2010) 037,
  [\href{http://xxx.lanl.gov/abs/0912.3524}{{\tt arXiv:0912.3524}}].

\bibitem{Chen:2010ts}
C.-M. Chen, J.~Knapp, M.~Kreuzer, and C.~Mayrhofer, {\it {Global SO(10)
  F-theory GUTs}},  {\em JHEP} {\bf 1010} (2010) 057,
  [\href{http://xxx.lanl.gov/abs/1005.5735}{{\tt arXiv:1005.5735}}].

\bibitem{Knapp:2011wk}
J.~Knapp, M.~Kreuzer, C.~Mayrhofer, and N.-O. Walliser, {\it {Toric
  Construction of Global F-Theory GUTs}},  {\em JHEP} {\bf 1103} (2011) 138,
  [\href{http://xxx.lanl.gov/abs/1101.4908}{{\tt arXiv:1101.4908}}].

\bibitem{Knapp:2011ip}
J.~Knapp and M.~Kreuzer, {\it {Toric Methods in F-theory Model Building}},
  {\em Adv.High Energy Phys.} {\bf 2011} (2011) 513436,
  [\href{http://xxx.lanl.gov/abs/1103.3358}{{\tt arXiv:1103.3358}}].

\bibitem{Kreuzer:2001fu}
M.~Kreuzer, E.~Riegler, and D.~A. Sahakyan, {\it {Toric complete intersections
  and weighted projective space}},  {\em J.Geom.Phys.} {\bf 46} (2003)
  159--173, [\href{http://xxx.lanl.gov/abs/math/0103214}{{\tt math/0103214}}].

\bibitem{Blaszczyk:2011hs}
M.~Blaszczyk, S.~Groot~Nibbelink, and F.~Ruehle, {\it {Gauged Linear Sigma
  Models for toroidal orbifold resolutions}},
  \href{http://xxx.lanl.gov/abs/1111.5852}{{\tt arXiv:1111.5852}}.

\end{thebibliography}\endgroup

\end{document}